\documentclass[twocolumn]{aastex63}
\usepackage{mathtools}
\usepackage{graphicx}
\usepackage{natbib}
\usepackage{amsmath}
\usepackage{booktabs}
\usepackage{listings}
\newcommand\teff{$T_{eff}$}
\newcommand\logg{$\log g$}
\newcommand\logL{$\log L$}
\newcommand\msun{M$_\odot$}
\graphicspath{ {./images/} }

\begin{document}

\author{Richard Olney}
\affil{Dept. of Computer Science, Western Washington University, 516 High St., Bellingham, WA 98225-9165, USA}

\author[0000-0002-5365-1267]{Marina Kounkel}
\affil{Dept. of Physics and Astronomy, Western Washington University, 516 High St., Bellingham, WA 98225-9164, USA}

\author{Chad Schillinger}
\affil{Dept. of Computer Science, Western Washington University, 516 High St., Bellingham, WA 98225-9165, USA}

\author[0000-0002-0748-9115]{Matthew T. Scoggins}
\affil{Dept. of Physics and Astronomy, Western Washington University, 516 High St., Bellingham, WA 98225-9164, USA}

\author{Yichuan Yin}
\affil{Dept. of Computer Science, Western Washington University, 516 High St., Bellingham, WA 98225-9165, USA}

\author[0000-0002-0716-947X]{Erin Howard}
\affil{Dept. of Physics and Astronomy, Western Washington University, 516 High St., Bellingham, WA 98225-9164, USA}

\author[0000-0001-6914-7797]{K. R. Covey}
\affil{Dept. of Physics and Astronomy, Western Washington University, 516 High St., Bellingham, WA 98225-9164, USA}

\author[0000-0002-5537-008X]{Brian Hutchinson}
\affil{Dept. of Computer Science, Western Washington University, 516 High St., Bellingham, WA 98225-9165, USA}
\affil{Computing and Analytics Division, Pacific Northwest National Laboratory, 902 Battelle Blvd, Richland, WA 99354-1793, USA}

\author[0000-0002-3481-9052]{Keivan G. Stassun}
\affil{Department of Physics and Astronomy, Vanderbilt University, VU Station 1807, Nashville, TN 37235, USA}

\title{APOGEE Net: Improving the derived spectral parameters for young stars through deep learning}

\begin{abstract}

Machine learning allows efficient extraction of physical properties from stellar spectra that have been obtained by large surveys. The viability of ML approaches has been demonstrated for spectra covering a variety of wavelengths and spectral resolutions, but most often for main sequence or evolved stars, where reliable synthetic spectra provide labels and data for training. Spectral models of young stellar objects (YSOs) and low mass main sequence (MS) stars are less well-matched to their empirical counterparts, however, posing barriers to previous approaches to classify spectra of such stars.
In this work we generate labels for YSOs and low mass MS stars through their photometry. We then use these labels to train a deep convolutional neural network to predict \logg, \teff\ and Fe/H for stars with APOGEE spectra in the DR14 dataset.
This ``APOGEE Net'' has produced reliable predictions of \logg\ for YSOs, with uncertainties of within 0.1 dex and a good agreement with the structure indicated by pre-main sequence evolutionary tracks, \added{and correlate well with independently derived stellar radii}. These values will be useful for studying pre-main sequence stellar populations to accurately diagnose membership and ages.
\end{abstract}

\section{Introduction}

Spectroscopy is a powerful technique for measuring stellar properties, and in recent years, large surveys such as SDSS APOGEE \citep{Abolfathi2018}, RAVE \citep{kunder2017}, and GALAH \citep{buder2018} have observed 10$^{5-6}$ stars each. This necessitates an effective method to uniformly and efficiently process these spectra to extract the stellar properties (e.g., effective temperature (\teff), surface gravity (\logg), and metallicity (Fe/H). 

\replaced{Traditional}{A common} approach to spectral analysis rely on comparisons between the target spectrum and a grid of spectral standards (that may be difficult to come by for specific source types or wavelength range) or synthetic templates (that may systematically differ from the real data). Synthetic templates typically offer more regular coverage of parameter space, enabling individual targets' parameters to be inferred precisely by using a higher order function to interpolate the goodness of fit parameters between the points for which the grid is defined. Many surveys adopt this approach in constructing their stellar parameter pipelines \citep[e.g., ][]{garcia-perez2016}, but the process is computationally intensive, particularly when fitting multiple parameters simultaneously, as well as determining the corresponding uncertainties.  

Computational efficiency aside, the reliability of parameters determined via direct model fitting can also vary strongly as a function of target type, with young stellar objects (YSOs) representing a particularly challenging target class \citep[e.g.,][]{doppmann2005}. Spectral fits can return reasonably accurate estimates of \teff\ for YSOs, but \logg\ has proven more difficult to accurately constrain. This parameter is particularly valuable for YSOs, as it serves as a proxy for stellar age, and is therefore of great value for calibrating pre-main sequence evolutionary models or inferring star formation histories within a given star forming complex. The APOGEE survey has conducted extensive surveys of several nearby star forming regions, providing a valuable opportunity to infer \logg\ and age constraints for large samples of YSOs, but those constraints have been difficult to achieve in practice.  For APOGEE data in particular, obtaining reliable \logg\ values for dwarf or pre-main sequence stars have been challenging: neither APOGEE's primary stellar parameter pipeline, ASPCAP \citep{holtzman2015,garcia-perez2016}, nor the community-provided Payne model-fitting framework \citep{ting2019} released \logg\ estimates for dwarf stars, due to the presence of clear systematic errors in uncalibrated values and the lack of a densely sample comparison sample for deriving calibration relations. The IN-SYNC pipeline \citep{cottaar2014,kounkel2018a}, developed and optimized for YSO spectra, provided \logg\ values whose age \replaced{gradient}{dependence} agrees well with physical models \added{(i.e., older populations have higher \logg\ than the younger ones, see Section \ref{sec:results} for discussion)}, but the precise values also show unphysical systematics, likely due to mismatches between the empirical and theoretical spectra. 

Data driven analysis pipelines eliminate errors due to model mismatches, by training prediction systems with empirical spectra for stars with well determined stellar parameters. One data-driven method that has demonstrated considerable success in assigning labels to APOGEE spectra is The Cannon \citep{ness2015}, which uses a reference sample of APOGEE spectra to train a \added{generative model} that can then be used to infer stellar parameters for any object with an APOGEE spectrum. Using the full wavelength information within the spectrum, and training \added{the parameters of the generative model} on empirical standards \added{(i.e., the high-quality APOGEE spectra with known stellar labels} that eliminate the potential for model-data mis-fitting, the Cannon is able to provide parameters of comparable quality to ASPCAP's for APOGEE spectra with SNR $\geq$ 25.  As informed by a set of 60 dwarf calibrators in the Hyades, the 2015 Cannon results also included realistic \logg\ values for the upper main sequence, but became incomplete for dwarfs with T$_{eff} <$ 4500 K, similar to the limit reached by the Payne results. While data driven models offer important performance and calibration advantages, they are unable to overcome the limits of their training sets. 

Neural networks offer a promising data driven method for inferring accurate stellar parameters from spectra, and with a potentially greater flexibility in inference methods than offered by the polynomial formalism adopted in the Cannon. Neural networks are a common machine learning model, in which multiple non-linear transformations of the input features are performed before assigning an output classification. A number of studies have demonstrated the ability of neural networks to classify stellar spectra \added{and derive stellar parameters and abundances}, e.g., \citet{Bailer-Jones1997, Bailer-Jones2000, Bazarghan2008,fabbro2018, Sharma2019,Leung2019}. Neural networks also offer important efficiencies for processing large datasets. Direct fitting can only consider a single spectrum at a time, redoing the same operation regardless of how similar two target spectra may be. On the other hand, neural networks can process in excess of $10^6$ observations in under an hour. As with all data-driven methods, however, the network must be trained on a reliable reference sample whose parameters span the full range of interest, making the construction of a label-set as important as the construction of the network itself. 

In this paper we aim to train a deep neural network to accurately classify APOGEE DR14 spectra of dwarfs, giants, and pre-main sequence stars.  To realize this goal, we first supplement the stellar labels provided by the Payne with a set of labels inferred from Gaia, 2MASS and Pan-STARRS photometry and astrometry for YSOs and M dwarfs with APOGEE spectra. In Section \ref{sec:Obs} we describe the data and the procedure used to generate the labels. In Section \ref{sec:SpectralModel} we then construct a convolutional neural network (CNN) that predicts parameters from the APOGEE spectra, which we refer to as the APOGEE Net. In Section \ref{sec:results} we highlight some analysis that could be derived from these spectral properties. Finally, we summarize our results and discuss the implications in Section \ref{sec:conclusions}.

\section{Data \label{sec:Obs}}

\subsection{APOGEE}

Apache Point Observatory Galactic Evolution Experiment (APOGEE) is a high resolution (R$\sim$22,500) near infrared (1.51\added{--}1.7 $\mu$m spectrograph mounted on a \replaced{2.5 Sloan Digital Sky Survey (SDSS) telescope}{Sloan Foundation 2.5m telescope} \citep{gunn2006,wilson2010,blanton2017,majewski2017}. APOGEE is capable of observing up to 300 targets in a field of view \added{with the radius of} 1.5$^\circ$. Over the years, the survey and its targeting priorities has evolved. The primary objective of both APOGEE-1 and APOGEE-2 has been to observe red giants to trace the dynamical and the chemical patterns of the Galaxy \citep[e.g.,][]{hayden2015,bovy2016,anders2017,zasowski2017}. However, \added{among other programs,} it has also observed a number of star-forming regions, including Orion Complex \citep{da-rio2016,da-rio2017,kounkel2018a}, NGC 1333 \citep{foster2015}, IC 348 \citep{cottaar2015}, NGC 2264, as well as several more evolved clusters, and some of the nearby main sequence stars.

As of the public Data Release 14 \citep{Abolfathi2018}, over 263,000 stars in the bulge, disk, and halo have been observed. We restrict the current analysis only to these sources.

The sources in the catalog typically have been observed for multiple epochs. Therefore, the data are stored in two formats: `apVisit', which contains the raw spectrum at a particular epoch, and `apStar', in which the Doppler shift has been removed from the epochs, placing them all in a common rest frame with identical wavelength solution across all sources, and multiple visits for the same source are combined into one, increasing the resulting signal-to-noise \citep{Nidever2015}.




\subsection{Payne Labels}

\citet{ting2019} used their newly developed spectral interpolator, along with a new grid of Kurucz spectral models calculated with an improved line list \citep{cargile2019}, to identify best-fit models \& infer revised stellar parameters for 222,707 spectra within the APOGEE DR14 dataset. The labels inferred from this interpolator, dubbed `The Payne' in honor of Cecilia Payne-Gaposchkin's seminal work in physically-based stellar models, are comparable to the calibrated parameters provided in DR14 for stars along the Red Giant Branch. Moreover, the Payne provides realistic \teff\ and \logg\ labels for warmer (\teff$>4250$ K) main sequence stars, for which calibrated parameters are not available in the DR14 dataset. Typical uncertainties, both random and systematic, are $\sim$100 K in T$_{eff}$, $\sim$0.1 dex in \logg, and $\sim$0.03 dex in abundance space \citep{ting2019}.

However, the Payne has non-physical correlations between \logg\ and Fe/H towards cooler dwarfs. Indeed, these systematics dominated the information content of the Payne labels to such a degree that dwarfs stars with \teff $<$ 4000 K were intentionally removed from the Payne outputs, to avoid potential mis-interpretation of the spurious correlation between the inferred log g and [Fe/H] values.



In training APOGEE Net, we adopt labels \added{for \teff, \logg, and Fe/H} from the Payne's outputs for stars with \teff $>$ 4000 K or \logg $<$ 3.5\added{.} \deleted{, and} \added{As the Payne does not yet produce reliable labels for stars with lower \teff\ and higher \logg\ than these limits, in the next section} we \added{use empirically calibrated photometric relationships} to generate new labels for 4,480 low mass main sequence stars and 2446 YSOs \deleted{through their photometry}.

\subsection{Deriving Alternate Labels for pre-main sequence \& low mass main sequence stars \label{sec:ourLabels}}
\subsubsection{MS stars}

\added{To derive labels of the low mass main sequence stars, we rely on various empirical photometric relationships.} We begin by identifying bona fide lower main sequence stars with potentially erroneous Payne labels, using 2MASS \& Gaia DR2 photometry, as well as Gaia's parallax measurements. We selected low mass MS stars by requiring 2MASS photometry of 0.7$<J-K<$1.05, 3.37$<g-K<$8.46, and 5$<M_K<$10. To ensure accurate $M_K$, we further restricted the sample to only those sources in which the \textit{Gaia} DR2 measured $\pi/\sigma_\pi>3$; this removed any giants with erroneous M$_K$ values due to spurious, low-quality parallaxes. Since the empirical relations used in this section were calibrated for main sequence stars, we flag all spectra within APOGEE fields  covering known star-forming regions and young clusters -- namely the Orion Complex, Perseus clusters, NGC 2264, and the Pleiades -- for a separate label generation procedure, which is described in the next section.

 We infer metallicities for these stars using the relation by \citet{Hejazi2015}
\begin{multline}
[Fe/H] = A_1 + A_2(g-K) + A_3(J-K) + A_4(g-K)^2 + \\ A_5(J-K)^2 + A_6(g-K)(J-K)
\end{multline}
in which
\begin{align*}
A = [-14.259, 0.0519, 29.5926,\\ -0.0529, -17.6762, 0.7032]
\end{align*}
This calibration is \added{derived from a sample of 73 M dwarfs with high quality SDSS+2MASS photometry and robust metallicity estimates inferred from high resolution spectroscopy of a common proper motion companion (18) or from moderate resolution spectra of the M dwarf itself (53).  The derived calibration is} applicable for stars between K6 and M6.5, with $-0.73 \leq [Fe/H] \leq +0.3$ dex, $3.3 \leq g - K \leq 8.46$ and $0.71 \leq J - K \leq 1.01$.  By design, these limits nearly exactly match the color-mag cuts used to select candidates for our alternate, pre-main sequence focused label generation procedure. We do allow a modestly broader range of $g-K$ colors, by a few hundredths of a magnitude in each direction, as the extrapolated metallicities will nontheless likely be more accurate than the Payne parameters, which clearly suffer from systematic errors in this space.

\deleted{To calculate mass, we use the Mass-Luminosity Relation from \citet{Benedict2016}
$M = B_0 + B_1(M_K - x_0) + B_2(M_K - x_0)^2\\ + B_3(M_K -x_0)^3 + B_4(M_K - x_0)$
with the magnitude offset $x_0 = 7.5$ being the absolute magnitude offset and $B_i$ given by
B = [0.2311, -0.1352, 0.0400, 0.0038, -0.0032]
which holds for $M_V \leq 19$ and $M_K \leq 10$.}

To estimate $T_{eff}$, we use the relation \added{derived} from \added{a sample of 183 M dwarfs with accurate ($\sigma_{\pi} <$ 5\%) pre-Gaia parallaxes and reliable spectrophotometric data by} \citet{Mann2015}
\begin{multline}
T_{eff} = B_0 +B_1(r-J) +B_2(r-J)^2 +B_3(r-J)^3 + \\ B_4(r-J)^4 +B_5(J-H) + B_6(J-H)^2
\end{multline}

with

\begin{align*}
B = [2.151, - 1.092, 0.3767, - 0.06292, \\ 0.003950, 0.1697, - 0.03106]
\end{align*} 
which hold for $4.6<M_K<9.8$, 2700$<$\teff$<$4100 K,  and -0.6 $<$[Fe/H]$<$ 0.5.


Finally, to estimate \logg\, we use the relation \added{calibrated} from \added{the analysis of Y band spectra of 29 M dwarfs} \citet{Veyette2017}
\begin{multline}
\log g = 7.912 -0.1880\times[Fe/H] - 1.334\times10^{-3}\times T_{eff} +\\ 1.313\times10^{-7}\times T_{eff}^2
\end{multline} 
which hold for 3200$<$\teff$<$4100 K and -0.7 $<$ [Fe/H] $<$ 0.3.


\subsection{YSOs}

\begin{figure*}
\epsscale{1.0}
 \centering
		\gridline{
  \fig{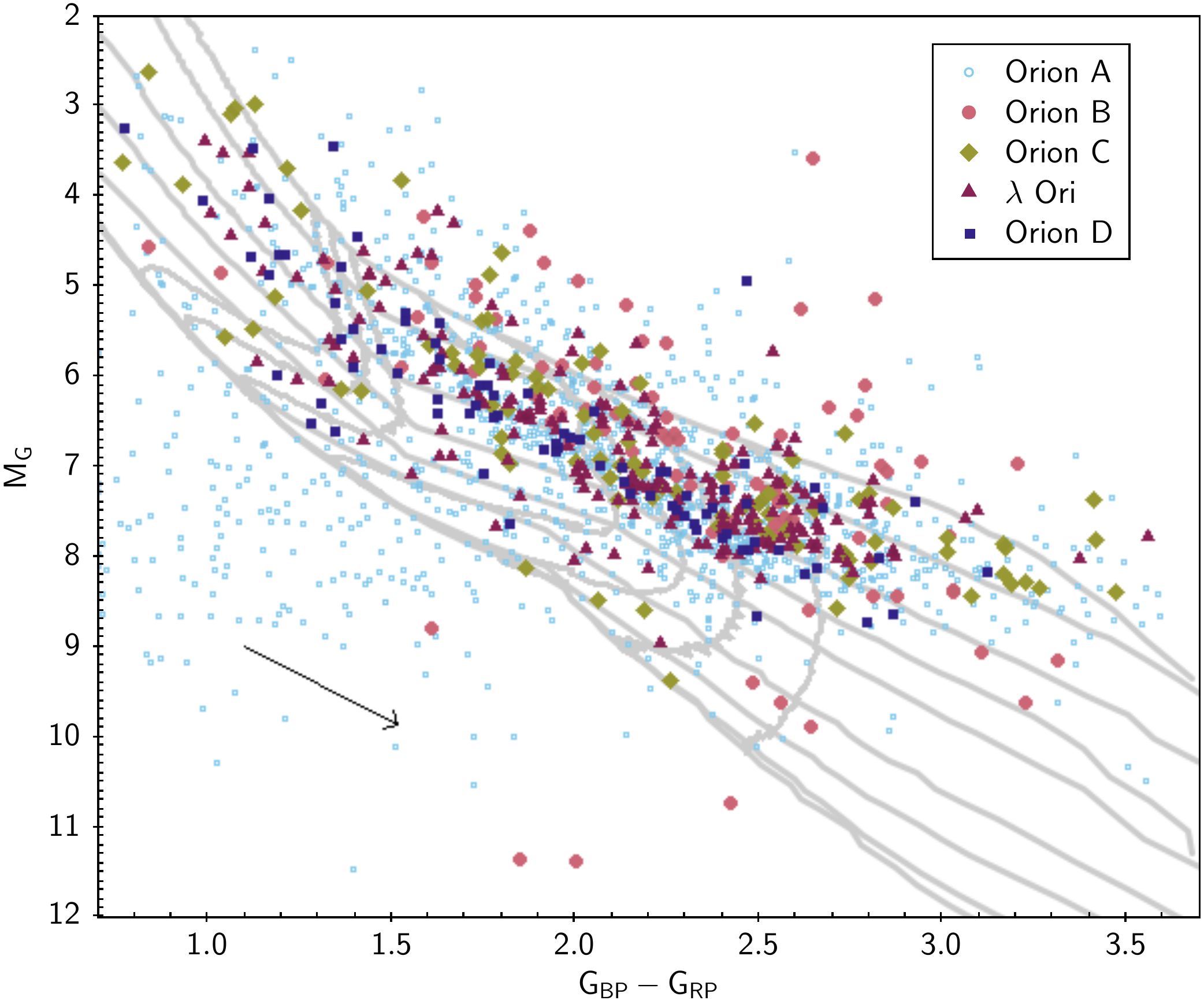}{0.5\textwidth}{}
   \fig{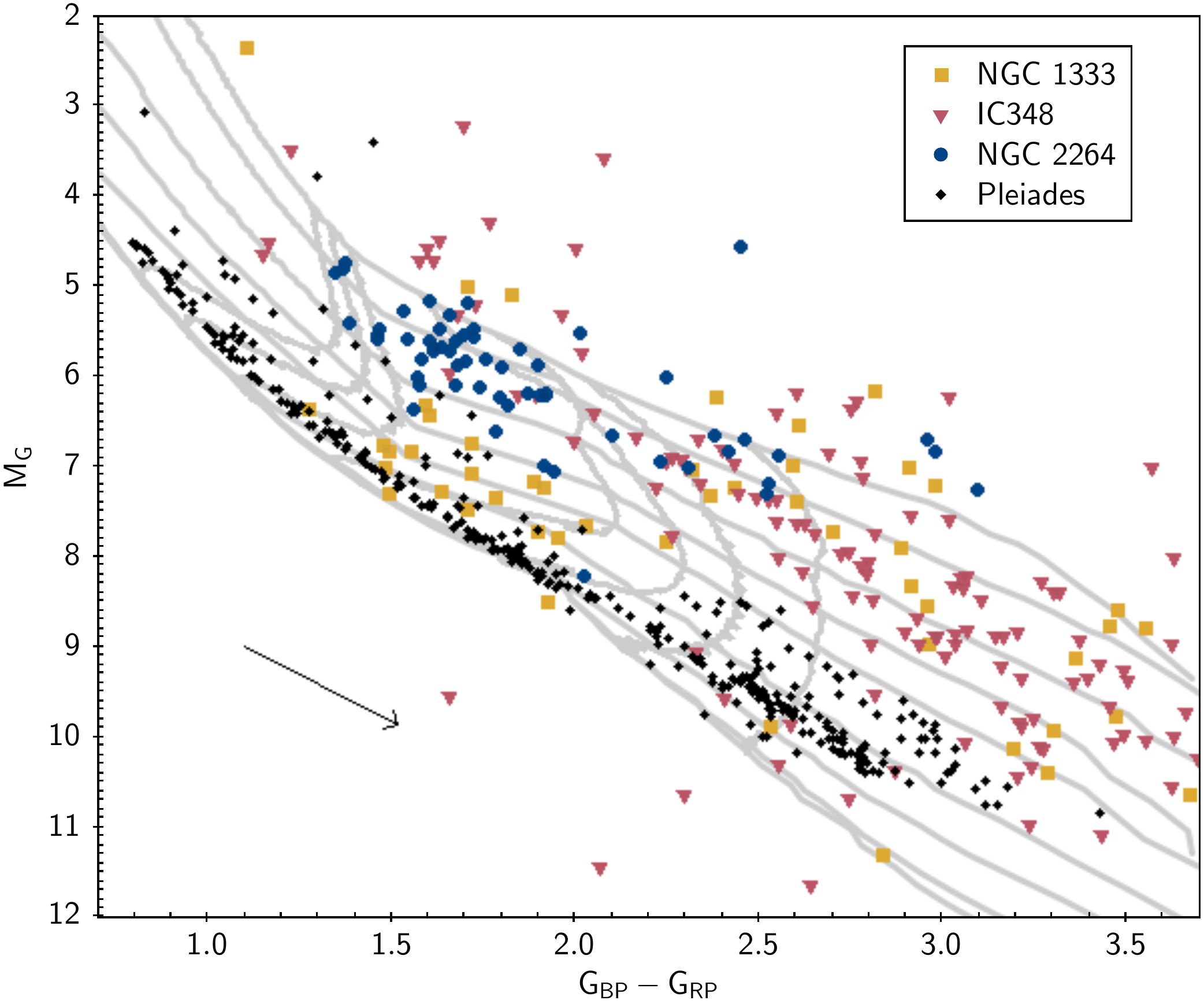}{0.5\textwidth}{}
 }\vspace{-1 cm}
		\gridline{
  \fig{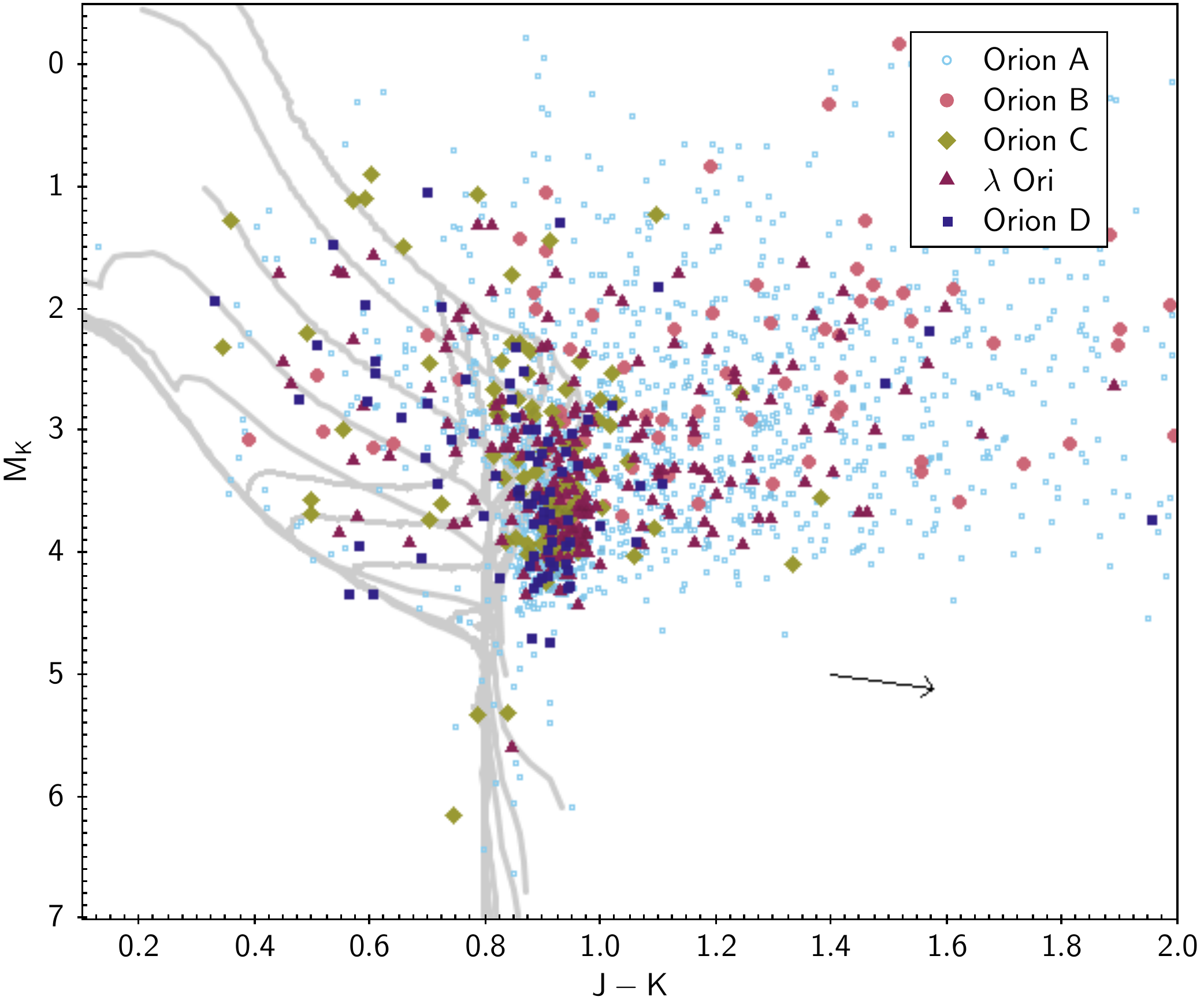}{0.5\textwidth}{}
   \fig{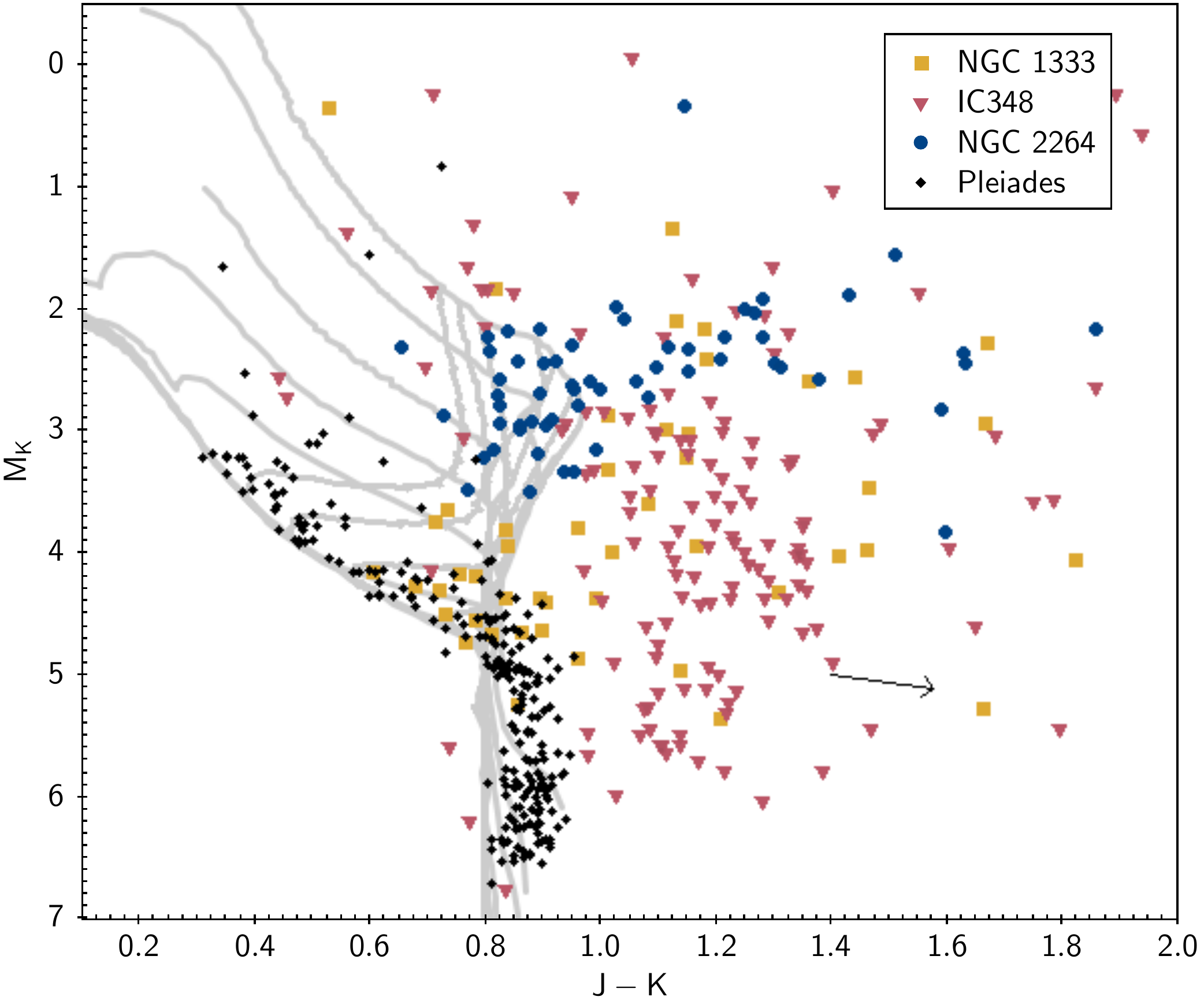}{0.5\textwidth}{}
 }\vspace{-1 cm}
\caption{\added{Color-magnitude diagrams constructed using \textit{Gaia} and 2MASS photometry. for the YSOs across the various regions The grey lines show the isochrones at ages of 1, 2, 5, 10, 20, 50, 100, 200, and 300 Myr, and 8.5 dex, as well as the evolutionary tracks for 0.4, 0.5, 0.6, 0.7, 0.8, 0.9, and 1 \msun\ stars from the PARSEC isochrones \citep{marigo2017}. The black arrow shows the extinction vector corresponding to 1 $A_V$.}
\label{fig:photometric}}
\end{figure*}

\begin{figure}
\epsscale{1}
\plotone{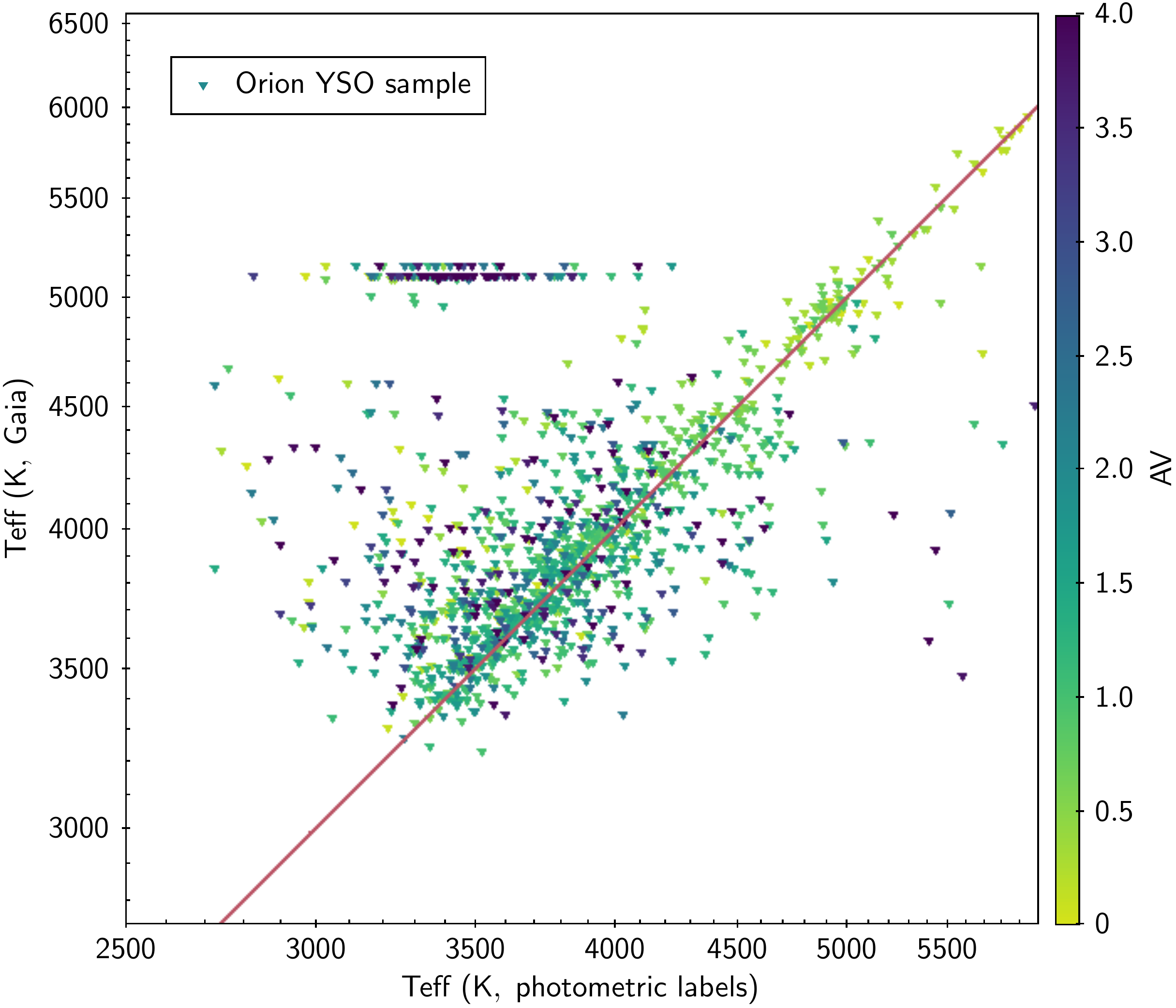}
\caption{Comparison of the \teff\ derived from the photometric CNN versus \teff\ from \citet{gaia-collaboration2018} for the YSOs.  \added{Each point is color coded with the $A_V$ from \citet{kounkel2018a}.}
\label{fig:compteff}}
\end{figure}

The APOGEE fields designed to target star-forming regions nonetheless include sources other than YSOs. To calculate parameters for a pure YSO sample, we restricted the sample to likely cluster members tabulated by \citet{Kounkel2019}, with spectra publicly released in DR14 \citep{Abolfathi2018}.

We attempted multiple approaches to interpolate each YSO's photometry onto a grid of isochrones to \deleted{measure} \added{generate initial} \teff \ and \logg\ \added{labels (Figure \ref{fig:photometric})}. We considered various isochrones for this purpose, including those from \citet{Baraffe2015}, PARSEC \citep{marigo2017}, and MIST \citep{choi2016}. We assessed combinations of various photometric bands from 2MASS, WISE, and \textit{Gaia}, and explored the ability to assign reliable stellar labels using standard isochrone fits (as implemented via the 'isochrones' python package, developed by Timothy Morton). 

\added{Ultimately, we were unable to achieve astrophysically realistic parameters for the majority of YSOs in our sample using standard interpolation methods.  This is due to several factors that reinforce one another:

\begin{itemize}

\item{\textbf{Extinction:} Many of the YSOs in our sample possess non-trivial extinctions due to the interstellar or circumstellar dust that is typically found in star forming regions. A YSO's photometry can be plausibly explained by a degenerate set of parameters that lie along an extinction vector in a single color-magnitude space.  This degeneracy can typically be broken, however, by simultaneously fitting multi-band photometry to leverage differences in the slopes of the isochrones and extinction vectors in different projections of color-magnitude space.  The isochrones package provides a multi-band fitting capability, which we utilized to infer a maximum likelihood A$_V$ value along with each YSO's other stellar parameters (i.e., mass \& age). The quality of these estimates, and the resolution of degeneracies between extinction and stellar properties, depend on the underlying agreement between the YSO's observed colors and those expected from reddened stellar models, the factor we consider next.}

\item{\textbf{Systematic offsets between isochrones \& empirical data:} As shown in Fig. \ref{fig:photometric}, many YSOs lie outside the bounds of an unreddened grid of isochrones.  This offset is most prominent in J-K, and larger for members of the youngest clusters (i.e., $\Delta_{(J-K)} \sim$ 0.3 mag. in NGC 1333, IC348, and NGC 2264), but still present in $\sim$120 Myr old stars (i.e., $\Delta_{(J-K)} \sim$0.05 mag for Pleiades members).  

In principle, extinction may contribute to this offset, but cannot explain it entirely. This is best illustrated by the J-K offset between the isochrones and the low-mass stars in the Pleiades, as the extinction measured towards the Pleiades is far too low to produce the necessary E(J-K) excess \citep[A$_V = 0.12$; E(J-K)$ = 0.178$]{Stauffer1998, Schlafly2011}.  This previous work has shown that these stars have offsets with respect to synthetic colors that are inconsistent with standard reddening vectors \citep{Bell2012, Covey2016}.  

Even worse, the youngest YSOs possess intrinsic infrared excesses due to emission by warm circumstellar dust, which are not included in standard isochrones and are inconsistent with color offsets predicted by standard extinction vectors. These astrophysically meaningful systematic offsets between isochrones and empirical photometry are not gracefully handled by standard isochrone fitting methods, which attempt to reproduce the offset with secondary model parameters (i.e., extinction), but self-consistent solutions are difficult to achieve when the offset is inconsistent with standard extinction curves and significantly larger than the YSOs photometric errors. }

\item{ \textbf{Error Normalization \& Grid Edge Effects:} The isochrones package fundamentally performs a chi-squared minimization of the residuals between each YSO's photometry and that predicted for synthetically reddened stellar models across a range of physical parameters.  In this process, the residuals in each band are normalized by the associated photometric errors to identify the global best fit solution.  This process works well for sources whose photometry is consistent with standard reddened stellar spectra, and where the model-data agreement is primarily disrupted by random photometric noise.  As outlined above, however, many of the YSOs in our sample do not meet this criteria: the quality of their fits are fundamentally limited by the systematic offsets with respect to the data, a factor which is decoupled from the precision of their photometry. In this case, the figure of merit drives the fit to areas of parameter space that minimize systematic model offsets, particularly in the photometric bands with the highest average precision.  This often drives the fit to select models with unrealistic combinations of \teff\, \logg\ and A$_V$, and at least one parameter (most often A$_V$) hitting the limits of the parameter space included in the search.   }

\end{itemize}

Ultimately, we were unable to overcome these limitations with the standard isochrone fitting approach. Our output labels, for example, typically featured a strong, but non-physical, correlation between \teff\ and A$_V$, and these effects did not substantially diminish when we eliminated the bands with the strongest model-data systematic offsets (i.e., the 2MASS photometry) or restricted the sample to sources with lower intrinsic reddening (i.e, Class III/Weak T Tauri stars vs. Class II/Classical T Tauri Stars). }

\deleted{Fundamentally, however, without an inclusion of spectroscopically measured parameters, none of them could produce an optimal distribution of the parameters for all sources. In part it was due to the difficulty of measuring the extinction, due to the photometric uncertainties that were too small, and possibly due to a slight offset between the various bands (especially those from different surveys) and the isochrones that prevented a convergence of \teff\ and \logg\ to a range of realistic parameters -- e.g., many sources would have parameters at the edge of the grid, or their parameters would be strongly correlated, as the fitting code would unsuccessfully try to correct any deviation between the data and the isochrones through extinction}

As an alternative \added{to traditional isochrone fitting, and to more gracefully fit YSOs whose photometry lies beyond the edges of the standard isochronal grid, we used a neural network trained on synthetic photometry to infer labels for YSOs across a broad range of parameter space. It proved to be less biased to systematic effects of the individual bands, rather, through assigning different weight to the inputs and treating them as a whole, it resulted in more physically realistic solutions. Specifically}, we constructed a CNN with three convolutional layers using max pooling and two fully connected layers. This network was trained on parameters from synthetic stars drawn from the PARSEC isochrones \added{(they provided a better convergence compared to other isochrones we tested)}. The synthetic stars were generated using a uniform distribution of stellar masses from 0.08 to 3 \msun, ages from 1 to 100 Myr, extinction from 0 to 20 $A_V$, and distance from 50 to 1000 pc. \added{Only isochrones with } Fe/H \replaced{was set to 0 for all of them}{=0 were used}, which is consistent with the nearby star-forming regions \citep[e.g.,][]{Dorazi2009,Dorazi2011}. The empirical parameters the network used to evaluate the labels included 9 photometric bands, stellar radii $r_*$, stellar luminosities \logL, and the distance. In cases where the photometry in a particular band was too faint to be reliably detected in the real data, it was set to the limiting magnitude ($G<19$, $G_{BP}<20.5$, $G_{RP}<17.5$, $J<17$, $H<16$, $K<16$, $W1<16$, $W2<16$, $W3<13.5$), with only $G$ band being required. The additional two parameters $r_*$ and \logL, were drawn from the isochrone but modeled after those reported by \textit{Gaia} DR2, in that they were only reported if $r_*>0.5 r_\odot$, \logL$>-1.54$, and $M_G<10$. Similarly with the photometric bands they were set to the limiting cases if they were not detectable. No scatter due to uncertainties was applied to the synthetic parameters. 

The CNN was trained on $\sim$42,000 synthetic stars to predict ages, masses, $A_V$, \teff, and \logg\ based on these input parameters. Applied on a separate synthetic sample that was generated similarly to the one on which it has been trained, not accounting for any uncertainties or systematic offsets, the neural network could recover \logg\ with a precision of 0.01 dex, and $\log$ \teff\ with a precision of 0.003 dex.

When applied to the real data, the CNN produced \logg\ that are consistent with the isochrones corresponding to the typically accepted ages of the individual clusters covered by APOGEE (Section \ref{sec:results}). Additionally, it could generally recover \teff\ estimated by \textit{Gaia} DR2 \added{with no evidence for bias in the quality of fit as a function of independently inferred extinction values} (Figure \ref{fig:compteff}).

\section{APOGEE Net \label{sec:SpectralModel}}

The APOGEE Net was designed to take in the raw spectra, and to return the predictions on \teff, \logg, and Fe/H. The sources, labels of which were determined in Section \ref{sec:Obs}, were split into three different subsets: a training set on which the model is trained, a held out development set which is used to evaluate the model's generalization performance during training and to tune hyperparameters, and finally a held out test set which is used to evaluate the model's performance once it has completed training. For the Payne catalog, the split between the train, dev, and test sets was 80/10/10\%. Due to a smaller number of sources in other categories, to have a sufficient number of sources in the test set, M dwarf and YSO catalogs were split 60-20-20\%.
\\

\subsection{Feature and Target Preprocessing}
\replaced{While we experimented with using various normalization techniques on the input flux, training on the raw flux from the `apStar' files yielded the best performance.}{While we experimented with using various lossless normalization techniques for the input flux (i.e., not altering the underlying shape of the spectrum, merely scaling it) these standardized fluxes failed to converge in training, and we found training on the raw flux from the `apStar' readily converged to good results. We did not investigate the performance of the `apStar' spectrum normalized in a way that removes the underlying shape of the SED, as, depending on the spectral type, such normalization may be uncertain and result in the additional noise in the line profile. The main benefit of the normalization would be removal of the extinction signature from the spectrum, however, it should be possible for a neural network to learn to ignore reddening from the raw flux as well. Nonetheless, comparison in the performance between normalized and raw spectra could be a fruitful avenue for further investigation.}

In contrast, in order to predict \teff, \logg, and Fe/H simultaneously, it was necessary to normalize these target values; normalizing the targets put the losses (and gradients during training) onto a comparable scale. To normalize, we calculated the mean ($\mu$) and standard deviation ($\sigma$) of each target variable using the training set and then standardized all prediction targets across all sources and all sets (train, development, and test). Specifically, we normalized as follows:
\begin{equation}
    \frac{x_i-\mu_x}{\sigma_x},
\end{equation}
where $x$ denotes a target variables (\teff, \logg\ or Fe/H) and $i$ denotes a specific datapoint.  Normalization values can be found in Table \ref{tab:norm-table}.

For evaluation purposes, the model's predictions are converted back to their physical units using the inverse relations from the above.

\begin{table}[]
\centering
\begin{tabular}{lr}
$\mu_{logg}$ & 2.88    \\
$\sigma_{logg}$  & 1.16    \\
$\mu_{Teff}$ & 4716.92 \\
$\sigma_{Teff}$  & 733.01  \\
$\mu_{Fe/H}$   & -0.22   \\
$\sigma_{Fe/H}$    & 0.30   
\end{tabular}
\caption{Normalization values for \teff, \logg\ and Fe/H}
\label{tab:norm-table}
\end{table}

\subsection{Convolutional Spectral Model}
Our model is a one-dimensional CNN, inspired by the VGG16 CNN architecture \citep{simonyan2014}.  It consists of 12 convolutional layers separated every two layers by a max-pooling layer, followed by 2 fully connected layers (See Appendix \ref{sec:NeuralNetwork} for term definitions and details). The model was implemented in PyTorch \citep{pytorch}. The architecture of the network is defined precisely in Appendix \ref{sec:Code}.

\subsection{Training and Tuning}

We trained the APOGEE Net using stochastic gradient descent to minimize mean squared error (MSE) loss. To improve the model's ability to generalize to new data, we employed early stopping; i.e., we stopped training when performance on the development set begins to decrease, which is indicative of overfitting to the training set at the expense of generalizability.
After each full pass through the training data, the model's performance is evaluated on the development set. If the development set performance has improved, as measured by a decrease in loss, then the model is saved. If, however, the loss on the development set does not improve after five consecutive evaluations, training is stopped. If the loss improves before the fifth evaluation, the model is saved, the counter resets and training resumes. 

After a modest amount of hyperparameter tuning, we settled upon  the following hyperparameter configuration: learning rate of 0.001,  dropout rate of 0.1, and a training batch size of 128.

\subsection{Model Adaptation}
In order to obtain high model performance on our set of interest (YSOs and M-Stars), we explored various strategies for adapting a model trained on a larger set of data to our smaller set of YSOs and M-Stars. \added{Training on the full set of stars allowed to achieve a good performance on the Payne subsample, but because it is much larger than the other two, APOGEE Net struggled to achieve acceptable loss for YSOs and M-stars} Another strategy was to first train on all stars, and then further train exclusively on the subset of interest. Unfortunately, this approach \added{also} proved suboptimal: while the model performance did improve significantly for M-Stars and YSO stars, it resulted in a dramatic degradation in performance for the red giants in the Payne catalog.

Instead, \added{we used a stratefied sampling strategy to balance the sizes of these three subsets in the training sample.} After initially training APOGEE Net to convergence on the entire dataset, we continued training on the YSO and M-Star samples plus a random 5\% of the Payne catalog. Each time the model stopped from early stopping, the last best performing model was be reloaded and another random 5\% of the Payne catalog was selected to train on. Through this method, we were able to focus training on the M-Star and YSO data without losing performance on the broader Payne data set. This process continued until the performance on M dwarfs and YSOs did not show continuing improvement. Final normalized MSE loss performances for training, development and test after tuning are reported in Table \ref{tab:mse-table}, and the resulting performance for each parameters in the native units is shown it Table \ref{tab:rmse-table}.

\begin{figure}
\epsscale{1.1}
 \hspace{-1.5cm}
 \includegraphics[width=.35\textwidth]{taror1.pdf}
 \plotone{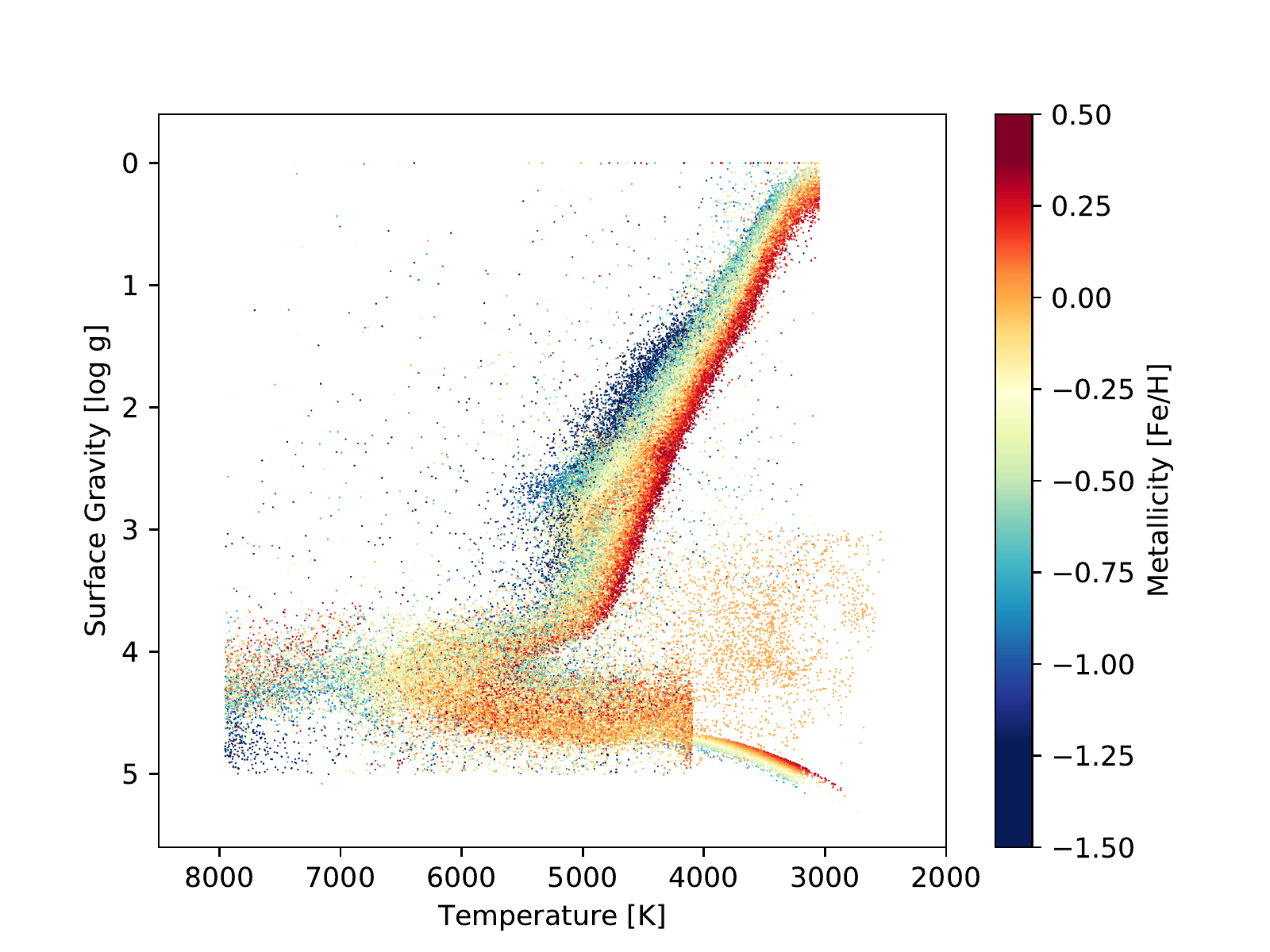}
 \plotone{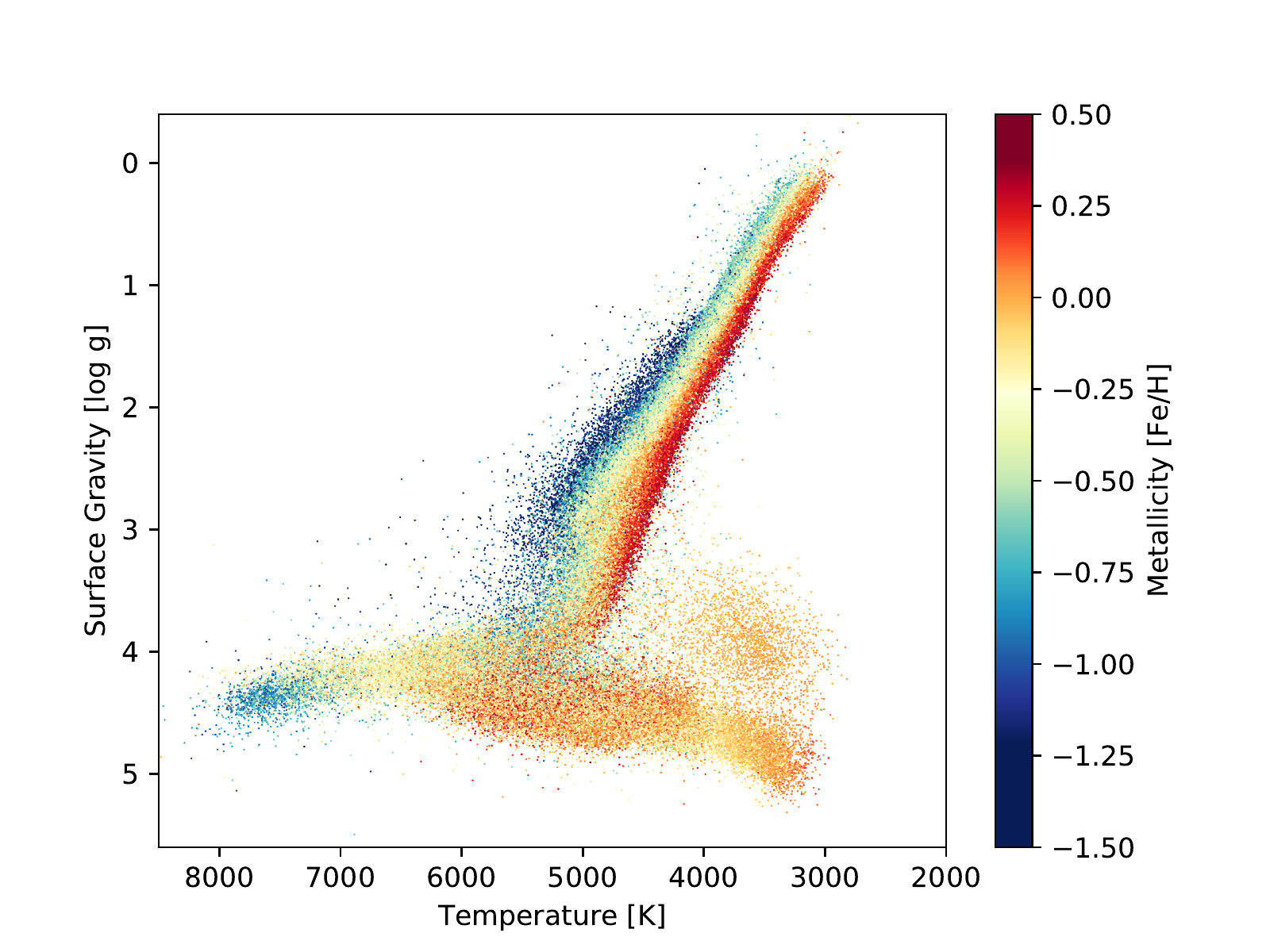}
 \caption{\teff\ and \logg\ distribution of the sample. Top: Categorization of the input parameters based on the origin of the measurements. Middle: The input parameters used in training. Bottom: The resulting predictions from the APOGEE Net.} 
 \label{fig:full}
\end{figure}


\subsection{Uncertainties}

Fundamentally, the predictions of a CNN are deterministic: after a model is trained, passing the same set of inputs always results in the same outputs. As is, the CNN is unable to realize the uncertainties in either the original data, or in its predictions (other than through a difference relative to the input labels).

However, given that the data themselves are uncertain, it is possible to vary inputs within the errors and retain the same underlying information. Each one of the realizations of the same spectrum would be perceived by the CNN as a distinct input, and would produce a slightly different prediction from the original. Measuring the scatter in these predictions can give an estimate of the uncertainties on the per source basis, akin to a Markov chain Monte Carlo. Although, unlike MCMC, this analysis is not particularly costly in terms of the computational time.

APOGEE has measured per pixel errors in flux. Therefore, at every pixel we generated a random value drawn from a normal distribution, multiplied it by the corresponding uncertainties, and added this noise profile to the flux. Some pixels (such as those near the chip gaps, or those that correspond to the telluric lines) had abnormally high uncertainties, to prevent them from skewing the model, we capped the maximum allowed error at 5 times the mean in the spectrum.

This procedure was repeated to generate 100 different realizations for each spectrum, and all of them were passed through the APOGEE Net. The mean and standard deviation values were then measured for each parameter for each source.

\subsection{Validation of the stellar parameters}\label{sec:validation}

\begin{table}
 \begin{tabular}{lccc}
 & \multicolumn{3}{c}{\textbf{MSE loss}} \\
 & \textbf{Train} & \textbf{Dev.} & \textbf{Test} \\ \cline{2-4} 
 \multicolumn{1}{l|}{\textbf{Full}} & \multicolumn{1}{c|}{0.044} & \multicolumn{1}{c|}{0.049} & \multicolumn{1}{c|}{0.063} \\ \cline{2-4} 
 \multicolumn{1}{l|}{\textbf{MStar}} & \multicolumn{1}{c|}{0.08} & \multicolumn{1}{c|}{0.098} & \multicolumn{1}{c|}{0.196} \\ \cline{2-4} 
 \multicolumn{1}{l|}{\textbf{YSO}} & \multicolumn{1}{c|}{0.153} & \multicolumn{1}{c|}{0.168} & \multicolumn{1}{c|}{0.217} \\ \cline{2-4} 
 \end{tabular}
 \caption{Combined Standardized MSE loss}
 \label{tab:mse-table}
\end{table}

\begin{table}
 \begin{tabular}{lccc}
& \textbf{Train} & \textbf{Dev.} & \textbf{Test} \\
& & \textbf{Full} & \\
\textbf{log g} & 0.189 & 0.203 & 0.216 \\
\textbf{Teff {[}K{]}} & 144.82 & 158.72 & 183.74 \\
\textbf{Fe/H} & 0.0769 & 0.0792 & 0.0914 \\
& & \textbf{MStar} & \\
 \textbf{log g} & 0.186 & 0.216 & 0.346 \\
 \textbf{Teff {[}K{]}} & 129.69 & 167.15 & 273.7 \\
 \textbf{Fe/H} & 0.129 & 0.137 & 0.18 \\
 & & \textbf{YSO} & \\
 \textbf{log g} & 0.366 & 0.364 & 0.4 \\
 \textbf{Teff {[}K{]}} & 413.92 & 436.87 & 490.99 \\
 \textbf{Fe/H} & 0.0616 & 0.0669 & 0.0879 
 \end{tabular}
 \caption{The typical scatter between the input labels and the predictions for each parameter in each group}
 \label{tab:rmse-table}
\end{table}

\begin{figure*}
\epsscale{1.0}
 \centering
		\gridline{\fig{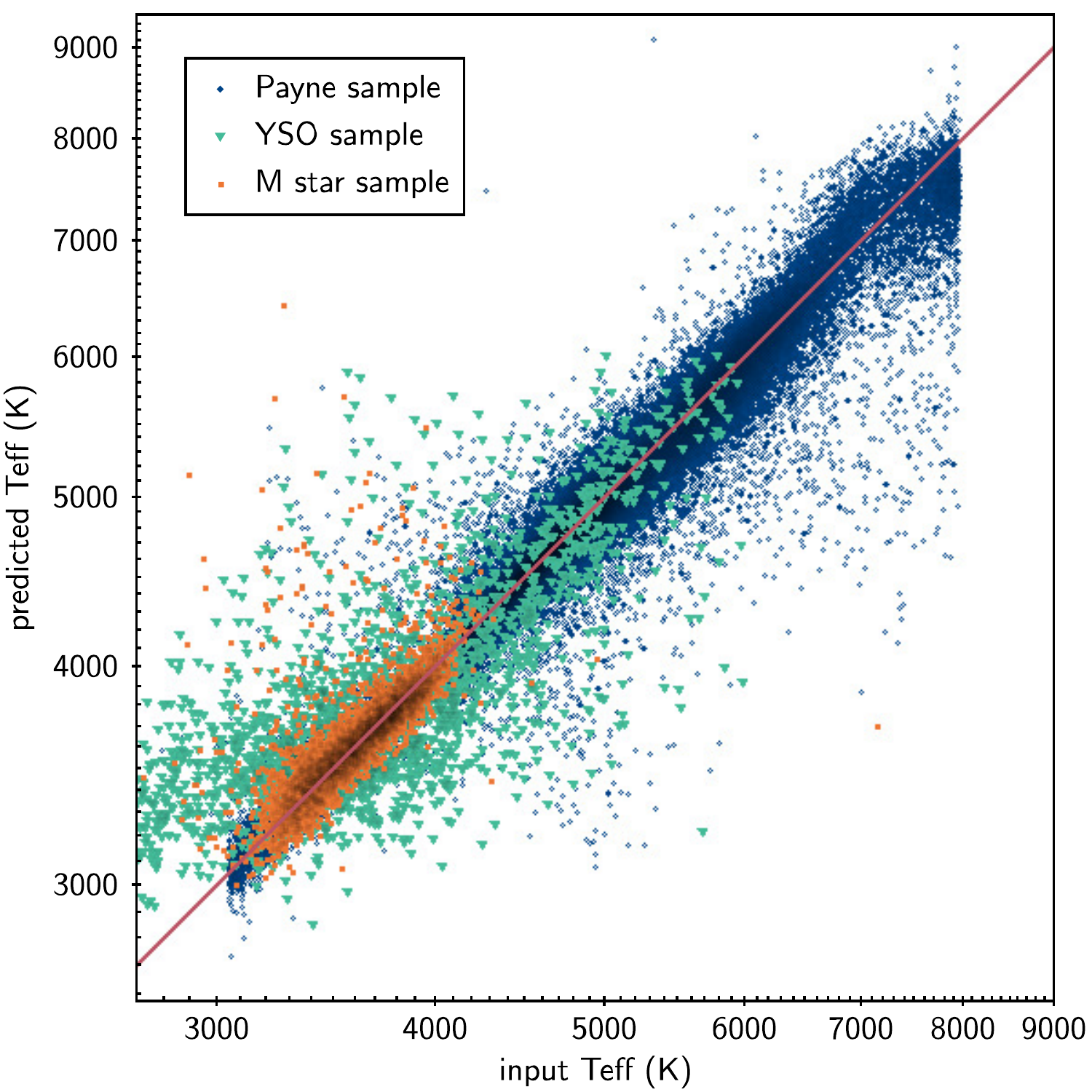}{0.4\textwidth}{}
		\fig{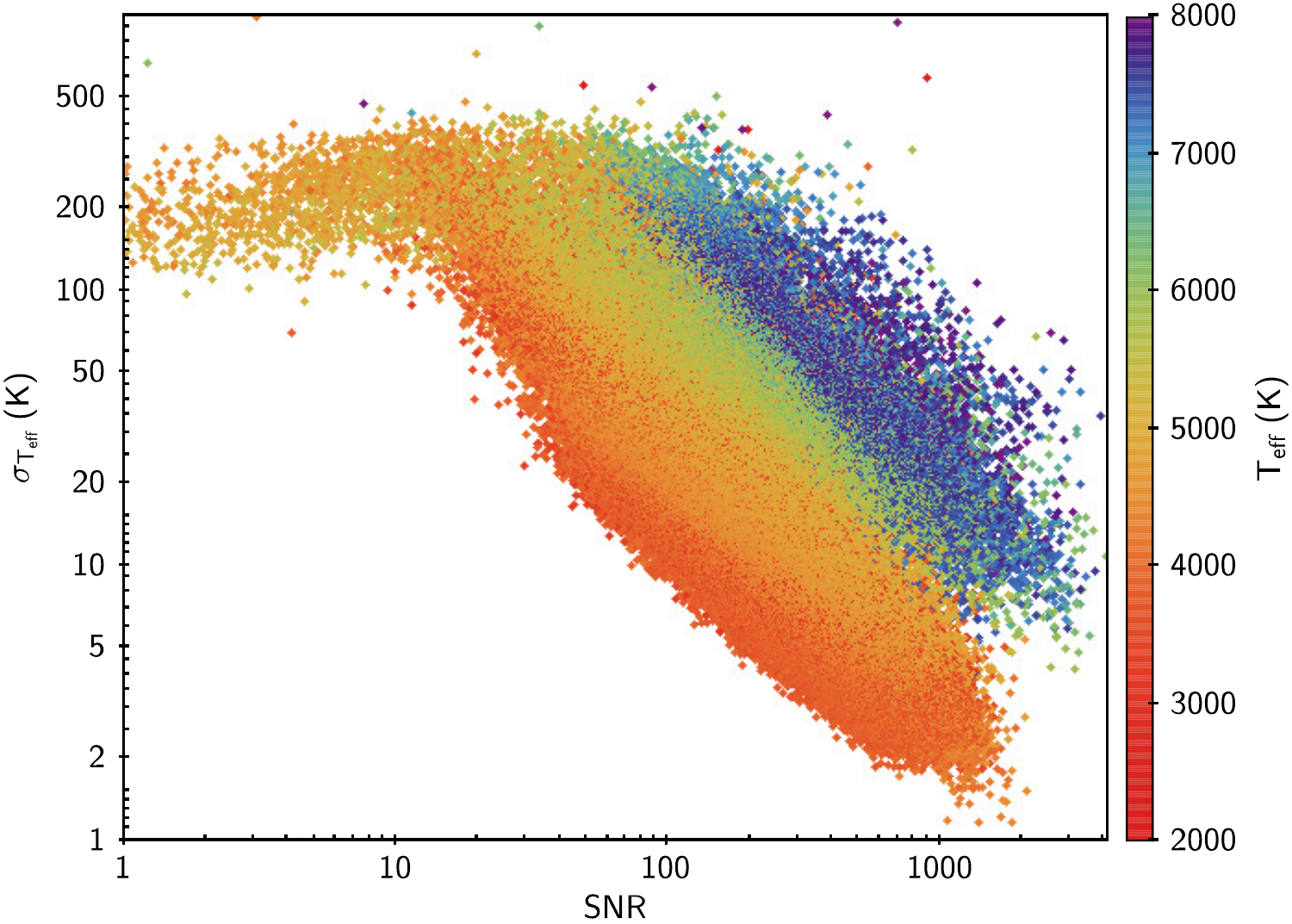}{0.56\textwidth}{}
 }\vspace{-1 cm}
		\gridline{\fig{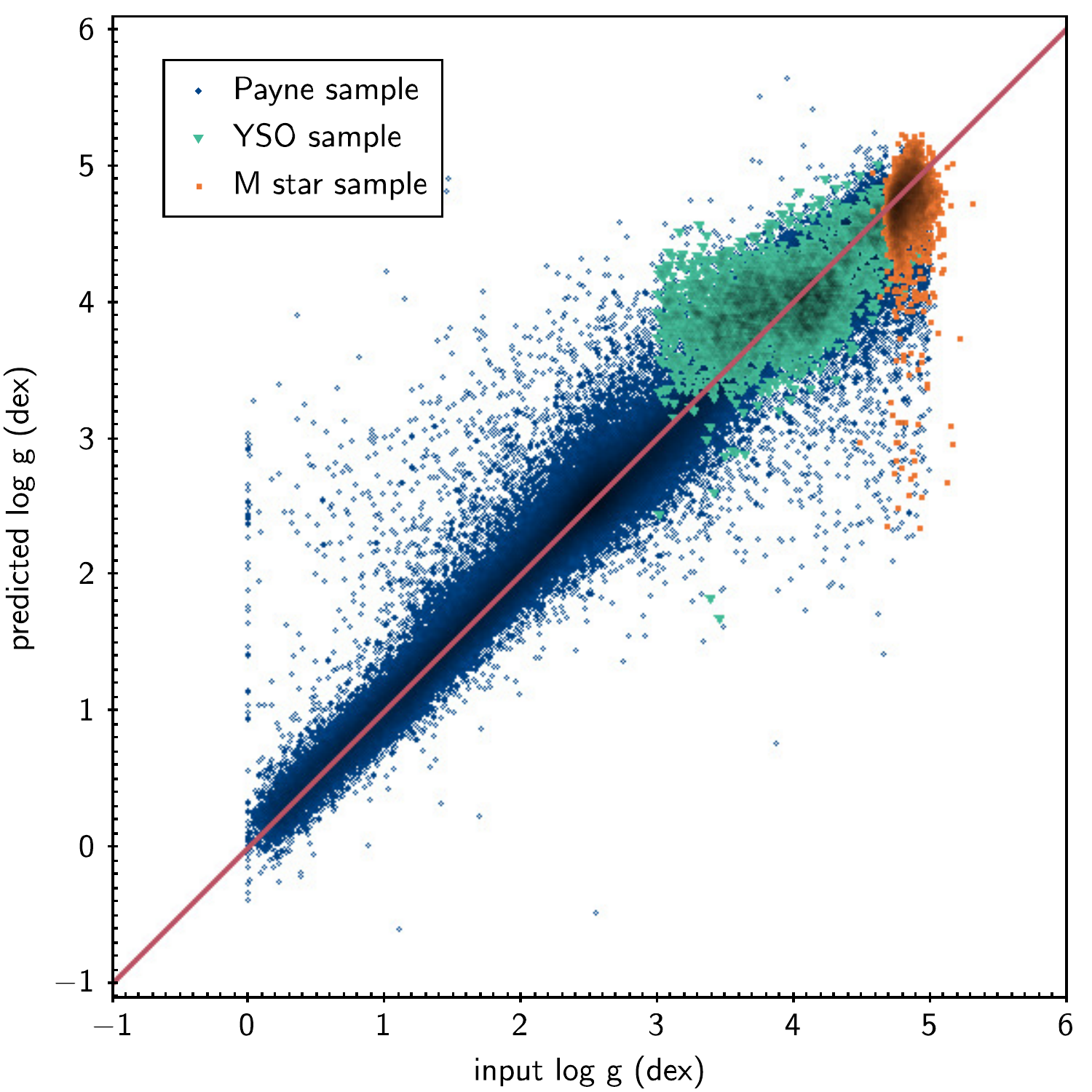}{0.4\textwidth}{}
		 \fig{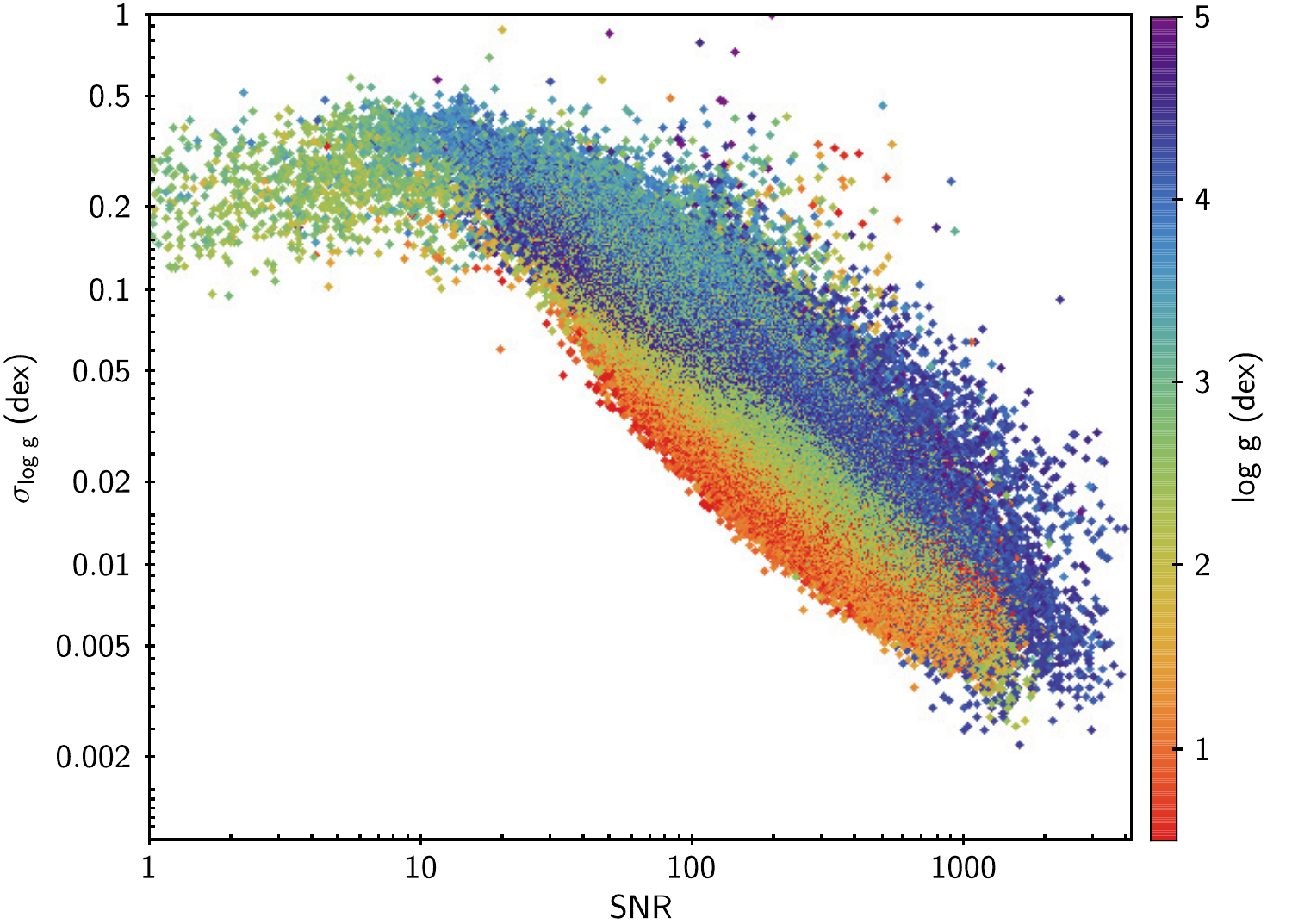}{0.56\textwidth}{}
 }\vspace{-1 cm}
		\gridline{\fig{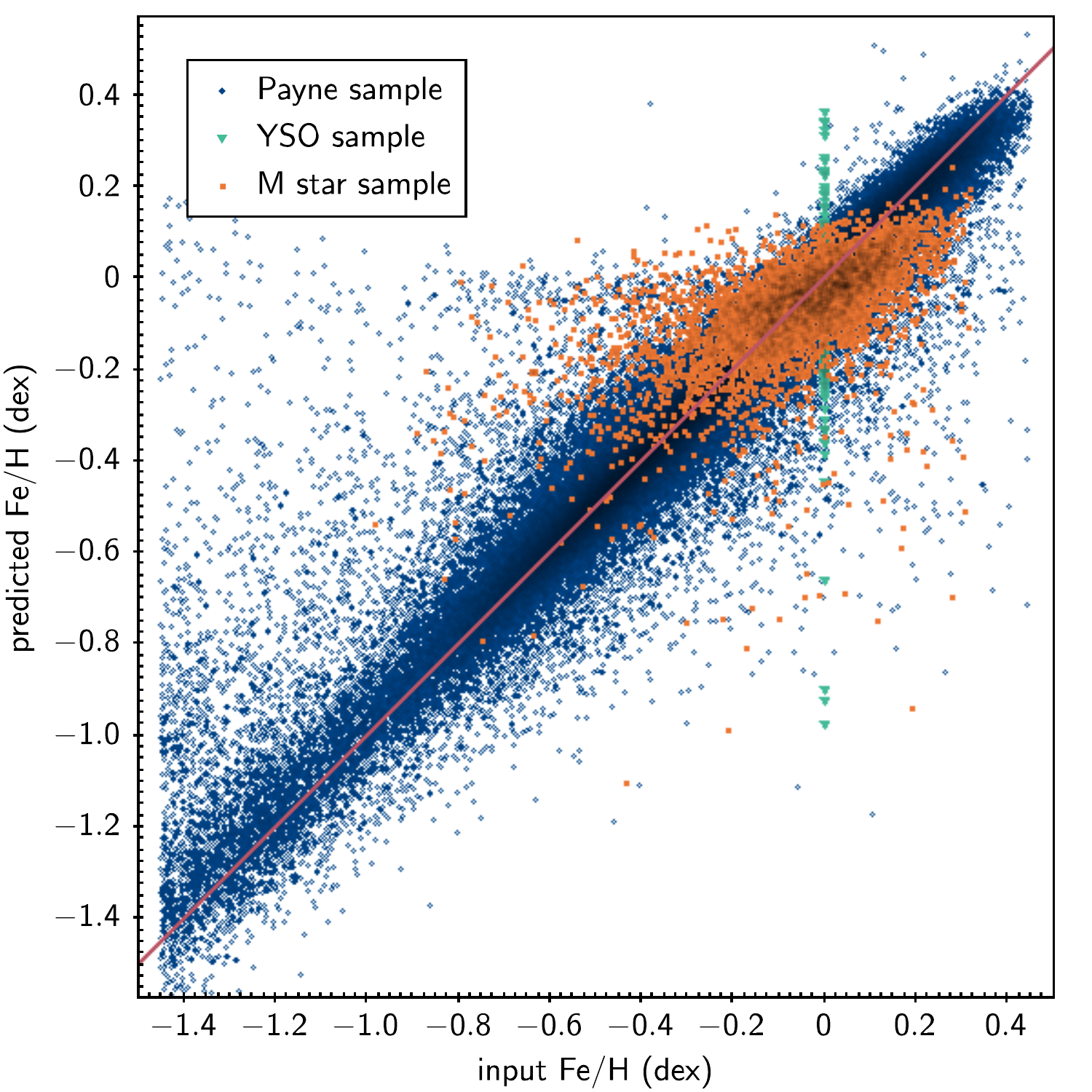}{0.4\textwidth}{}
		\fig{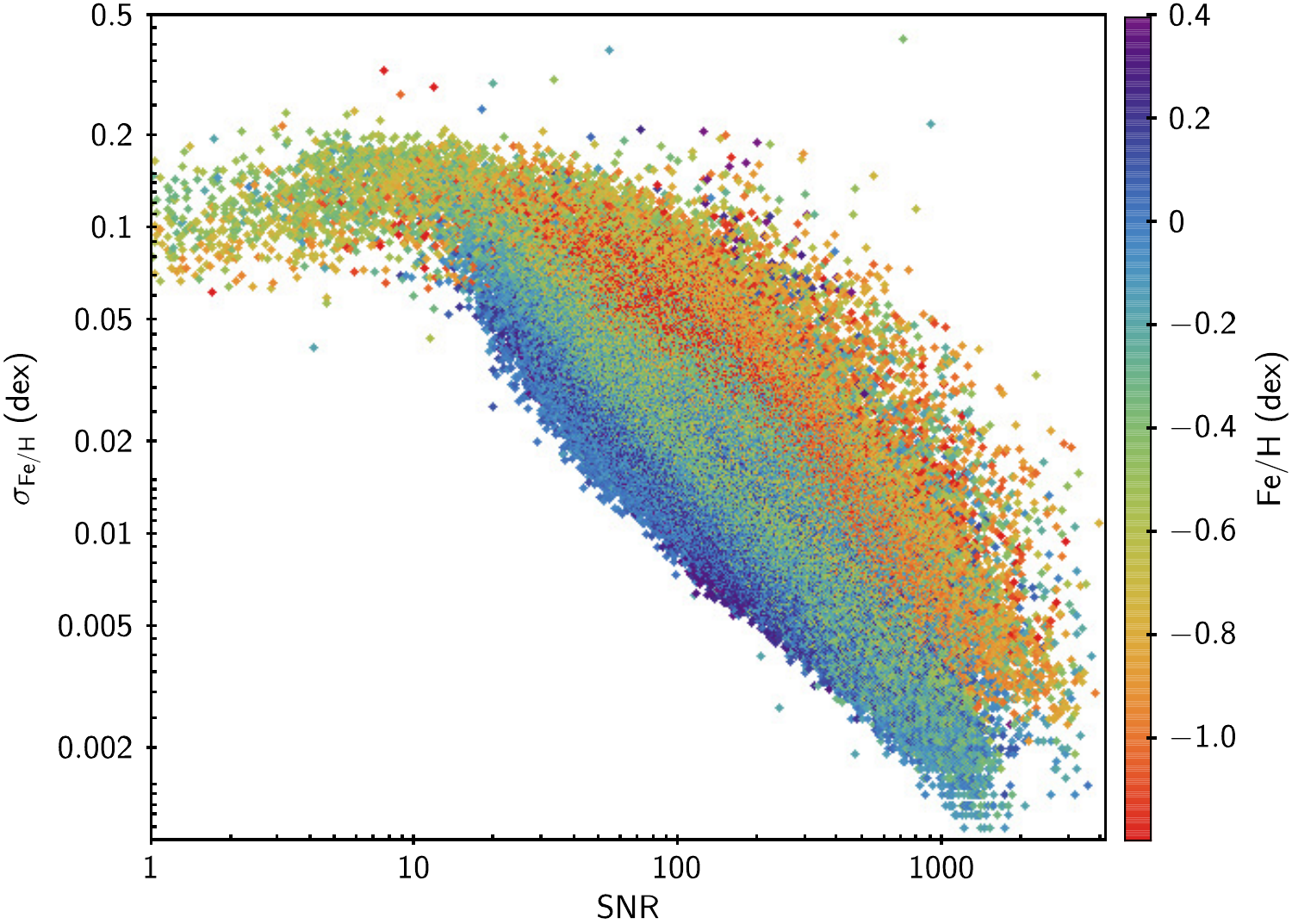}{0.56\textwidth}{}
 }
\caption{Left: Comparison between the input labels and the predicted spectroscopic parameters. Right: Uncertainties as a function of SNR.
\label{fig:sigma}}
\end{figure*}

We report on the resulting predictions with the corresponding uncertainties in Table \ref{tab:CombinedEvalResults}.

The typical agreement between the input labels and the resulting predictions is 100 K in \teff, 0.15 dex in \logg, and 0.07 dex in Fe/H (Table \ref{tab:rmse-table}, Figure \ref{fig:sigma}, left). The scatter in \teff\ and \logg\ is slightly higher for the YSOs, as it improved on some of the systematic issues the photometric labels had, which were originally derived somewhat crudely, fine-tuning them based on the overall grid. For the M-stars, comparison between the labels and the predictions for Fe/H is slightly offset from the line of unity, with predictions somewhat compressing the range of Fe/H offered by the labels, having fewer sources as metal rich, and fewer as metal poor, but showing a good linear agreement overall.

The typical reported uncertainties are 25 K in \teff, 0.04 dex in \logg, and 0.015 dex in Fe/H, thus they underestimate the scatter between the labels and the predictions by approximately a factor of 4. In comparison to other pipelines, the uncertainties for the same sources are not strongly correlated, but they are generally comparable to the errors reported by the IN-SYNC pipeline, and approximately a factor of 2 smaller than those reported by ASPCAP pipeline. The reported uncertainties strongly depend on the SNR of the spectrum, as well as on the spectral parameters (Figure \ref{fig:sigma}, right). As expected, the parameters of hotter or more metal poor stars would be more uncertain due to a fewer number of lines that can be used to determine stellar properties. Similarly, sources with higher \logg\ would have shallower lines, resulting in more uncertain predictions. Although the APOGEE Net had no information on the uncertainties in the labels, and the errors were generated from slightly perturbing the input spectral fluxes and taking an rms of the resulted predictions, it was able to reproduce physically expected trends.

At low SNR ($<$10---20), the uncertainties in all parameters reach a ceiling of $\sim$200 K in \teff, $\sim$0.2 dex in \logg, and $\sim$0.1 dex in Fe/H. \deleted{Few of the sources with low SNR spectra have reasonable parameters in the range that is appropriate, i.e., located far outside of the bounds of the Figure \ref{fig:full}. However,} This ceiling suggests that the uncertainties for these sources are underestimated, furthermore, that the CNN does not derive any meaningful information in the low SNR spectra. This brings into question the reliability of other parameters (e.g., $v \sin i$, radial velocity) derived by other means in these low SNR spectra. \added{Some of the sources with SNR$\sim$0 have unphysical parameters, i.e., located far outside of the bounds of the Figure \ref{fig:full}. These sources were removed from the catalog.}

\begin{figure*}
\epsscale{1.0}
 \centering
		\gridline{\fig{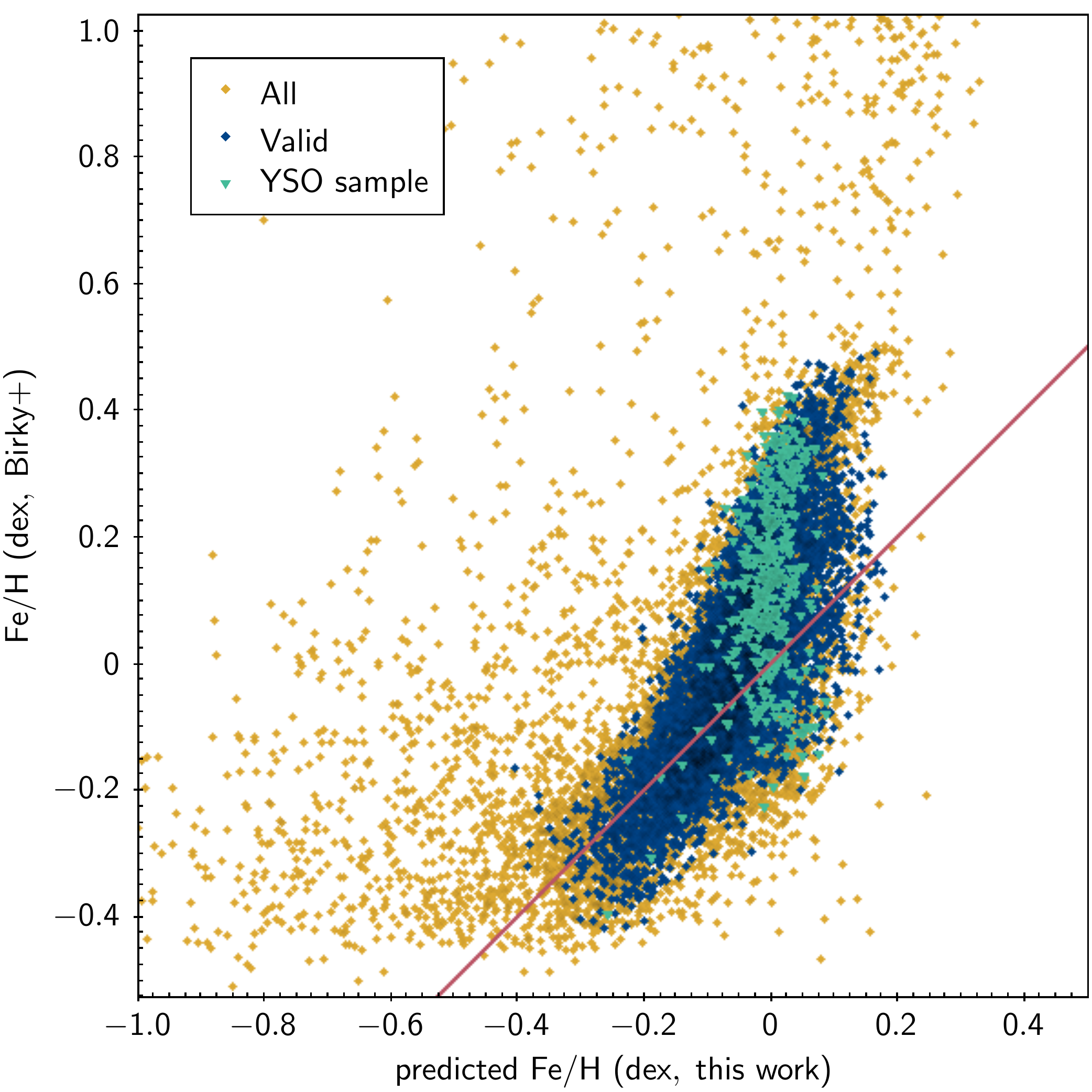}{0.3\textwidth}{}
		\fig{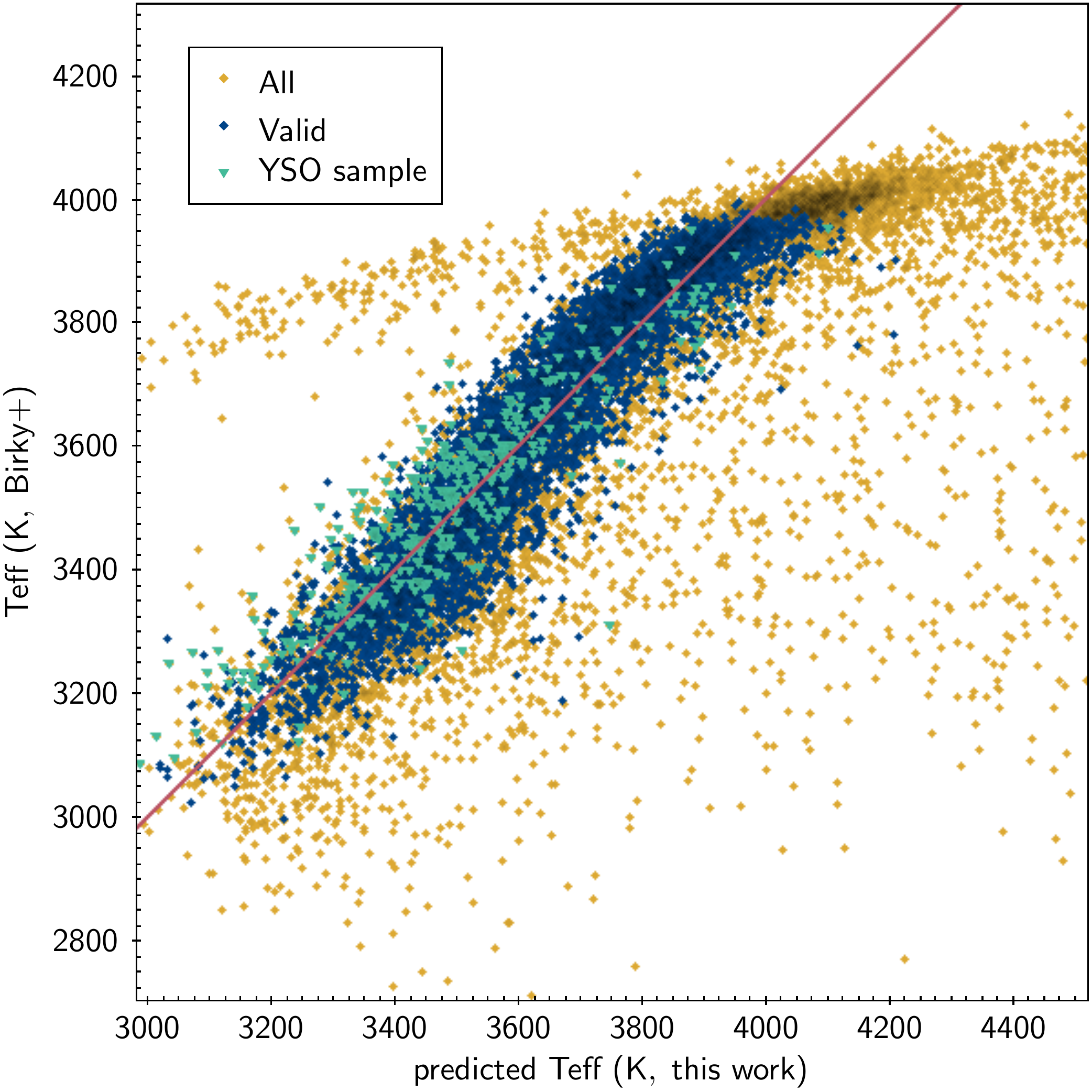}{0.3\textwidth}{}
		\fig{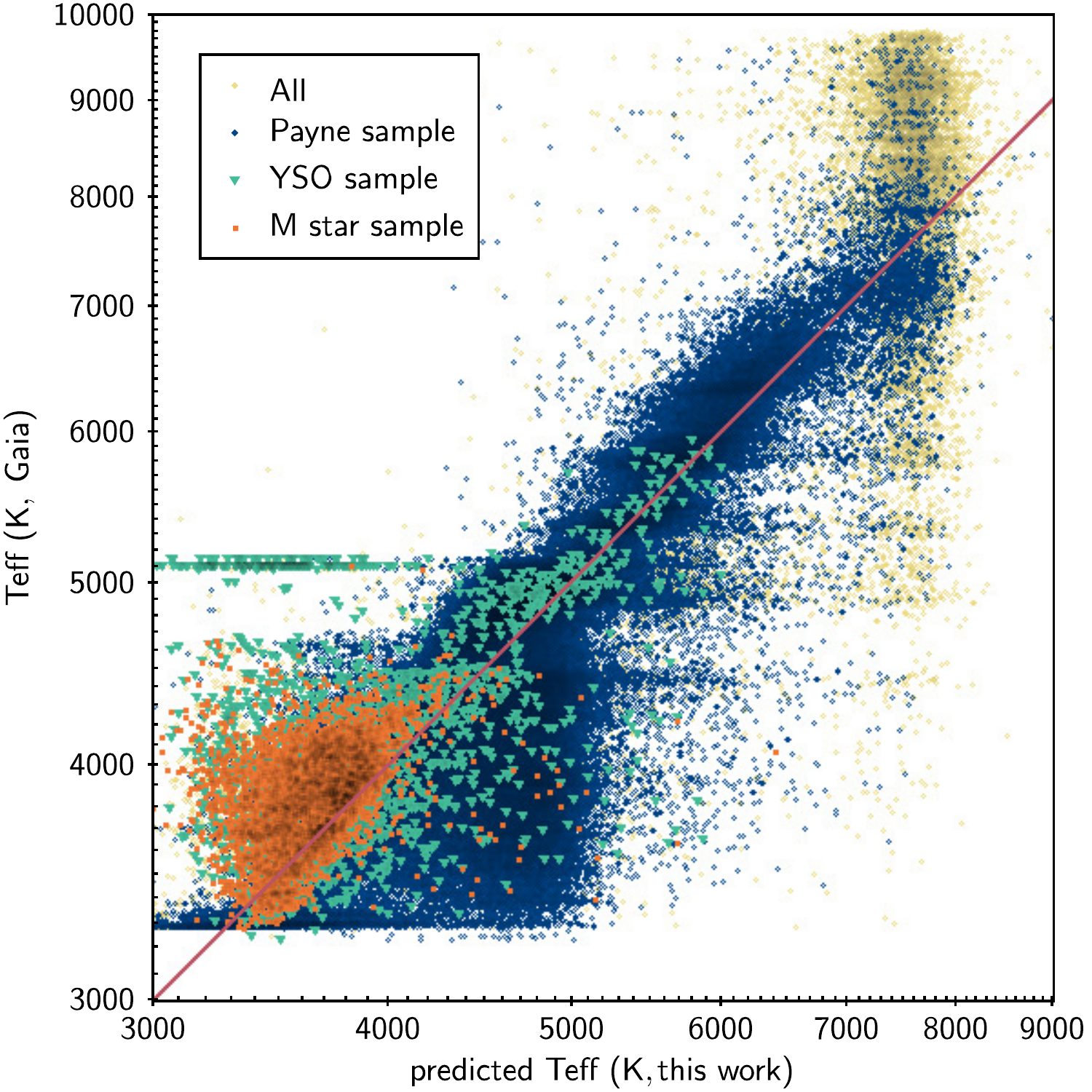}{0.3\textwidth}{}
 }\vspace{-1 cm}
\caption{Left and center: Comparison of Fe/H and \teff\ derived in this work compared to those in \citet{birky2020}. The sources in the Valid sample satisfy the quoted ranges of the parameter space from Birky+ over which their extra extrapolation is applicable, combined with having uncertainties in the predicted parameters in our work of $\sigma_{T_{eff}}<$100 K and $\sigma_{\log\ g}<0.1$ dex (typically corresponding to SNR$>20$), restricting the comparison to 5893 stars. The YSO sample is a subset of the Valid sample, with the overlapping sources from \citet{Kounkel2019}.  Right: Comparison of the derived \teff in this work and \teff\ from Gaia DR2. Note the edge effect at \teff$\sim$4,000 K for the sample from Birky et al., and 8,000 K for the sample in this work.
\label{fig:birky}}
\end{figure*}

A similar approach to determining spectroscopic parameters from the APOGEE spectra for the M-dwarfs was recently undertaken by \citet{birky2020}, based on the Cannon \citep{ness2015} data-driven spectral modeling code (Figure \ref{fig:birky}). While some YSOs are included in their sample, they tend to have higher $\chi^2$ for the fit compared to the rest of the sample. In particular, because their code did not train to distinguish between metallicity and \logg, it considered YSOs to be more metal rich than they are likely to be due to deeper spectral lines. In the M-dwarf sample, for the metallicity, there is a good agreement at $Fe/H<0$, but at $Fe/H>0$ they tend to systematically differ by a factor of 2; it is not clear why. The predicted \teff\ appears to be comparable between the two works, with a scatter of $\sim$100 K. However, because they have not included any sources hotter than $\sim$4,100 K, they tend to run into the edge effects at $\sim$4,000 K, with sources piling at the boundary. We note that a similar effect occurs in this work as well, at \teff$\sim$8,000 K -- any stars intrinsically hotter than are interpolated towards this value. This potentially explains the excess of metal poor stars in Figure \ref{fig:full}. For this reason we do not include any stars hotter than 6,700 K in Table \ref{tab:CombinedEvalResults}. In future, however, by generating more reliable labels for massive stars and including them in the training, it would be possible to minimize these edge effects.

\section{Results \label{sec:results}}
\subsection{Overall properties}

\begin{splitdeluxetable*}{cccccccccBcccccc}
\tabletypesize{\scriptsize}
\tablewidth{0pt}
\tablecaption{Results for combined evaluation  \label{tab:CombinedEvalResults}}
\tablehead{
\colhead{APOGEE ID}	&	\colhead{$\alpha$}	&	\colhead{$\delta$}	&	\colhead{\logg}	&	\colhead{\logg}	&	\colhead{$\sigma$ \logg}	&	\colhead{\teff}	&	\colhead{\teff}	&	\colhead{$\sigma$ \teff}	&	\colhead{Fe/H}	&	\colhead{Fe/H}	&	\colhead{$\sigma$ Fe/H}&	\colhead{SNR}	& 	\colhead{Data} &	\colhead{Data}	\\
\colhead{}	&	\colhead{(J2000)}	&	\colhead{(J2000)}	&	\colhead{(label, dex)}	&	\colhead{(prediction, dex)}	&	\colhead{(dex)}	&	\colhead{(label, K)}	&	\colhead{(prediction, K)}	&	\colhead{(K)}	&	\colhead{(label, dex)}	&	\colhead{(prediction, dex)}&	\colhead{(dex)}	&	\colhead{}	&	\colhead{Set} &	\colhead{Type}	}
\startdata
2M00003379+7940362 &  0.140803 &  79.676727 &4.94 &4.47 &0.06 &3307 &3357 &28 &-0.03 &-0.05 &0.01 &137.0 & train & Mstar \\
2M00013219+0016012 &  0.384140 &   0.267008 &4.74 &4.64 &0.08 &3997 &3998 &47 &-0.34 &-0.25 &0.02 &57.2 & train & Mstar \\
2M00024474+6158060 &  0.686431 &  61.968346 &4.74 &4.50 &0.10 &3787 &4062 &50 &0.02 &-0.15 &0.02  &57.7 & train & Mstar \\
2M00025988+0148410 &  0.749506 &   1.811404 &4.72 &4.65 &0.07 &3959 &3919 &32 &-0.20 &-0.19 &0.02 &82.1 & train & Mstar \\
2M00030930+0110025 &  0.788757 &   1.167374 &4.73 &4.66 &0.07 &3947 &3974 &33 &-0.21 &-0.24 &0.02 &88.1 & train & Mstar \\
\enddata
\tablenotetext{}{Only a portion shown here. Full table with all deconvolved parameters is available in an electronic form.}
\end{splitdeluxetable*}

\begin{figure*}
\epsscale{1.0}
 \centering
		\gridline{\fig{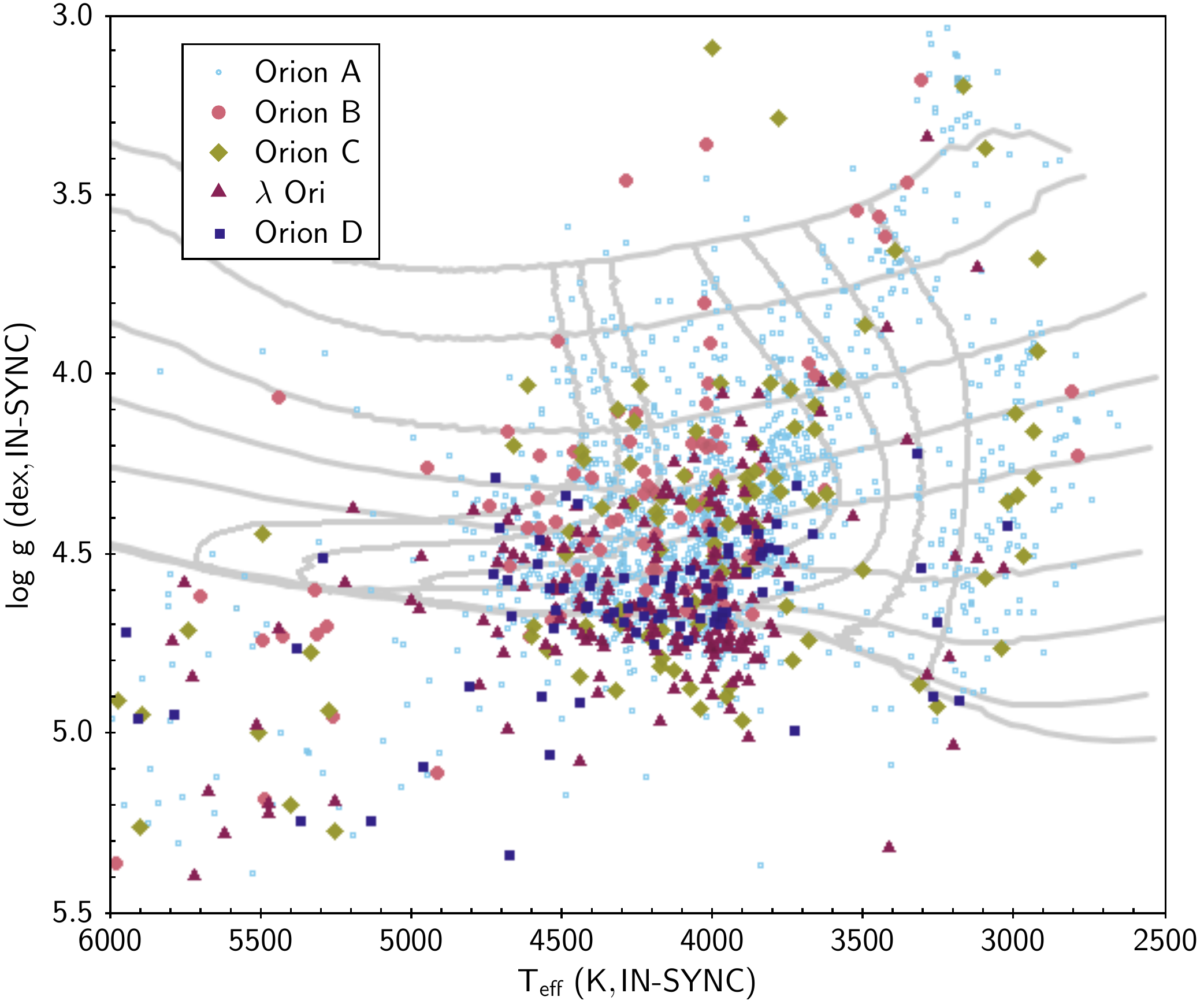}{0.5\textwidth}{}
 \fig{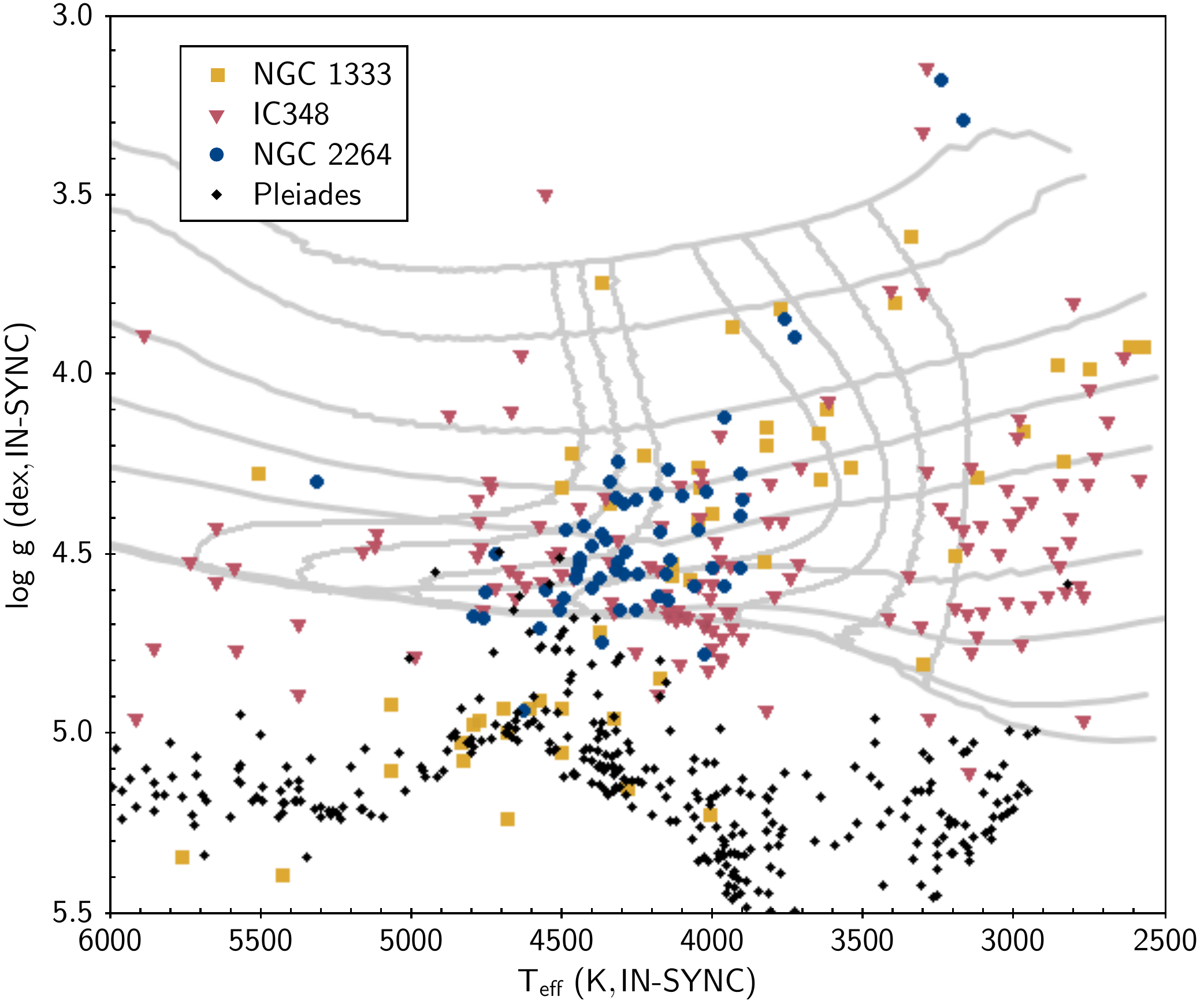}{0.5\textwidth}{}
 }\vspace{-1 cm}
		\gridline{\fig{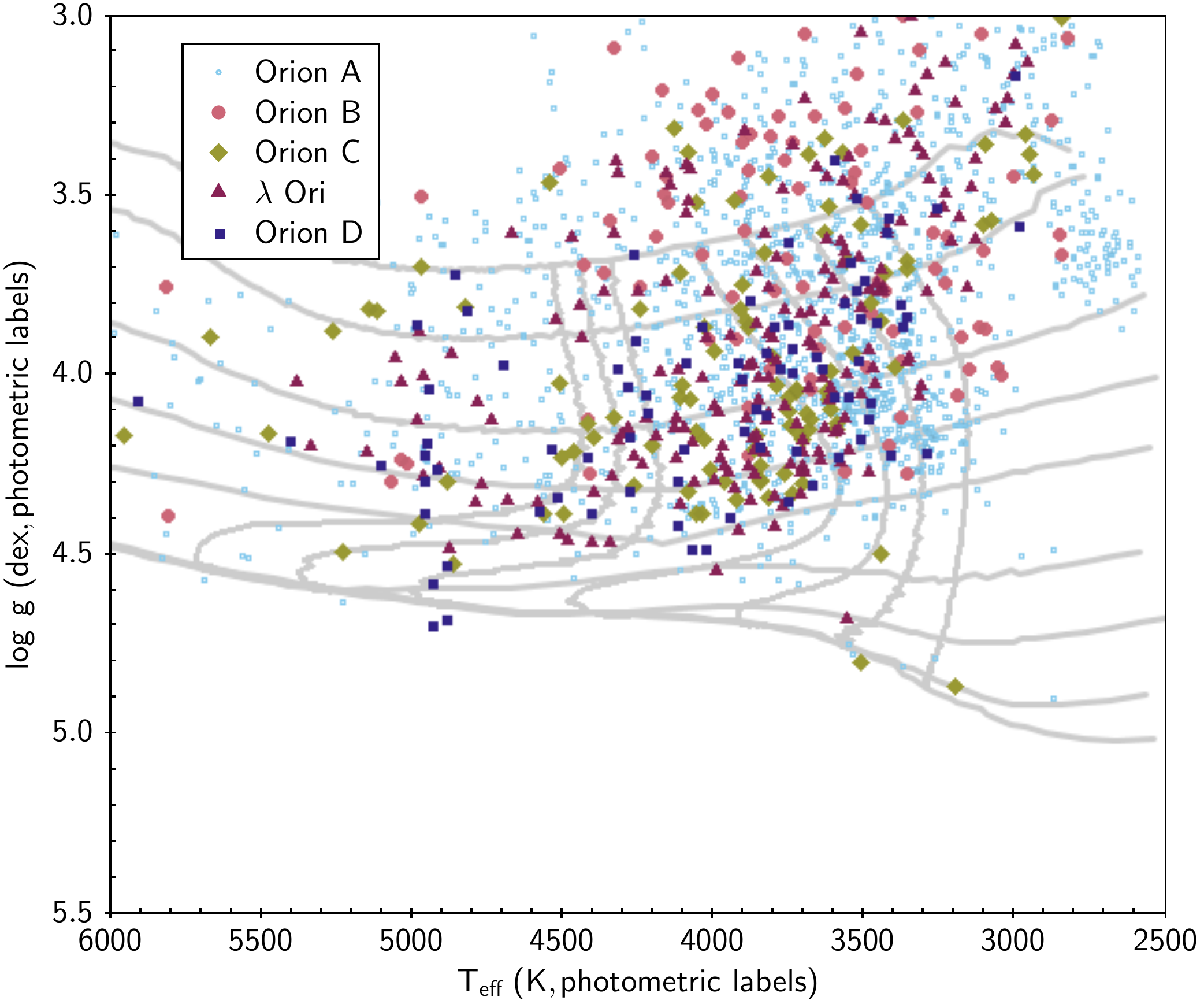}{0.5\textwidth}{}
 \fig{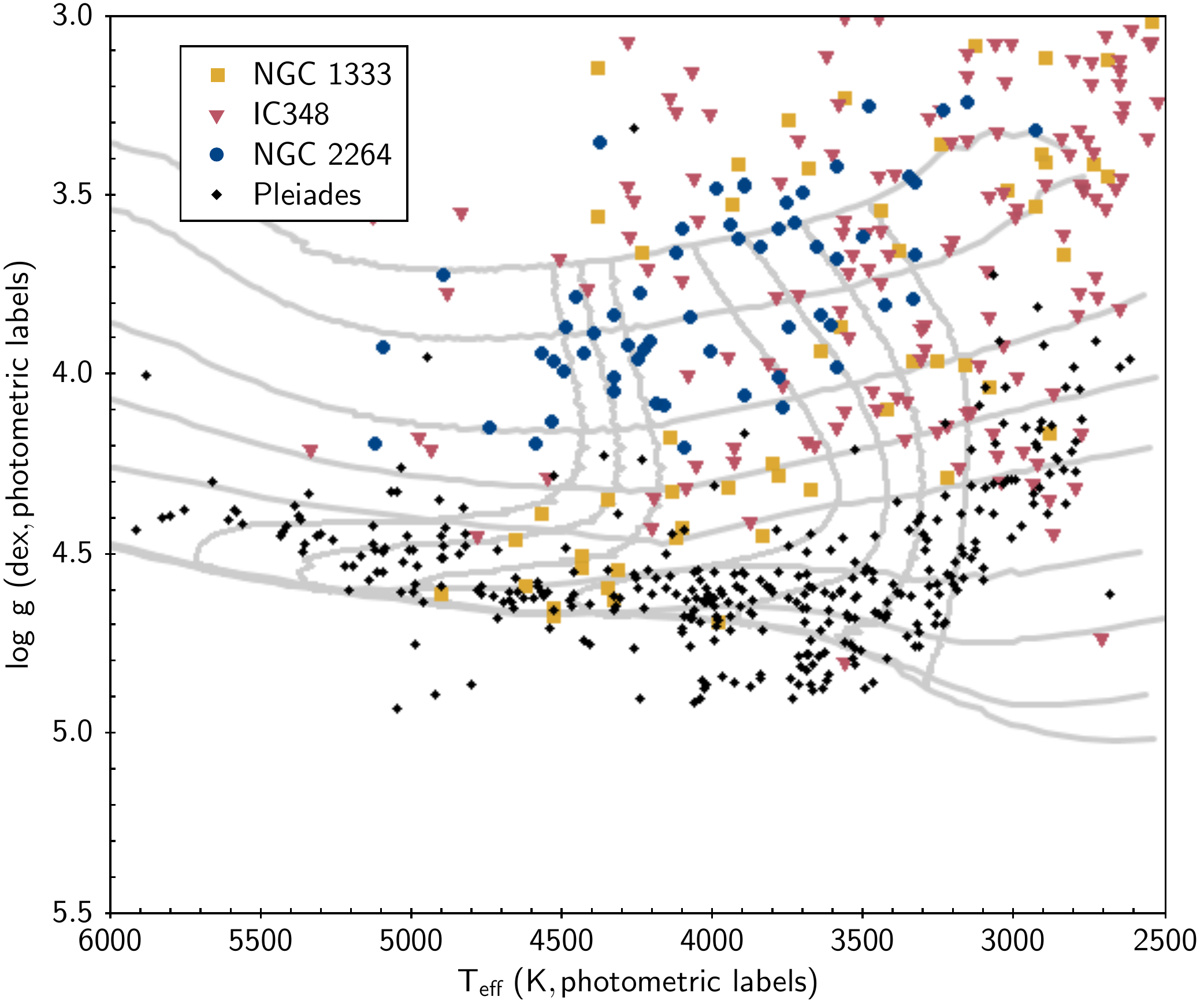}{0.5\textwidth}{}
 }\vspace{-1 cm}
		\gridline{\fig{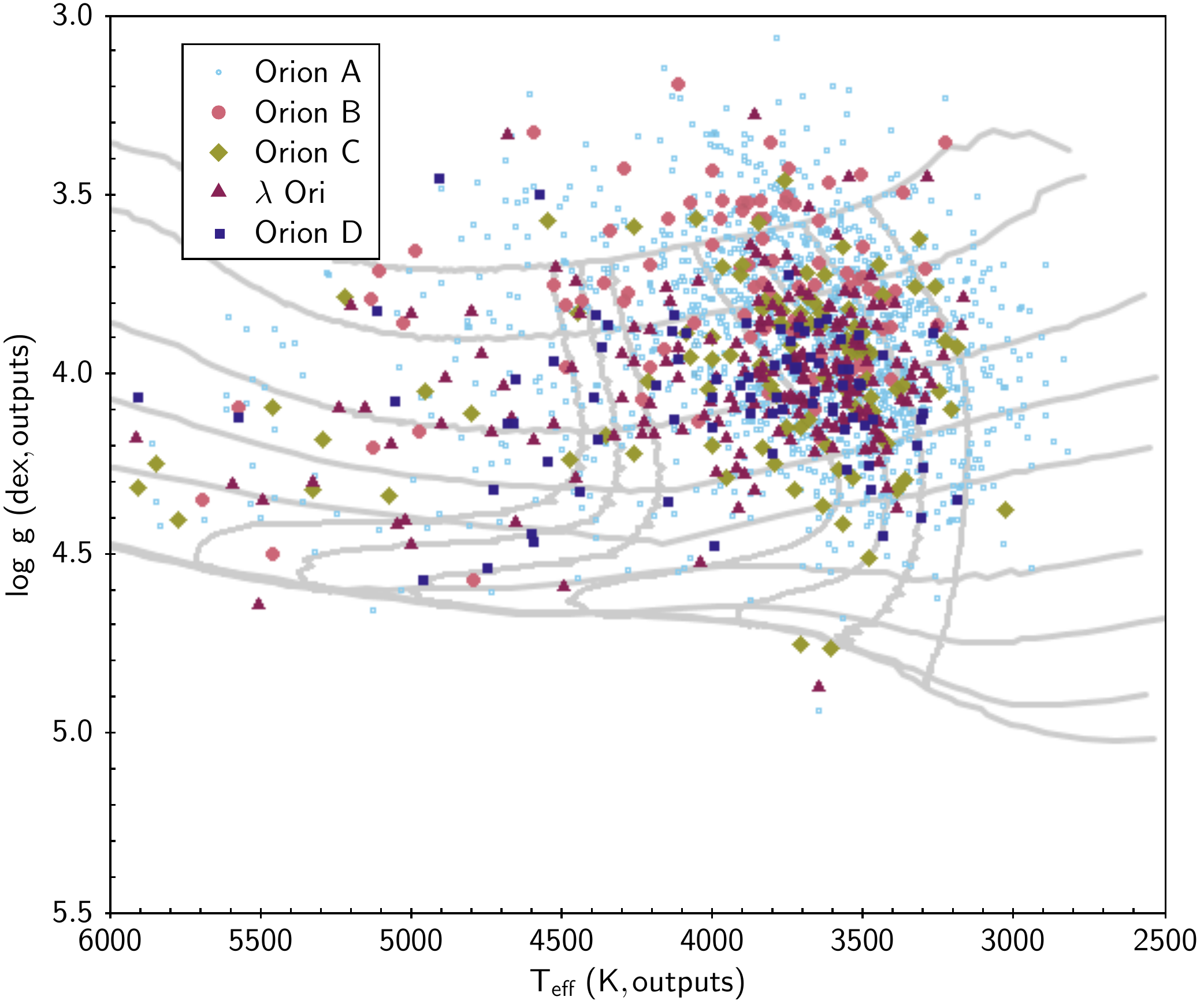}{0.5\textwidth}{}
 \fig{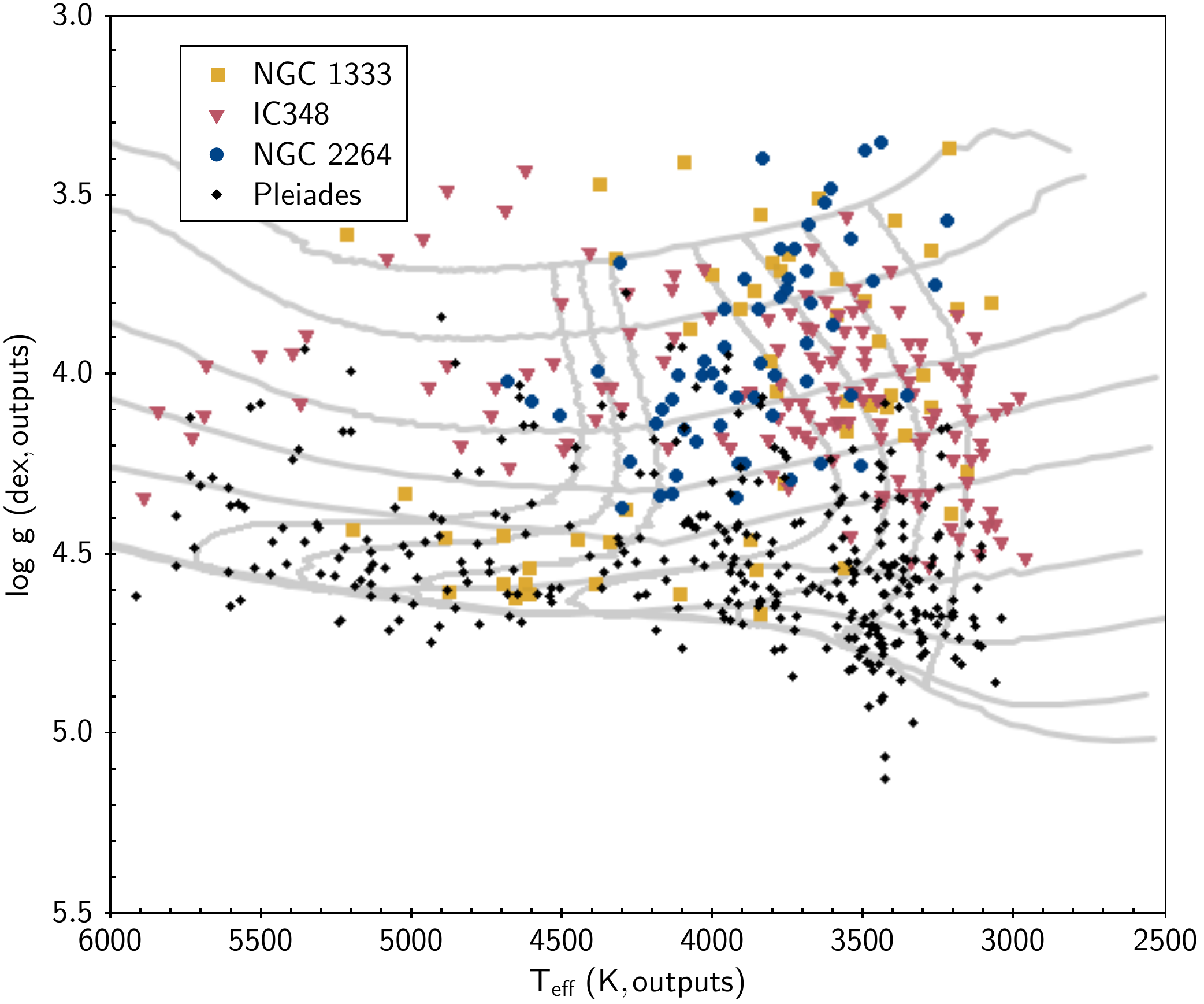}{0.5\textwidth}{}
 }\vspace{-1 cm}
\caption{Distribution of \teff\ and \logg\ values for the YSOs across the various regions. The compilation of the measurements from the IN-SYNC pipeline (top), photometrically derived input labels (middle), and the resulting output from the APOGEE Net (bottom) are shown. The grey lines show the isochrones at ages of 1, 2, 5, 10, 20, 50, 100, 200, and 300 Myr, and 8.5 dex, as well as the evolutionary tracks for 0.4, 0.5, 0.6, 0.7, 0.8, 0.9, and 1 \msun\ stars from the PARSEC isochrones \citep{marigo2017}.
\label{fig:ysos}}
\end{figure*}

\begin{figure*}
\epsscale{0.8}
\plottwo{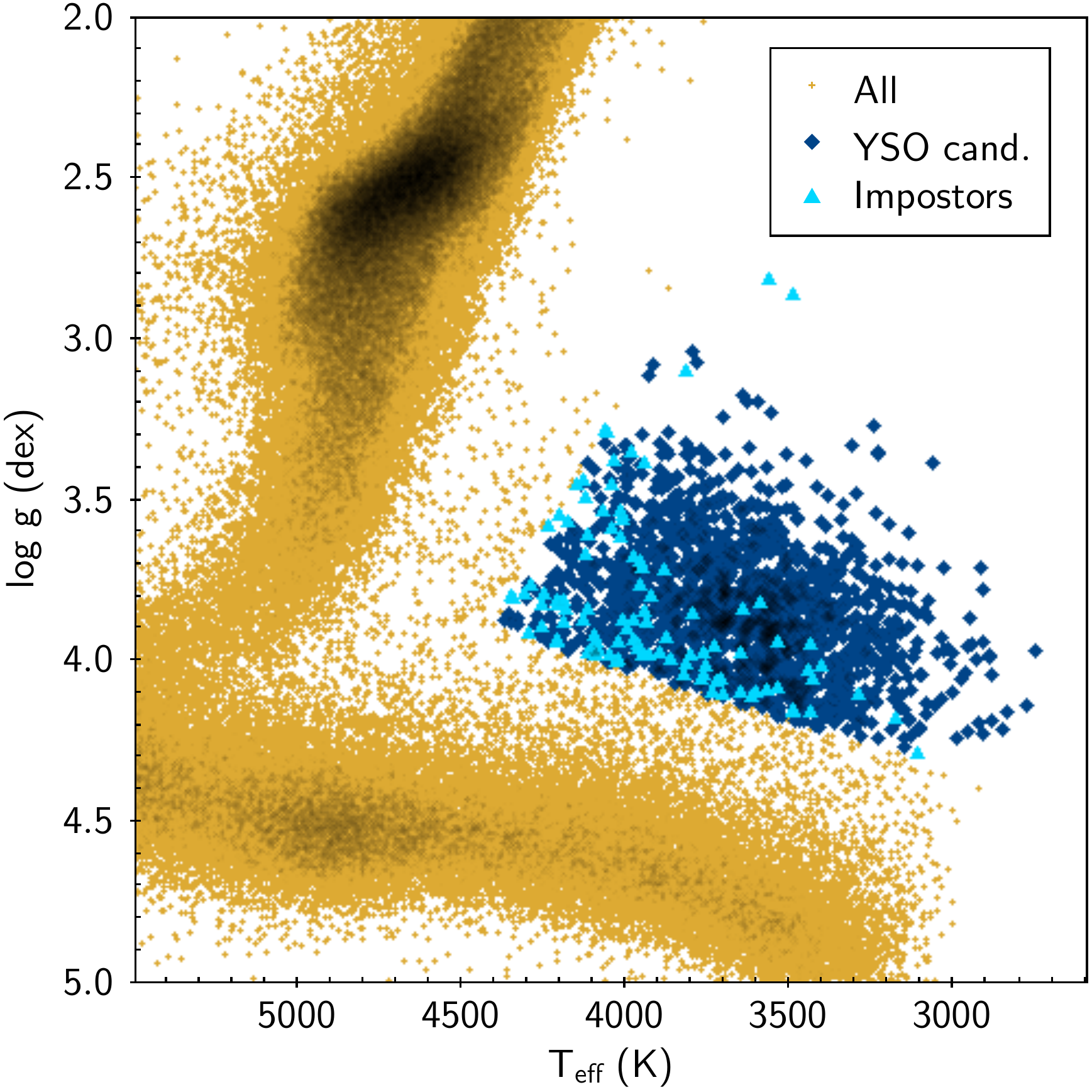}{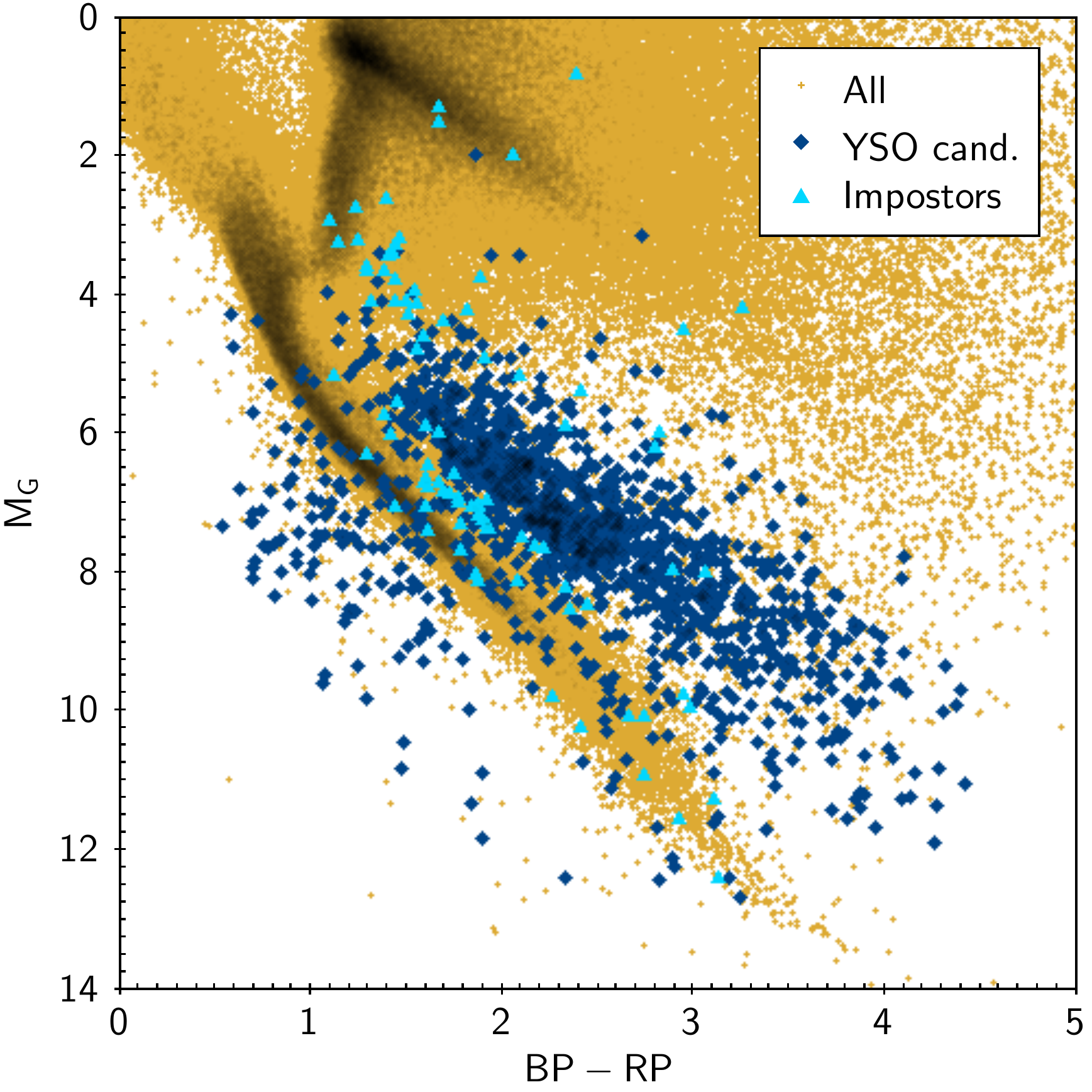}
\plotone{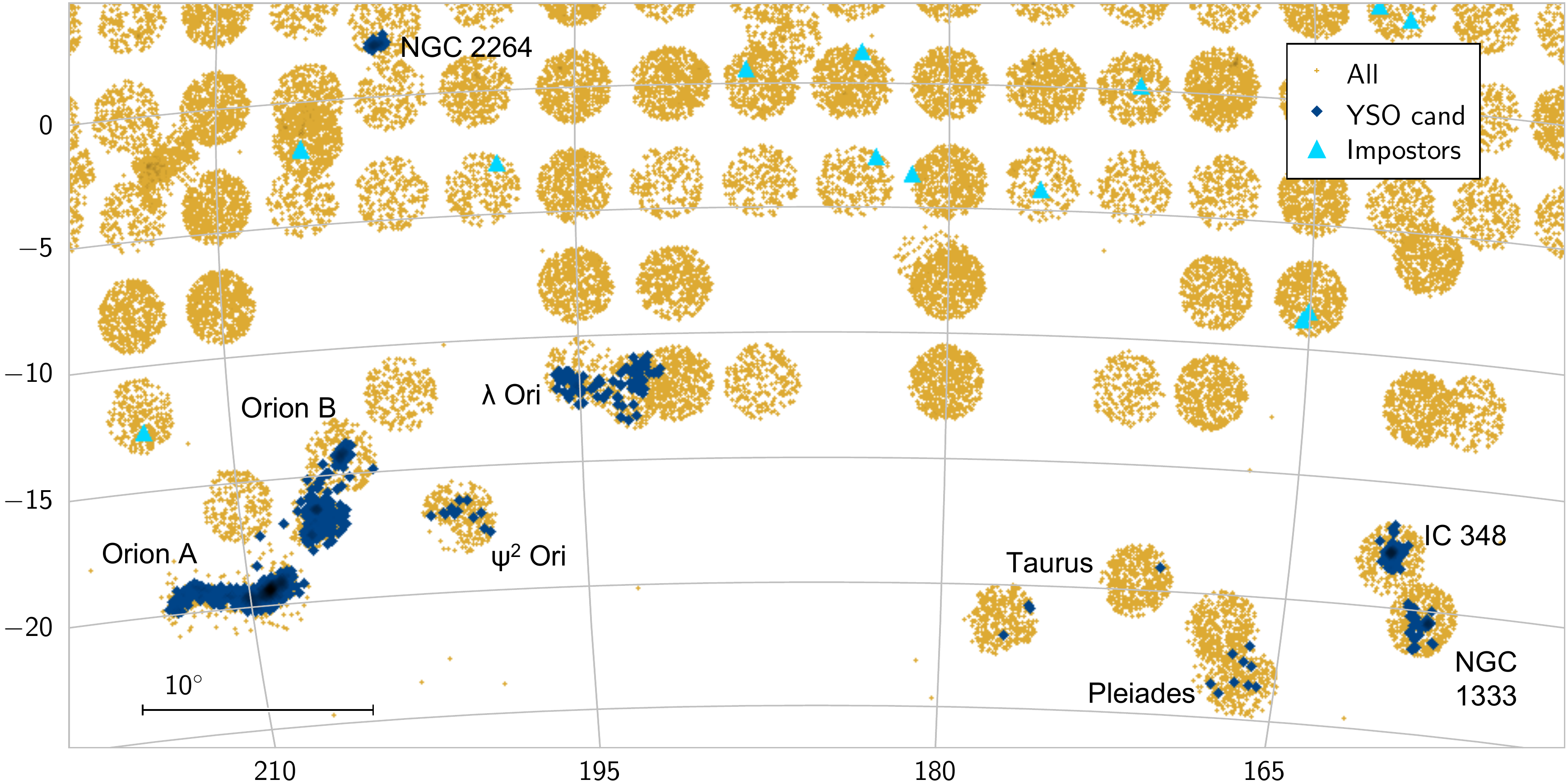}
\plotone{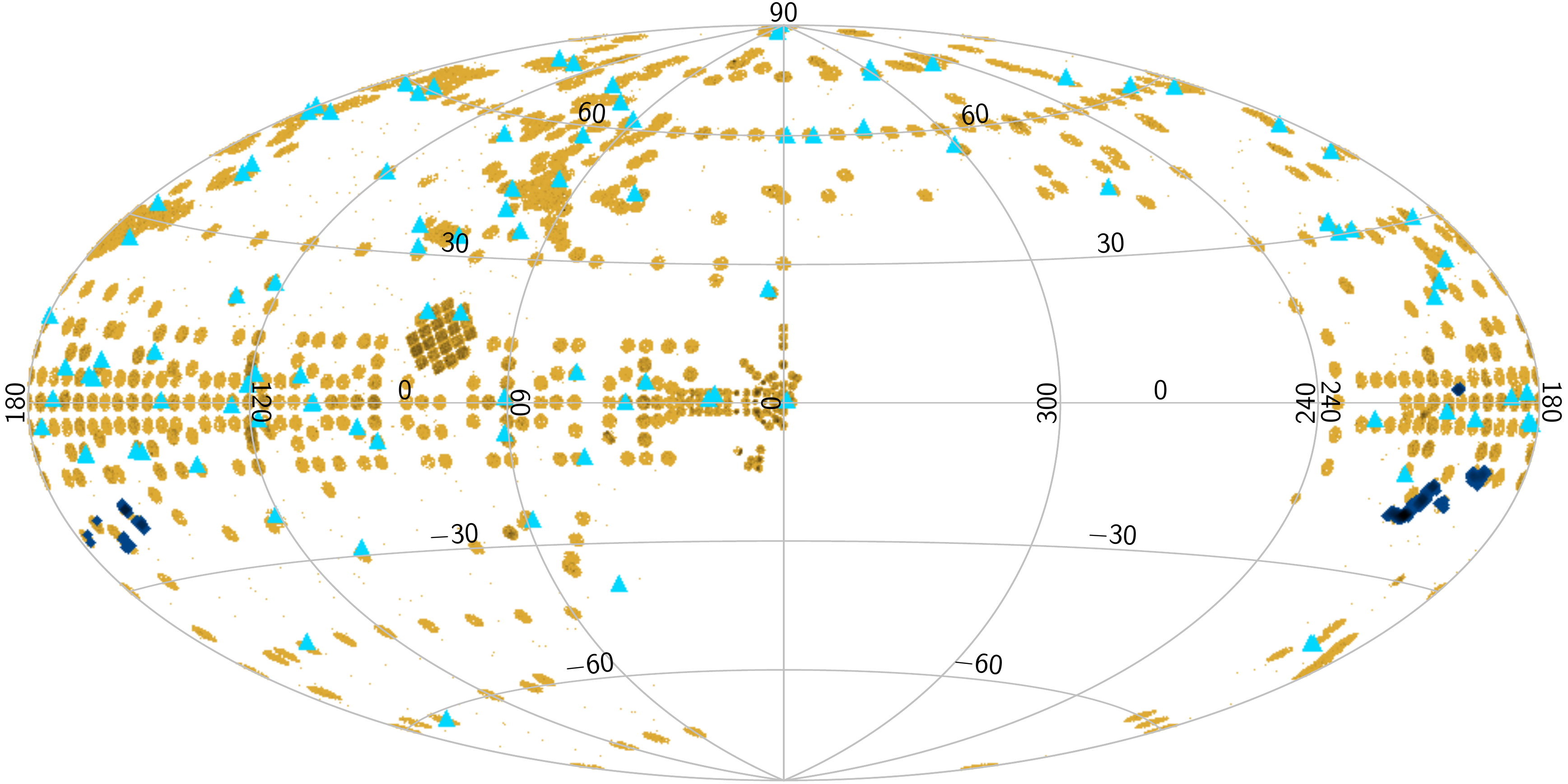}
\caption{Top: Selection of the YSOs based on the predicted stellar parameters. Objects are considered YSO candidates if they fall into the parameter space bounded by (\teff, \logg) of (2200,4.4) (4500,3.9), and (3200,1.7). Left panel shows the spectroscopic parameters, right panel shows the HR diagram constructed using \textit{Gaia} photometry \added{for the same sources. The sources in blue that fall bellow the photometric main sequence are known YSOs, most likely with poor \textit{Gaia} parallaxes and/or fluxes.}. Bottom: spatial distribution of the identified candidates in galactic coordinates. Sources that do not coincide with nearby star-forming regions are flagged as the potential impostors (6\% of the sources in the selected area).
\label{fig:ysoreg}}
\end{figure*}

As can be seen in Figure \ref{fig:ysos}, the quality of the derived \teff\ and \logg\ for YSOs improves with each \replaced{iteration}{method for extracting them}. The original parameters from the IN-SYNC pipeline are systematically offset from the isochrones, have an odd shape of the main sequence, and have various unphysical gaps, most notable of which is at \teff=3,600 K. The labels derived from the photometry show similar agreement to the isochrones in the sequences of the ages of the individual regions, and it renormalizes the derived parameters \replaced{to the space that shows a reasonable agreement with the isochrones, but it introduces significant scatter}{to the appropriate range, but it does show a somewhat peculiar behavior especially at low \teff\ as it attempts to reconcile the differences between various bands and the isochrones.}

Finally the APOGEE Net connects the derived parameters for the YSOs to those for the M dwarfs and the red giants, making it possible to interpret the resulting values, and removing all of the systematic gaps that persisted in the previous iterations. However, there may still be a weak systematic offset in the shape of the isochrones and the YSOs at \teff$<3500$ K, as at a particular age, the isochrone tends result in somewhat lower \logg\ at lower \teff. On the other hand, \logg\ traced by the individual stellar populations either remain flat throughout or slightly increase towards higher values at low \teff, oriented in parallel to the main sequence. Considering that the photometric labels showed the opposite trend at lower \teff, it is unclear how such a potential discrepancy could be better rectified in the future, or if the cause of the discrepancy is necessarily in the predicted parameters as opposed to the isochrones.

The spectroscopic parameter space that the low mass YSOs occupy can be rather cleanly separated from the other stellar objects, as they have \logg\ typically \replaced{higher}{lower} than the main sequence stars, and \teff\ typically cooler than the red giants. Thus, by selecting the sources in this parameter space, we can robustly identify young stars in the catalog and look at their spatial distribution (Figure \ref{fig:ysoreg}). With a simple cut, restricting the selection bound by (\teff, \logg) of (2200,4.4) (4500,3.9), and (3200,1.7) it is possible to recover all of the star forming regions in the sample: the Orion Complex, Perseus clusters NGC 1333 and IC348, as well as NGC 2264. Some of the sources from Pleiades end up in this selection as well. Surprisingly, some fields (K2\_C4\_172-20 and K2\_C4\_177-21) also appear to include sources from the Taurus Molecular Clouds. While Taurus has been observed with APOGEE, the dedicated observations of young stars in this region have not begun until after the release of DR14. Thus, the sources we see have been observed serendipitously.

The aforementioned selection does not recover all of the young stars that have been observed, particularly those that are hotter or those that are somewhat more evolved.  Additionally, approximately 6\% of the sources from this simple selection are scattered all across the sky, many at high galactic latitudes, which do not appear to be associated with any particular star forming region, suggesting that they are impostors, i.e, contamination.

The distribution of the selected sources in the spectroscopic space is comparable to the distribution of sources in the HR diagram. Even the impostors appear to be bona fide sub-giants \added{(i.e., appearing above the main sequence and below the red giant branch. It is unclear what mechanism may drive them, although these sources appear to be fast rotators)}. Furthermore, this comparison demonstrates that it is possible to use the spectoscopically derived parameters to effectively derive stellar properties, even for sources that are located in regions of high extinction, those have very uncertain parallaxes, or no parallax measurements at all.

\subsection{Ages}

\begin{figure*}
\plottwo{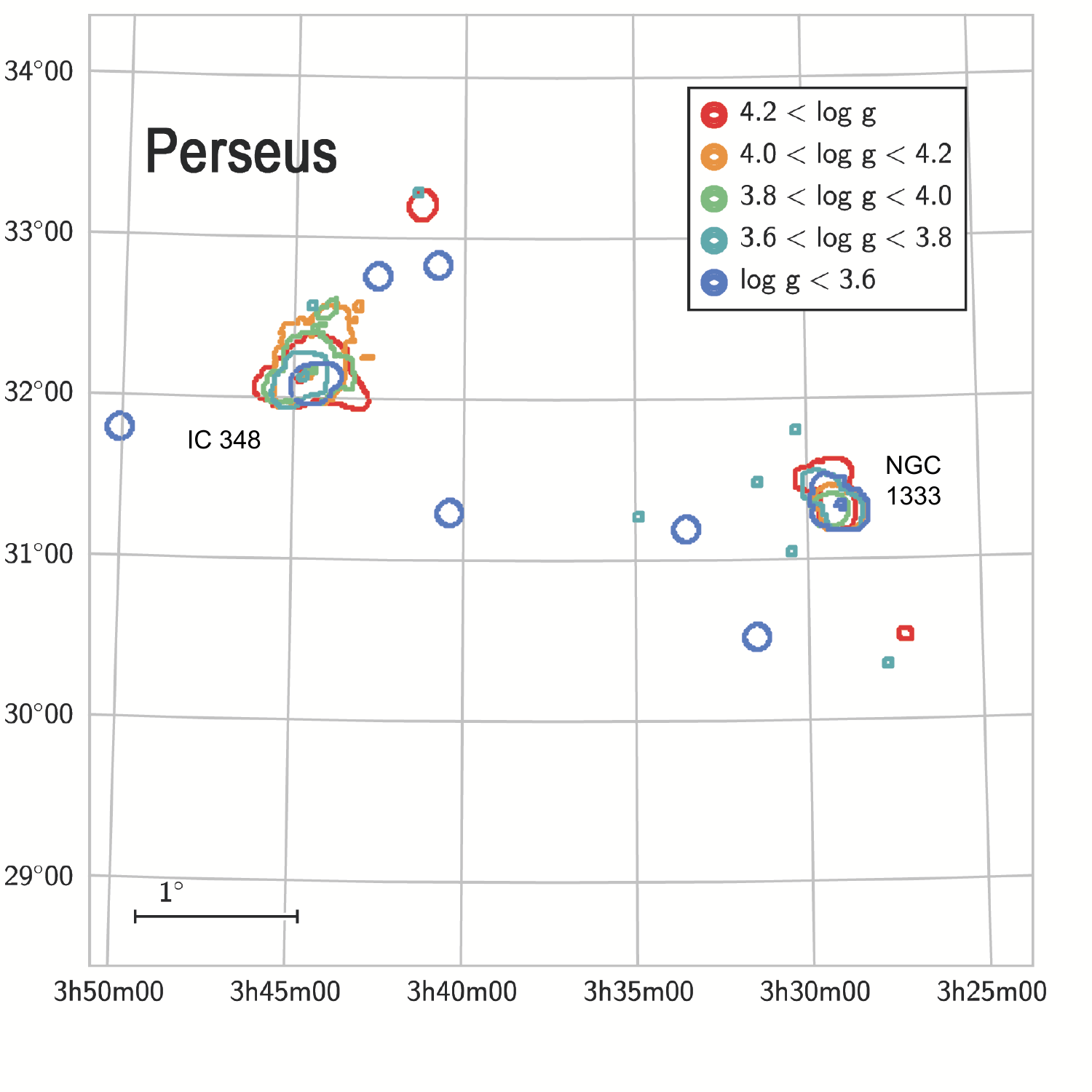}{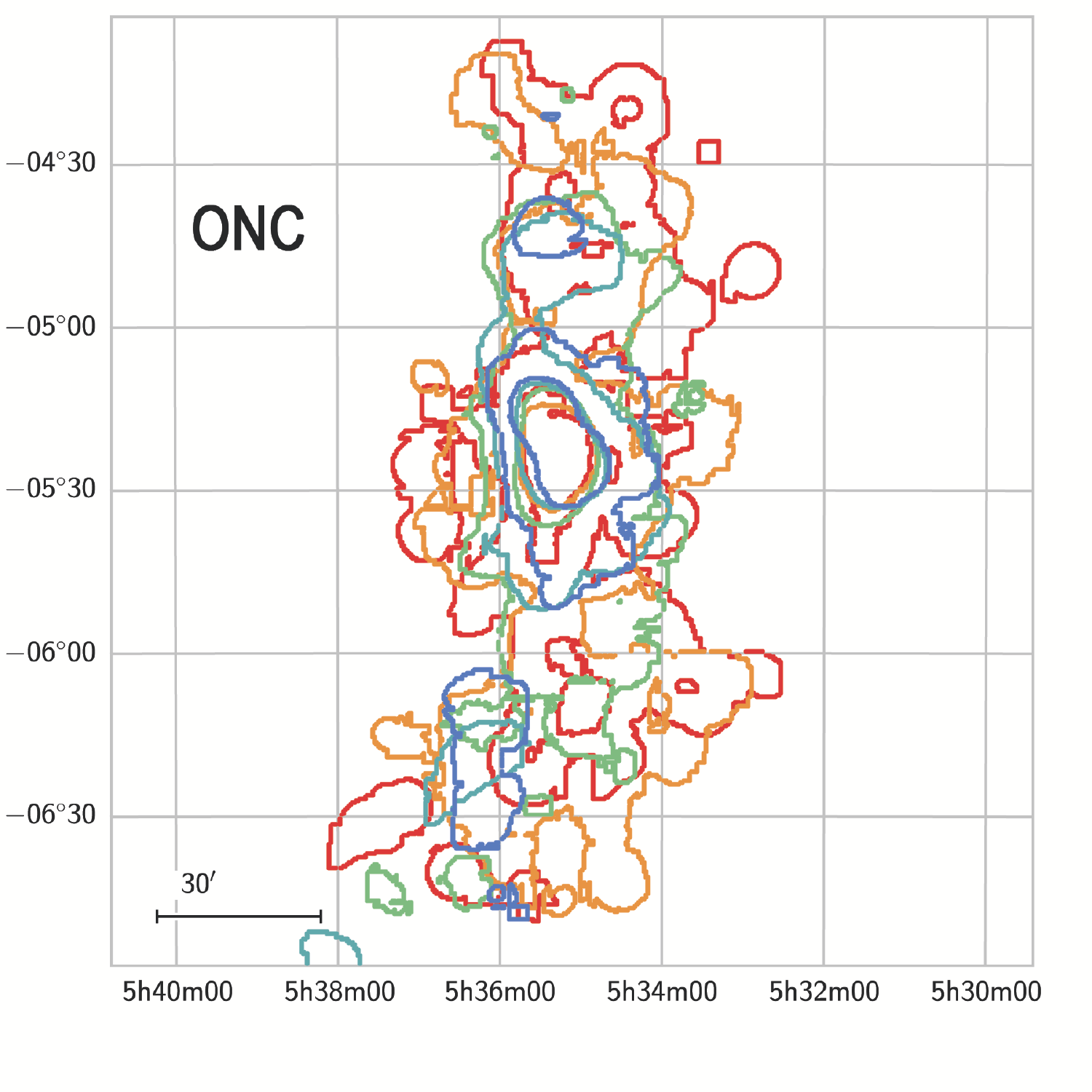}
\plottwo{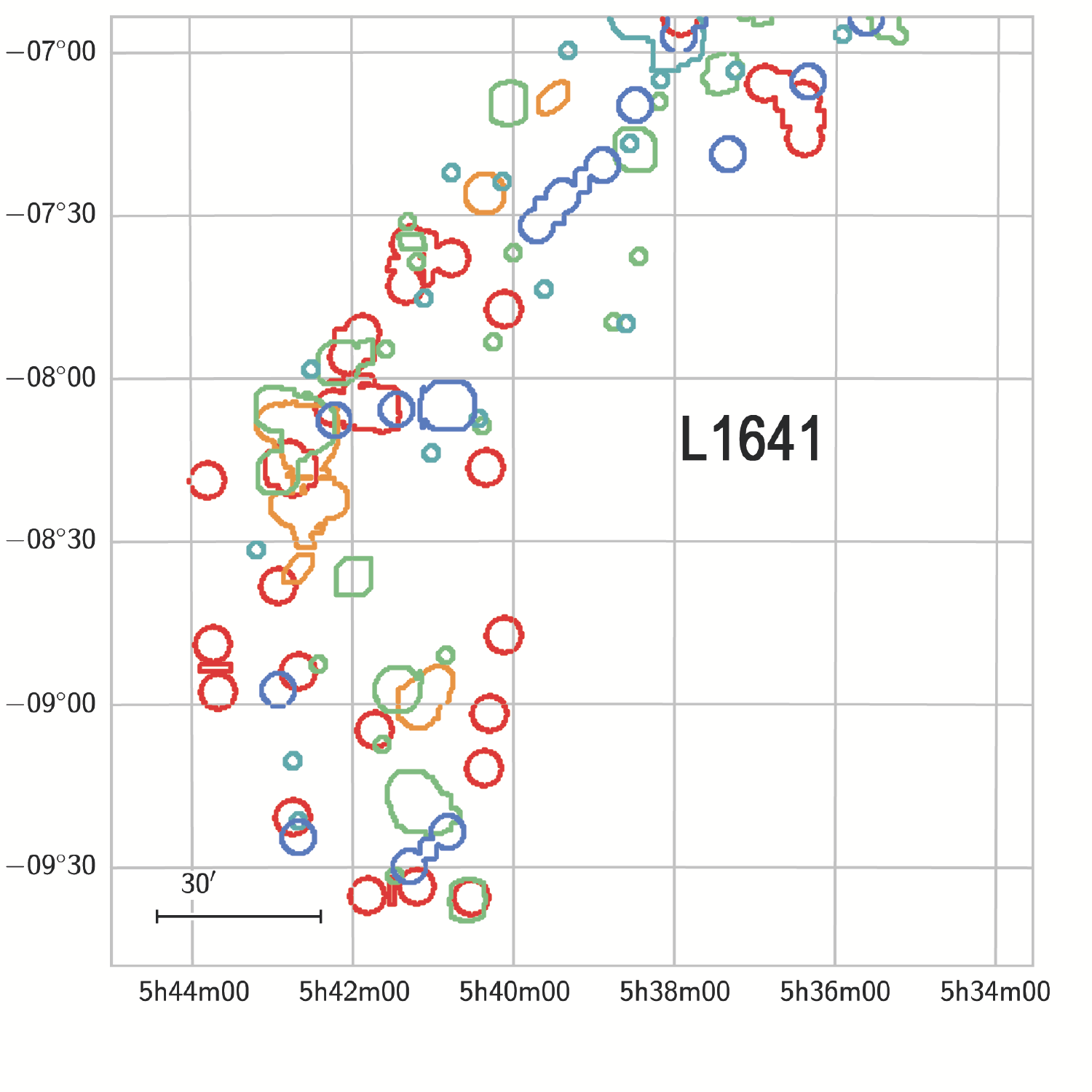}{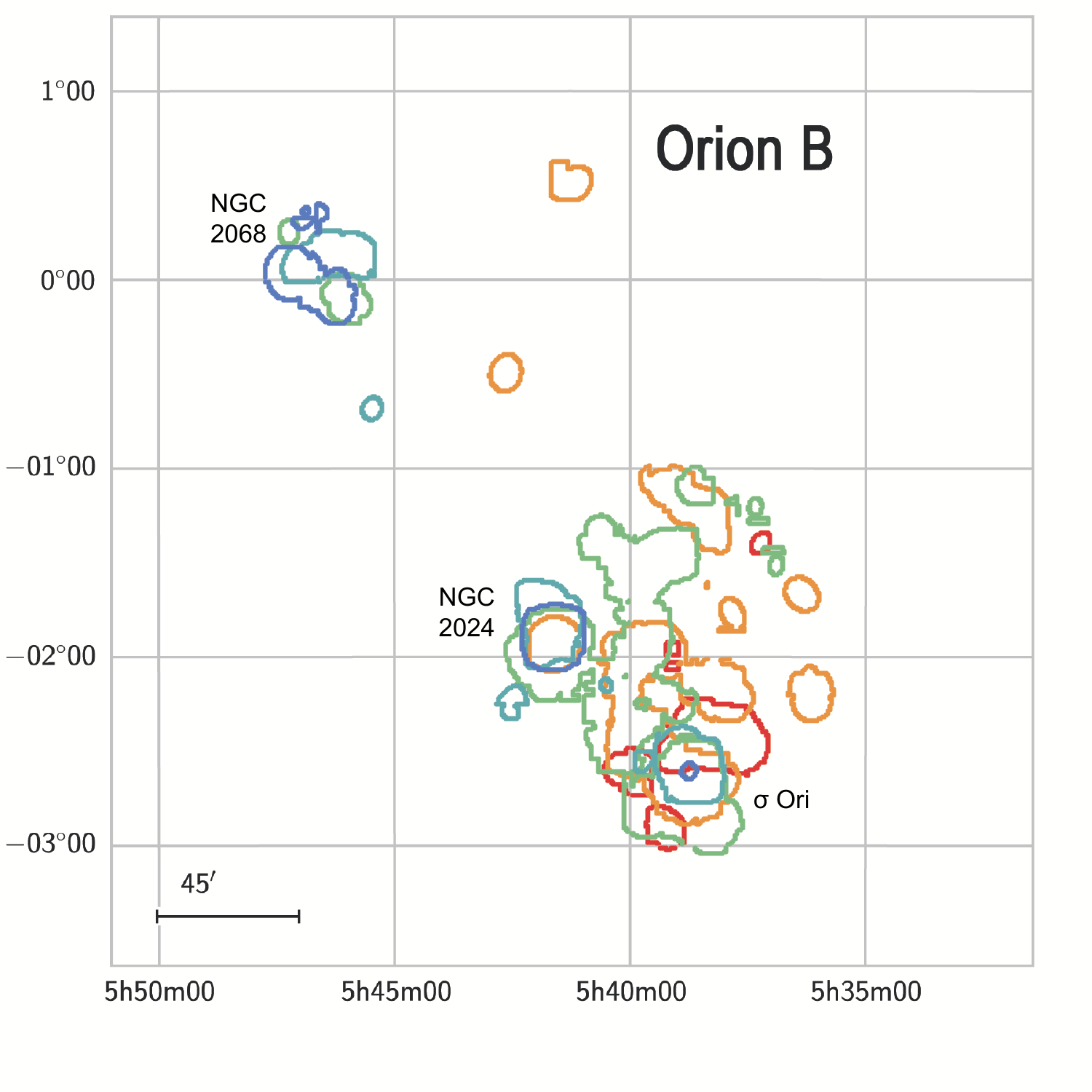}
\plottwo{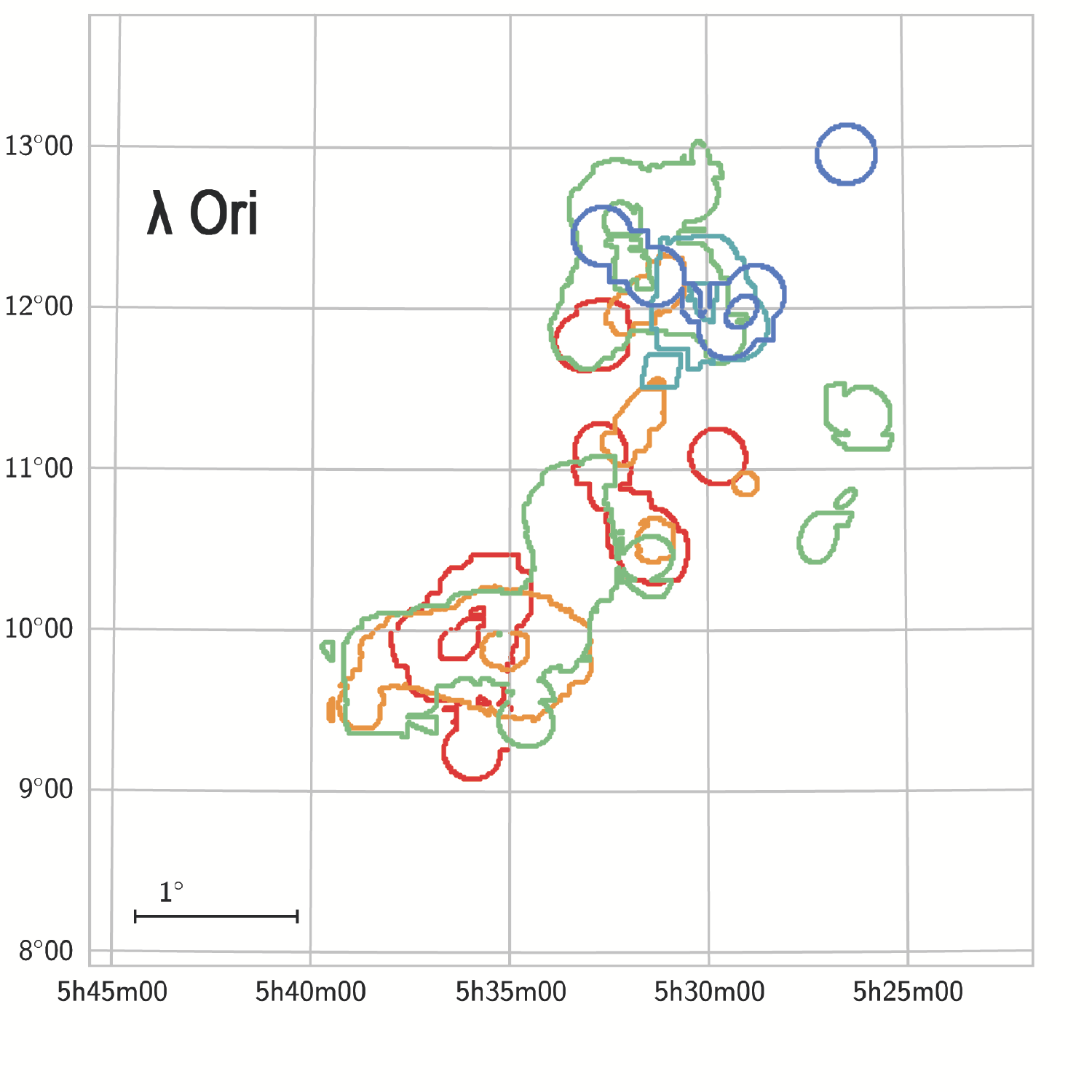}{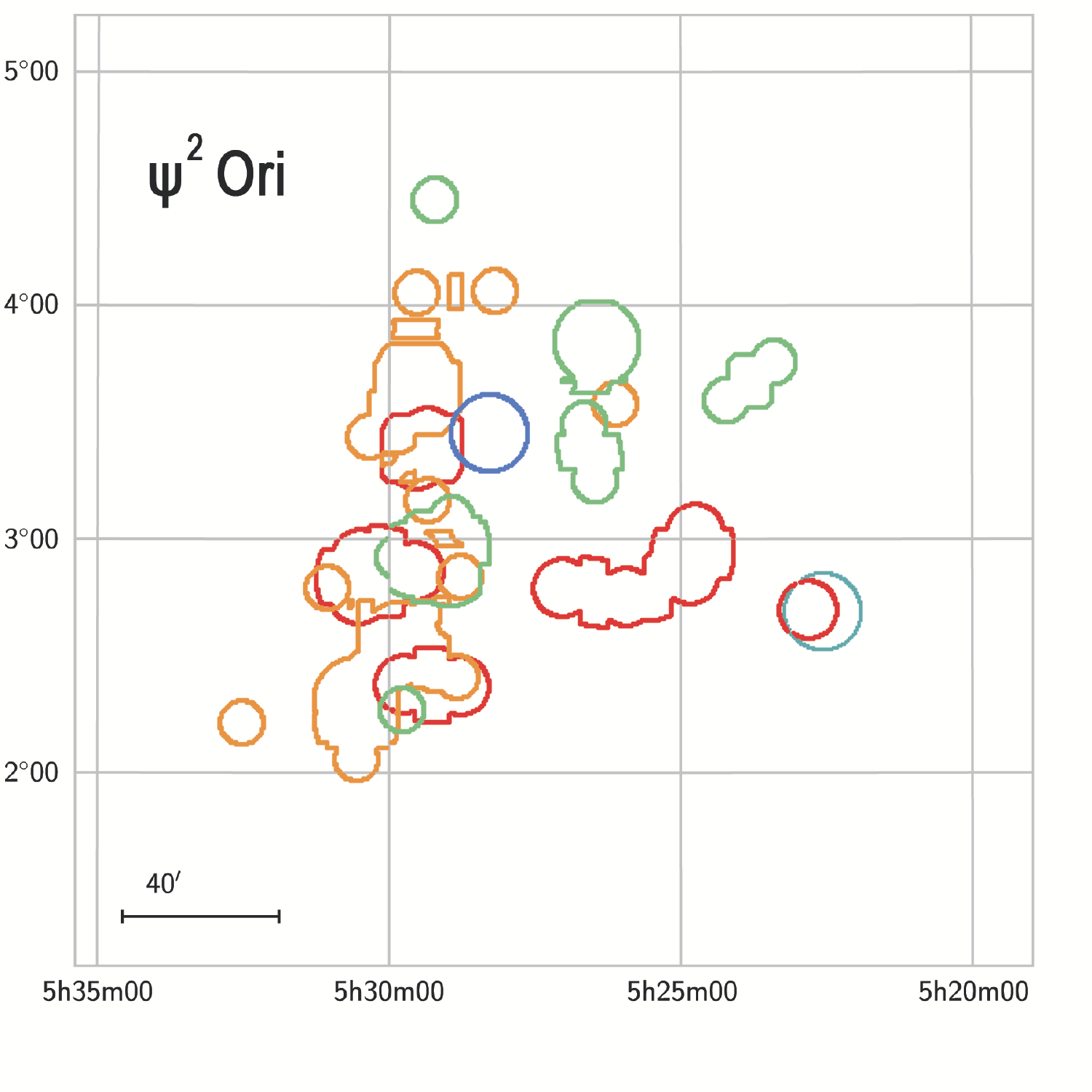}
\caption{Contour plots showing the density distribution of sources with different \logg\ values, as a proxy for age, color-coded from the youngest ($\lesssim$1 Myr, blue), to the oldest ($\gtrsim$5 Myr, red)}
\label{fig:density}
\end{figure*}

It is possible to use the derived \logg\ values as a proxy for age. We have not explicitly interpolated the \teff\ and \logg\ across the isochrones, however we divided the sample into 5 bins: \logg$<3.6$ dex ($\lesssim$1 Myr), $3.6<$\logg$<3.8$ dex (1$\sim$2 Myr), $3.8<$\logg$<4.0$ dex (2$\sim$3 Myr), $4.0<$\logg$<4.2$ dex (3$\sim$5 Myr), and $4.2<$\logg\ dex ($\gtrsim$5 Myr). We then constructed a density map of sources in each bin (Figure \ref{fig:density}). \added{This is done to further test the \logg\ on the more granular scale, in order to compare these age estimates to the known distribution of ages in various populations in the sample}.

In Orion, the distribution of ages is largely consistent what has been previously measured for each individual region. For example, in $\lambda$ Ori, the central cluster is $\sim$5 Myr, and the outer shell that has been triggered by a supernova, consistent with what has been previously measured by \citet{kounkel2018a}. In the vicinity Orion B, there is a clear separation between $\sim$1 Myr old clusters, NGC 2068 and NGC 2024, a somewhat older (2--3 Myr) cluster $\sigma$ Ori, and 3--5 Myr extended population associated with it. A similar agreement with previously measured ages can also be observed in $\psi^2$ Ori.

\begin{figure*}
\epsscale{1.1}
\plottwo{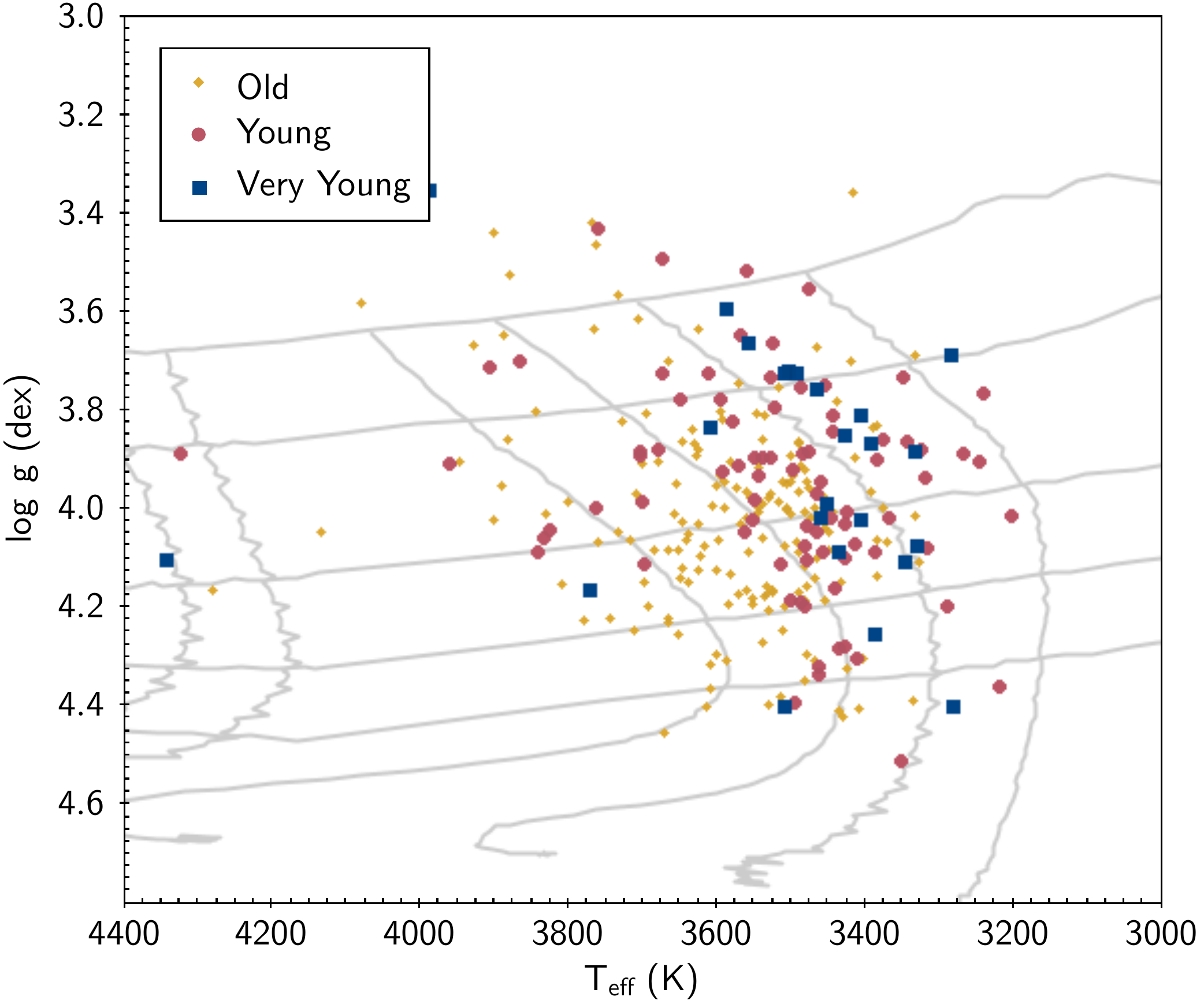}{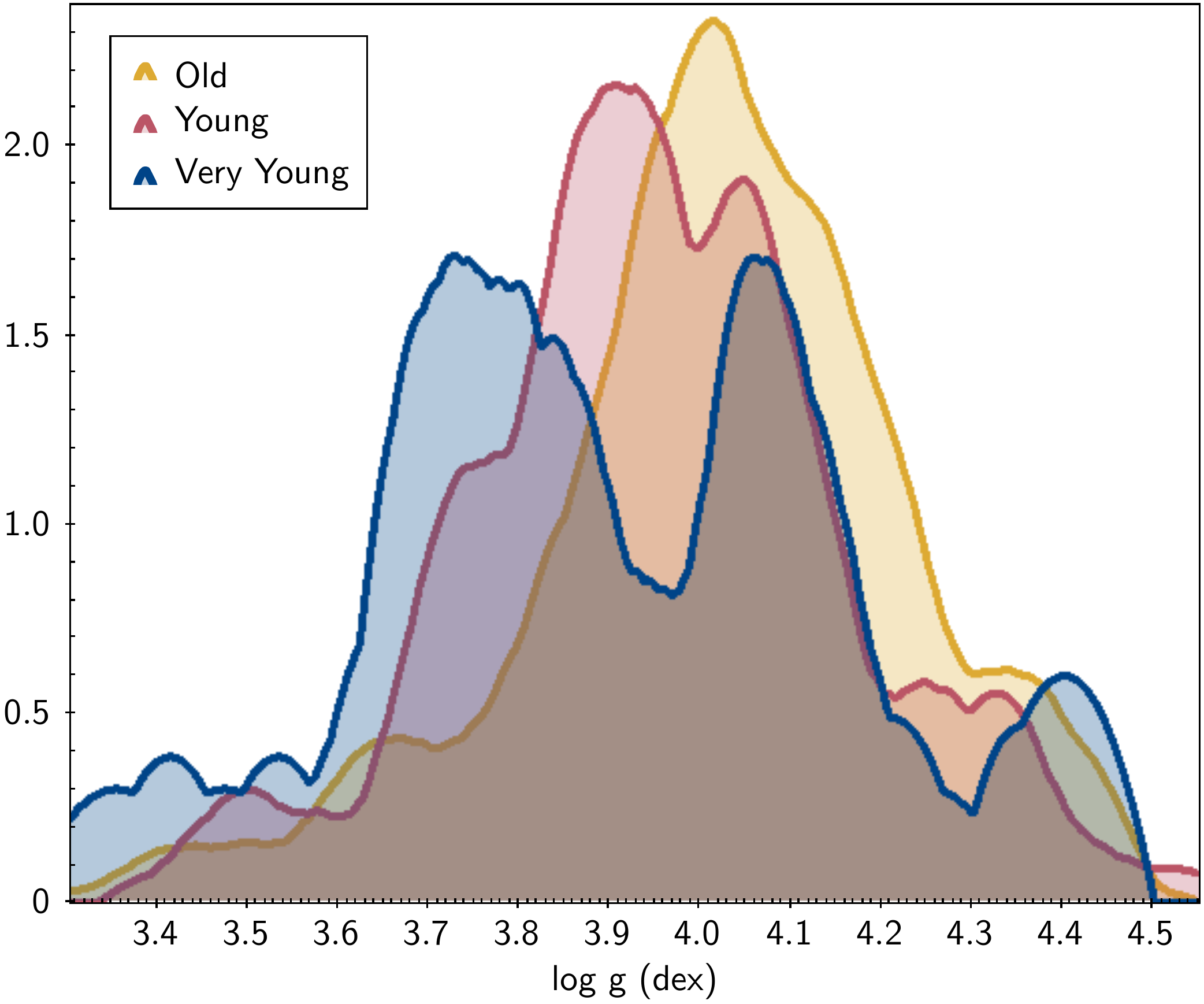}
\plottwo{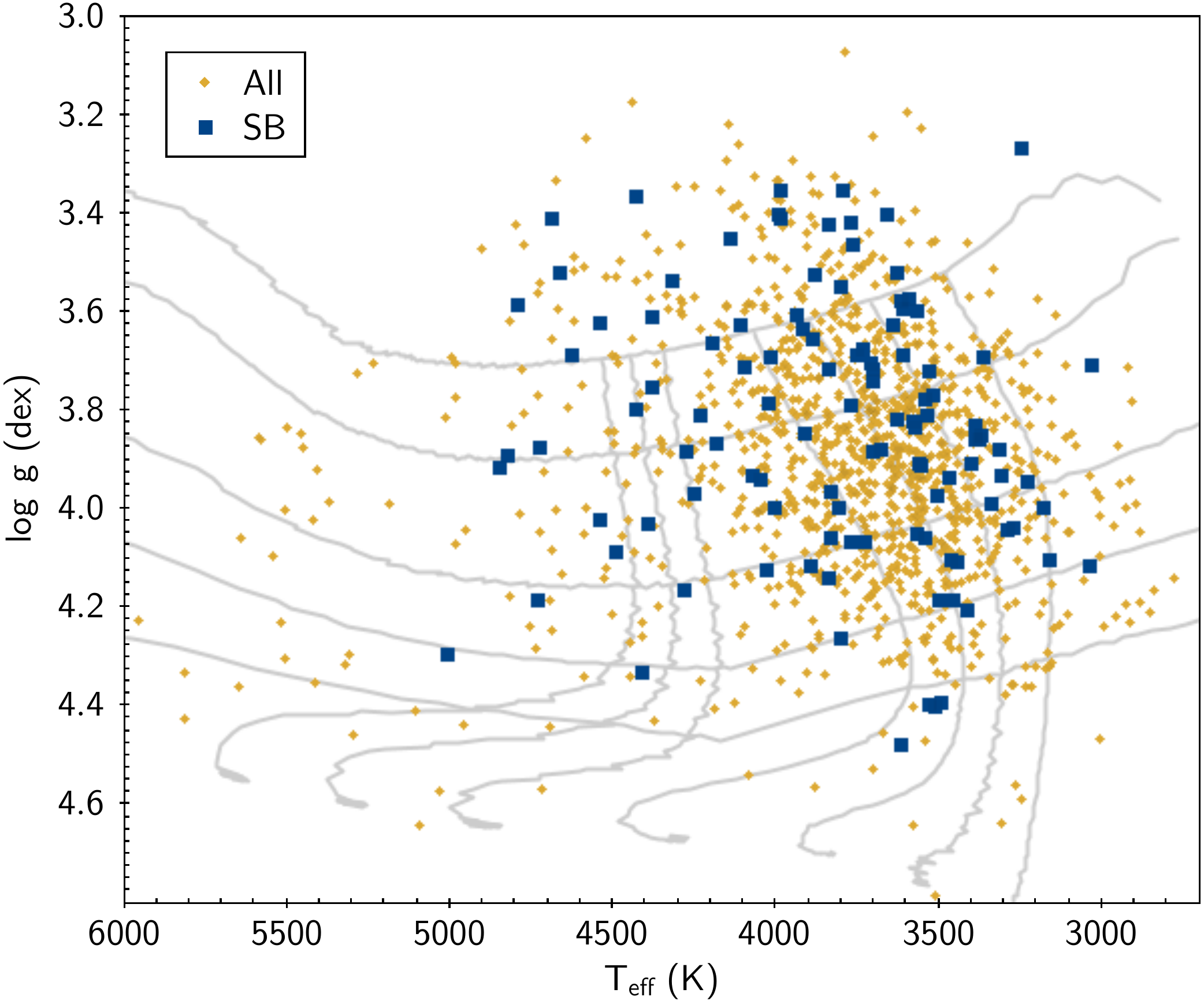}{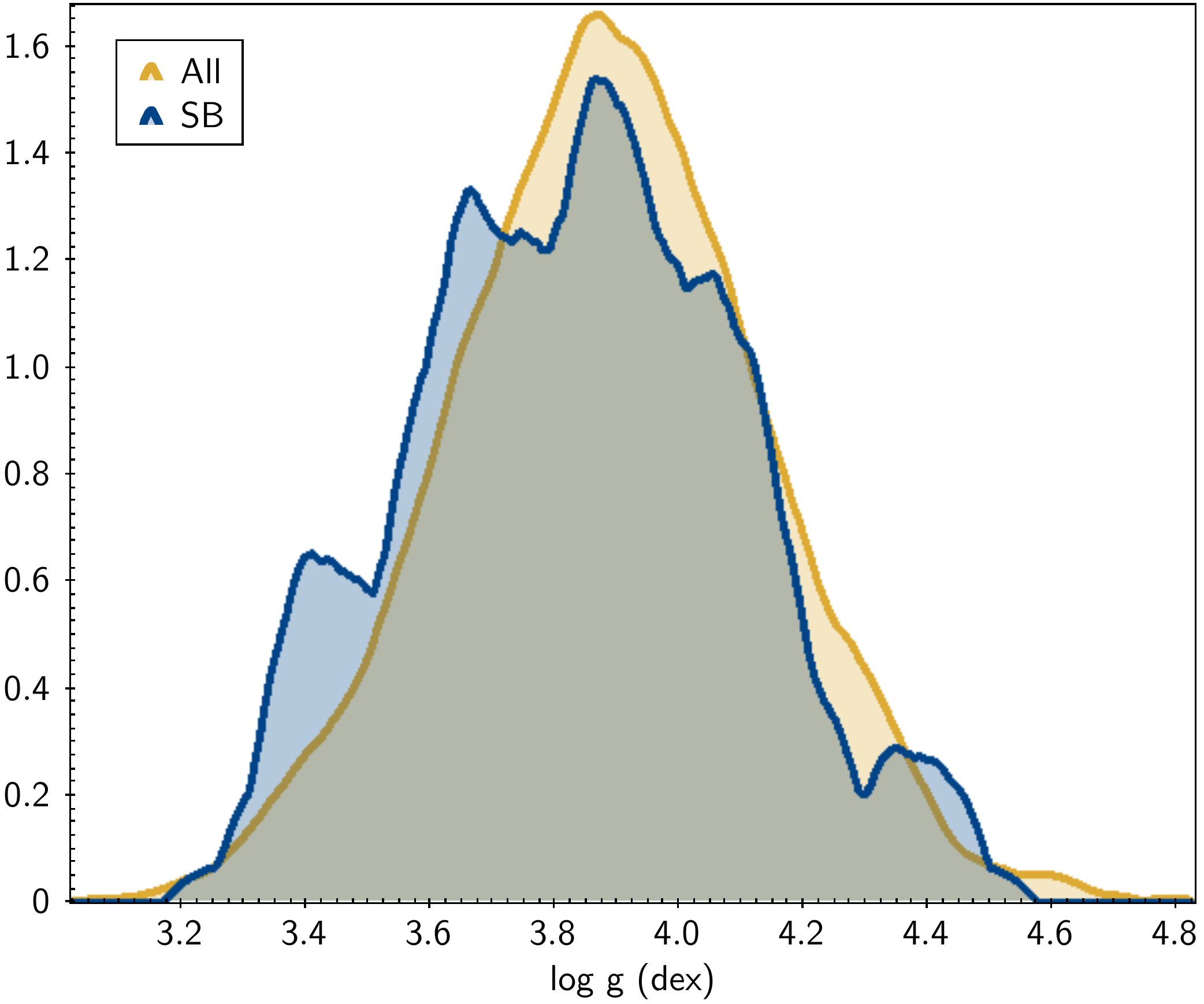}
\plottwo{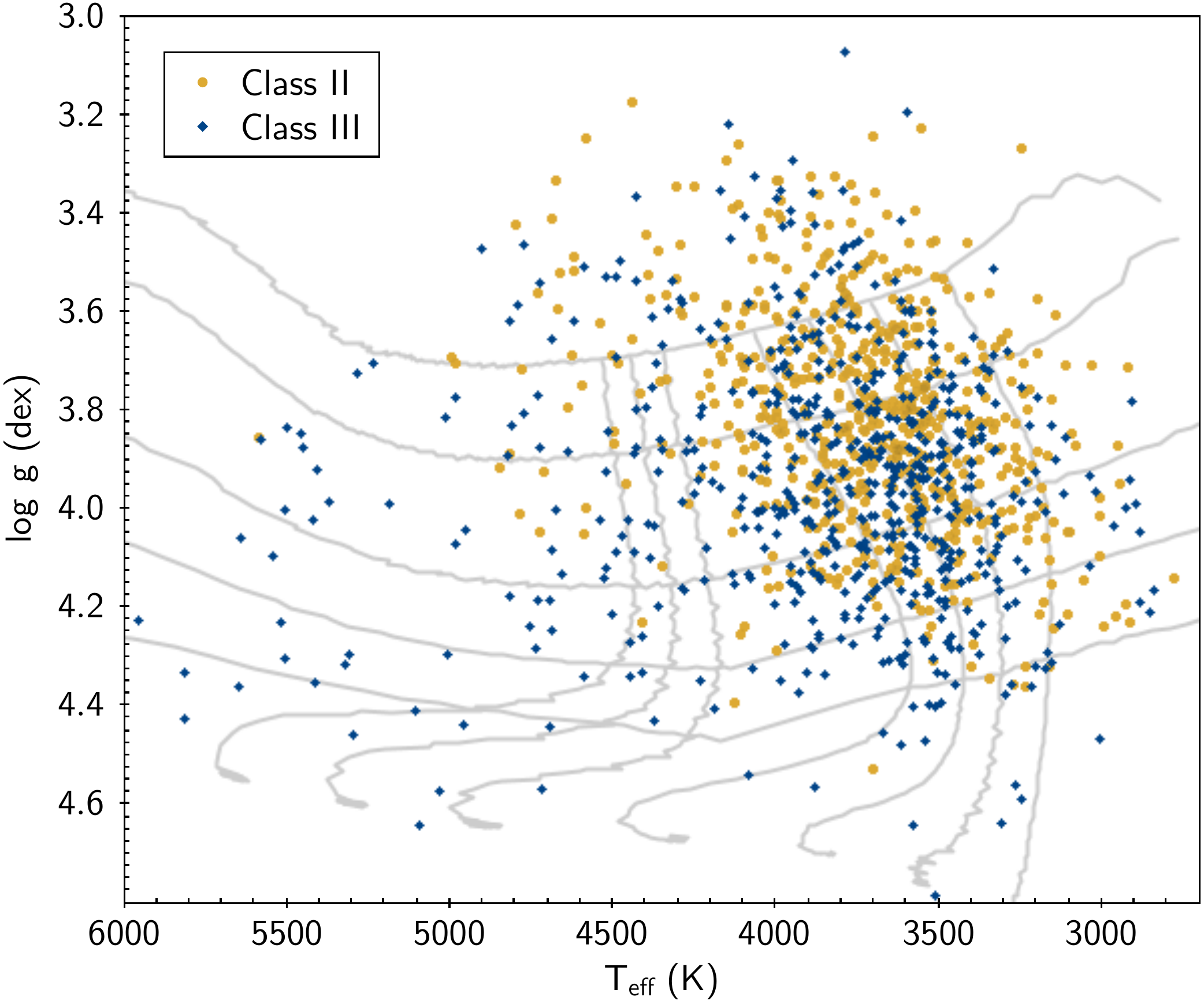}{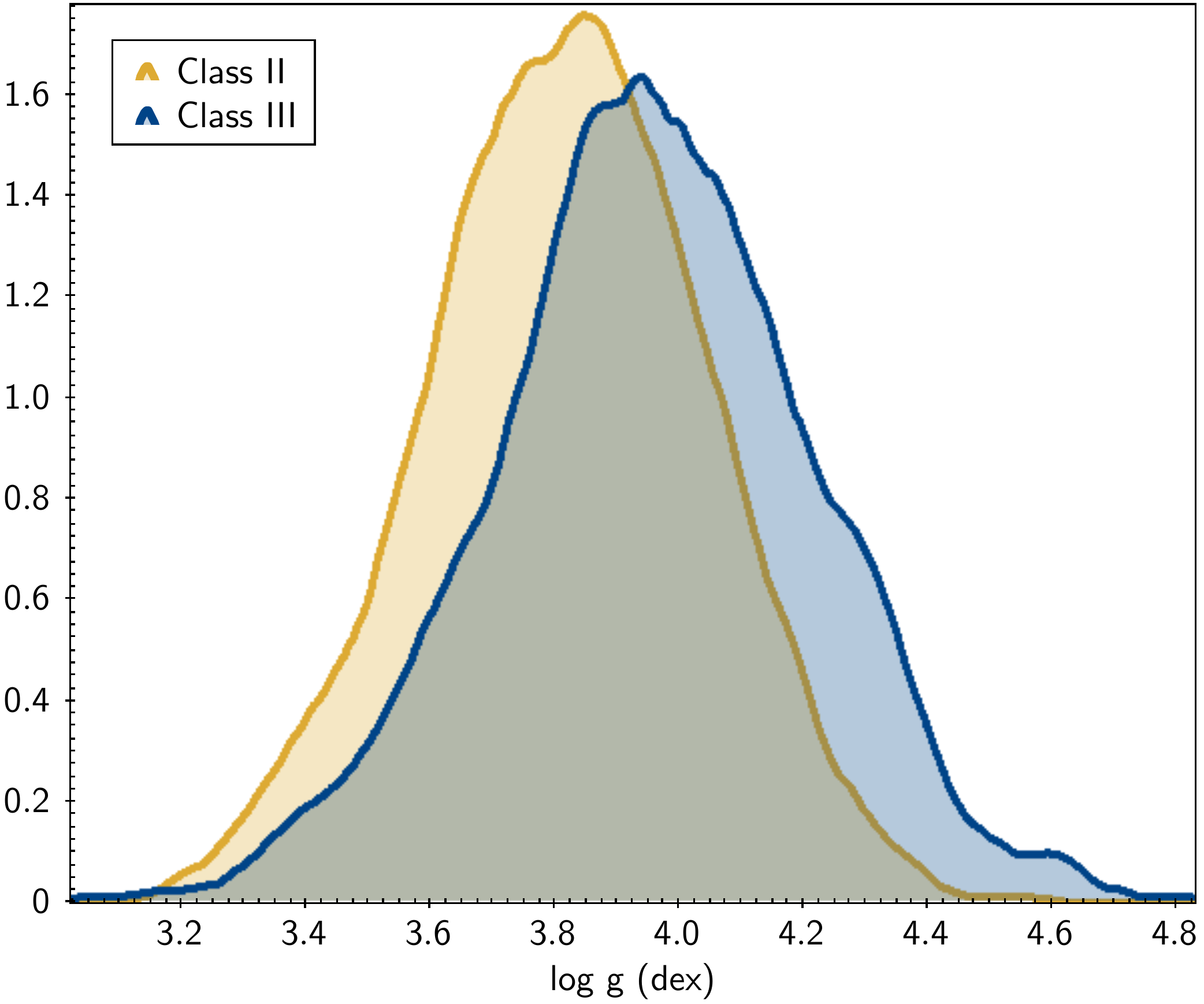}
\caption{Left: distribution of \teff\ and \logg\ for the sources in Orion A shown against 1, 2, 5, 10, and 20 Myr isochrones. Right: Kernel Density Estimation of the \logg\ with the kernel size of 0.1 dex. Top panels show the three populations identified by \citet{beccari2017}. Middle panels highlight spectroscopic binaries from \citet{Kounkel2019}. Bottom panels show the difference in the distribution of disk-bearing Class II YSOs, and diskless Class III photospheres.
\label{fig:ONC}}
\end{figure*}

In the ONC, \citet{beccari2017} have observed three distinct populations with the mean ages of 1.2, 1.9 and 2.9 Myr, and that these populations, while overlapping, cover different volume in the sky, with the youngest one being the most centrally concentrated, and the oldest one being the most diffuse. However, they could not rule out that these populations are not due to an unresolved binary and tertiary sequences. Crossmatching against their catalog of sources does indeed show that the these populations are separated in the \logg\ space, independently confirming different ages (Figure \ref{fig:ONC}, top). While there is some crossover between the sources, the ``Very Young'' sources peak at \logg=3.7 dex, corresponding to $\sim1$ Myr. The ``Young" sources peak at \logg=3.9 dex, or at the age of $\sim2$ Myr, and the ``Old" sources peak at \logg=4 dex, or at $\sim3$ Myr, consistent with the original age estimates.

It should be noted, though, that in the full spectroscopic sample, there do not appear to be distinct \added{age} sequences, rather, the distribution appears to be more continuous. This does not appear to be explained by the smearing of the sequences from the uncertainties. But, examining the density of sources at a given age slices in Figure \ref{fig:density}, similarly to \citet{beccari2017}, we do indeed find that the youngest sources are located primarily close to the center of the cluster, while the older stars are more distributed throughout the cluster. A similar trend can also be seen in Perseus, in IC348. 

Along the L1641 there are curious chains of coeval stars running in parallel to the cloud. \citet{hacar2018} have found a large network of gas fibers towards the ONC, any one of which is in process of forming stars along their length, and the stars that would form from the same gas fiber are approximately coeval. \citet{hacar2016} have previously searched for chains of stars in the APOGEE data in Orion A that are comoving in the position-position-velocity diagram, although the stellar ages have not been considered. Only a few of such chains have been found in L1641, and they did not exhibit any preferential orientation relative to the cloud. It is unclear whether the chains we see in this work is just a chance alignment, or if their arrangement could be thought of as significant.

The derived parameters are not strongly confused by multiplicity - the spectroscopic binaries identified by \citet{Kounkel2019} do not appear to occupy a systematically different \teff\ or \logg\ space from other single stars in their corresponding clusters (Figure \ref{fig:ONC}, middle). The uncertainties in the parameters also do not appear to be affected. Metallicity is the only parameter where there might be a slight, barely significant systematic shift, with spectroscopic binaries being on average more metal poor by 0.02 dex. \added{There still may also be non-systematic offsets that might affect the parameters of binary stars. A more advanced fitting of multiple templates to blended spectra would be needed to properly disentangle the companions in order to understand the magnitude of these potential offsets.}

However, there is a strong difference in \logg\ as a function of whether a YSO has a protoplanetary disk or not (Figure \ref{fig:ONC}, bottom), with the disk-bearing Class II sources having \logg\ 0.2 dex lower than their diskless Class III counterparts, although there is much overlap between the two populations. In a given cluster, this effect translates to 1--2 Myr systematic difference in age. A similar separation in \logg\ has previously been observed by \citet{yao2018} using the values derived from the IN-SYNC pipeline. It is notable that despite the fact that the aforementioned pipeline does not produce reliable absolute calibration, it does discriminate between the effects of veiling from disks and other spectral parameters. Thus, it is likely that the systematic difference we observe is real and not due to Class II sources being deviating from the model produced by APOGEE Net.

\begin{figure}
\epsscale{1.1}
\plotone{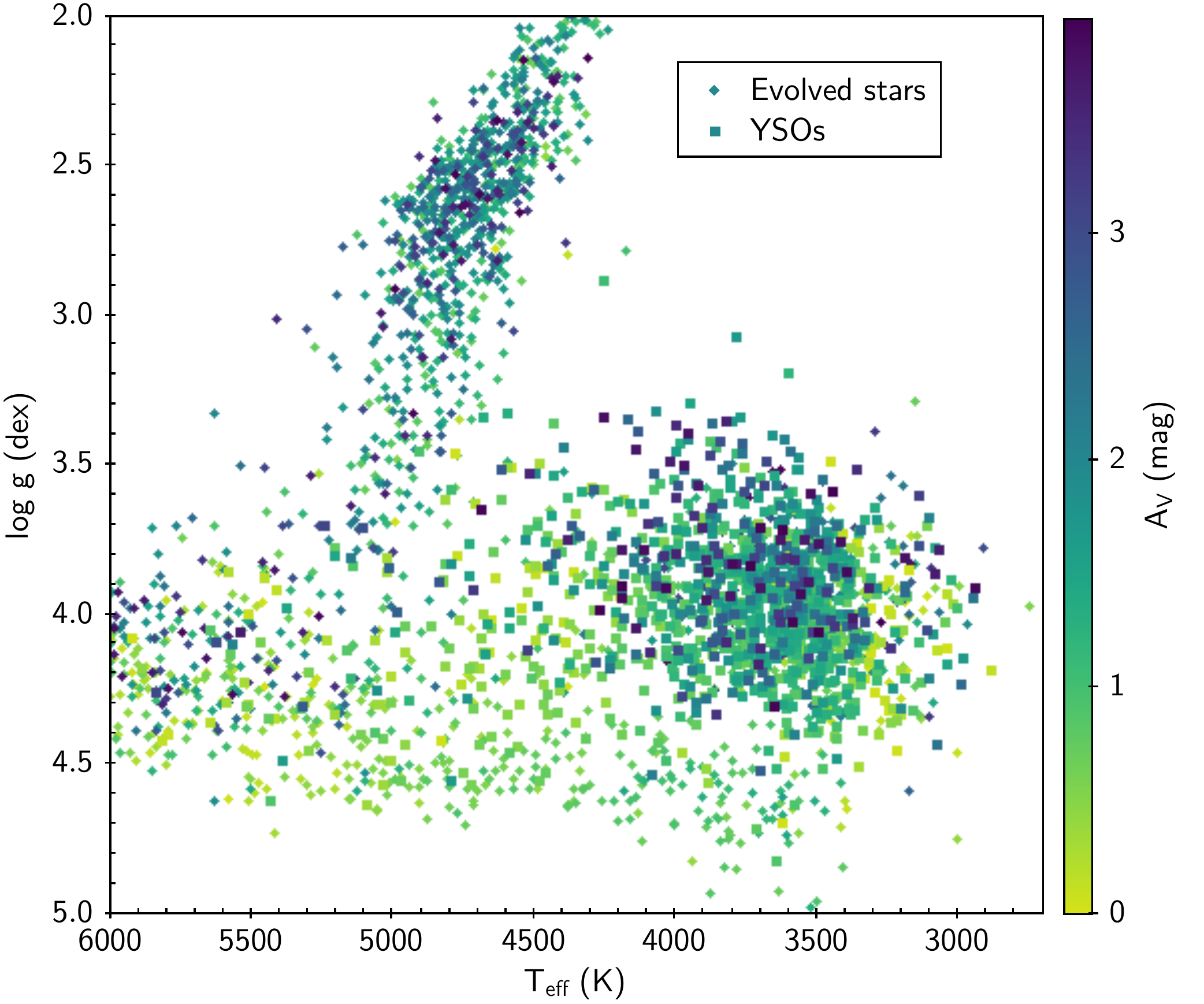}
\plotone{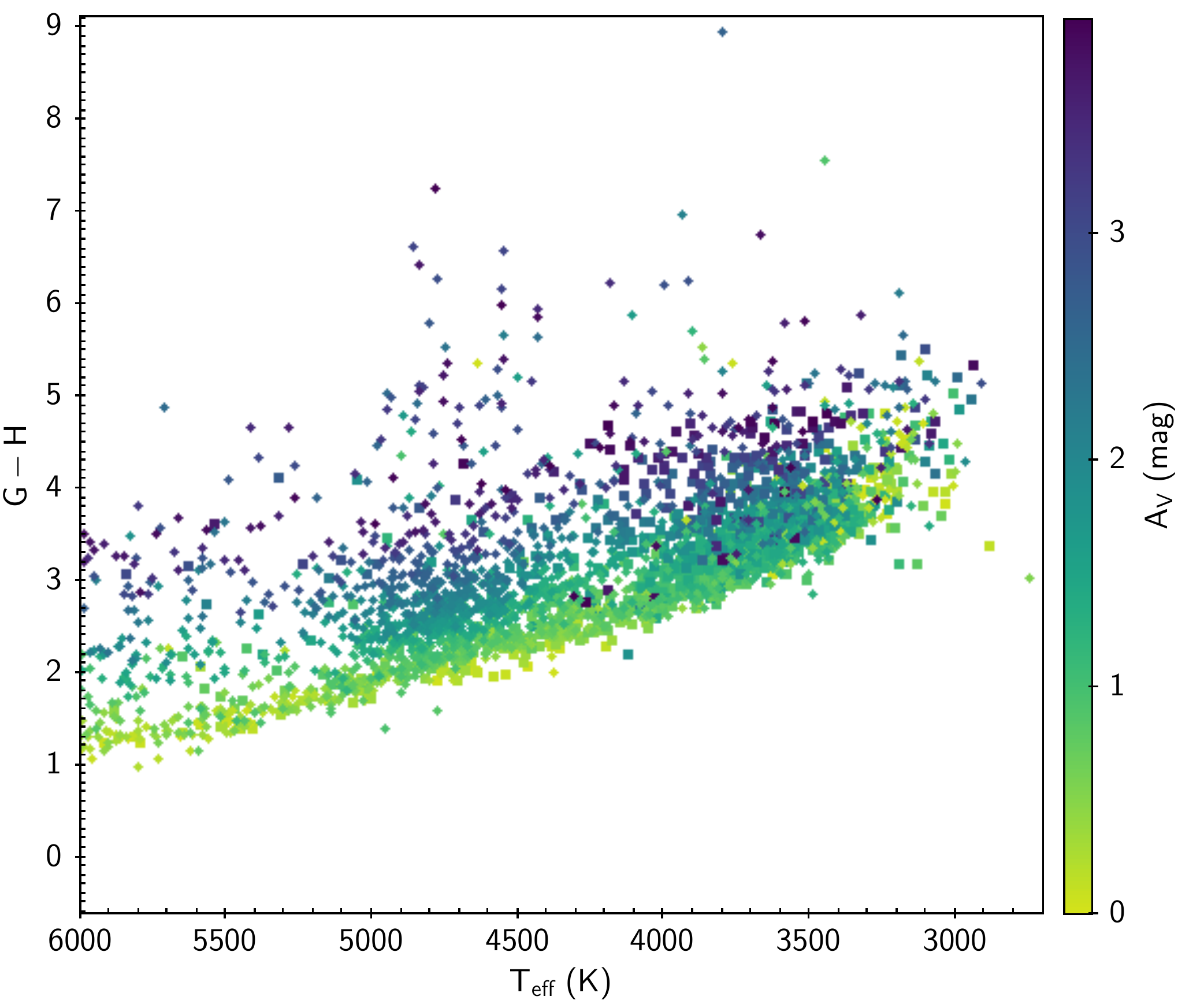}
\plotone{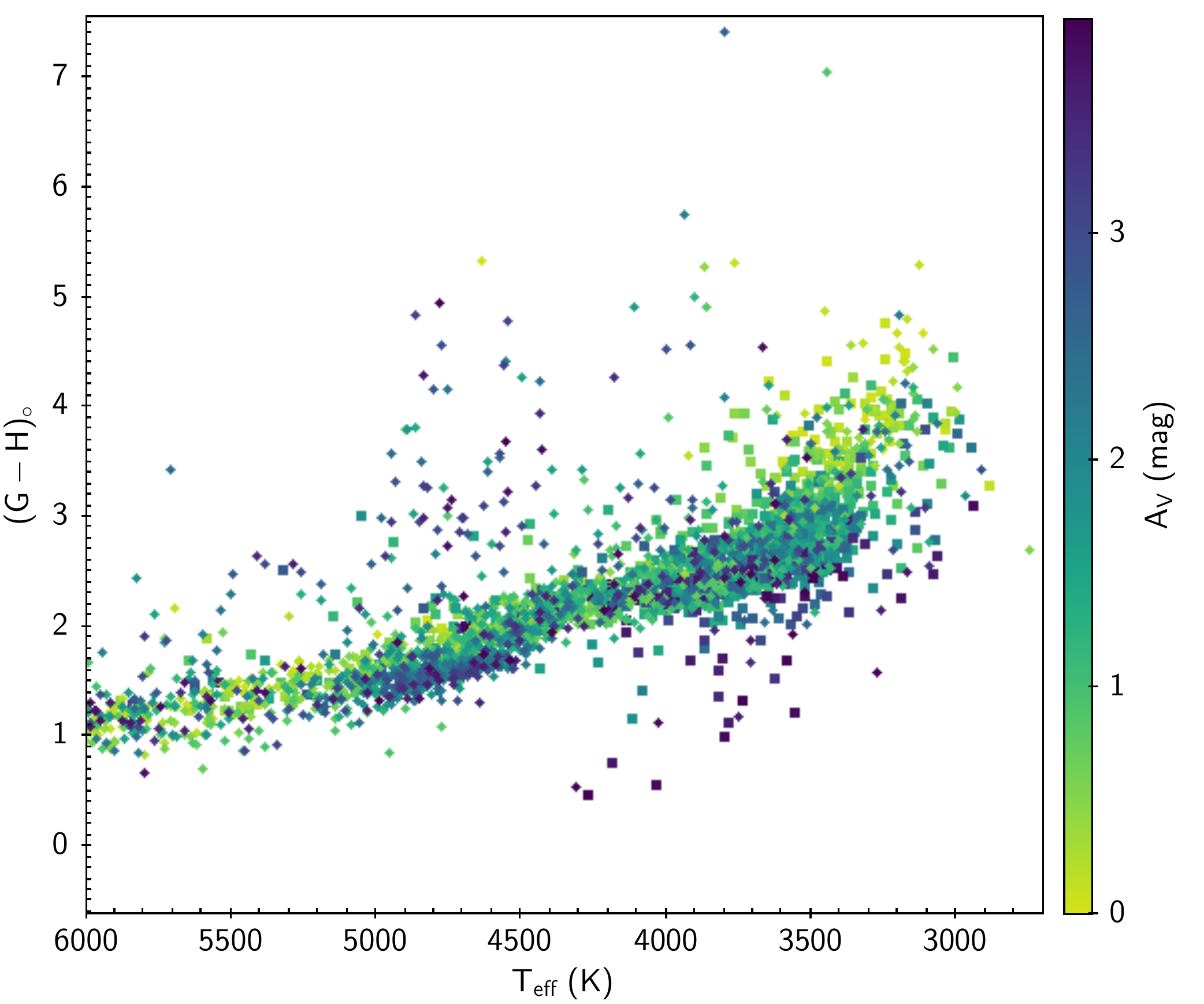}
\caption{\added{Top: Distribution of \teff\ and \logg\ of sources towards Orion, color coded by the measured $A_V$ from \citet{kounkel2018a}. Extinction tends to be higher in youngest stars, decreasing incrementally as they age. The dwarfs in the field are all nearby, without much dust along the line of sight. Red giants tend to be more extinct the further away they are. Middle: \teff\ vs $G-H$ color. Bottom: \teff\ vs the unreddened $G-H$ color}
\label{fig:av}}
\end{figure}

\added{This difference is also apparent when examining the distribution of A$_V$ (Figure \ref{fig:av}), as YSOs with higher degree of extinction from their parental cloud and protoplanetary disk are more likely to be somewhat younger compared to those whose extinction is driven only by the external foreground dust. It should be noted though that A$_V$ of the individual stars are available only for the Orion regions, and that they were computed using the previously measured \teff\ from the IN-SYNC pipeline \citep{kounkel2018a}.}

\begin{figure}
\epsscale{1.1}
\plotone{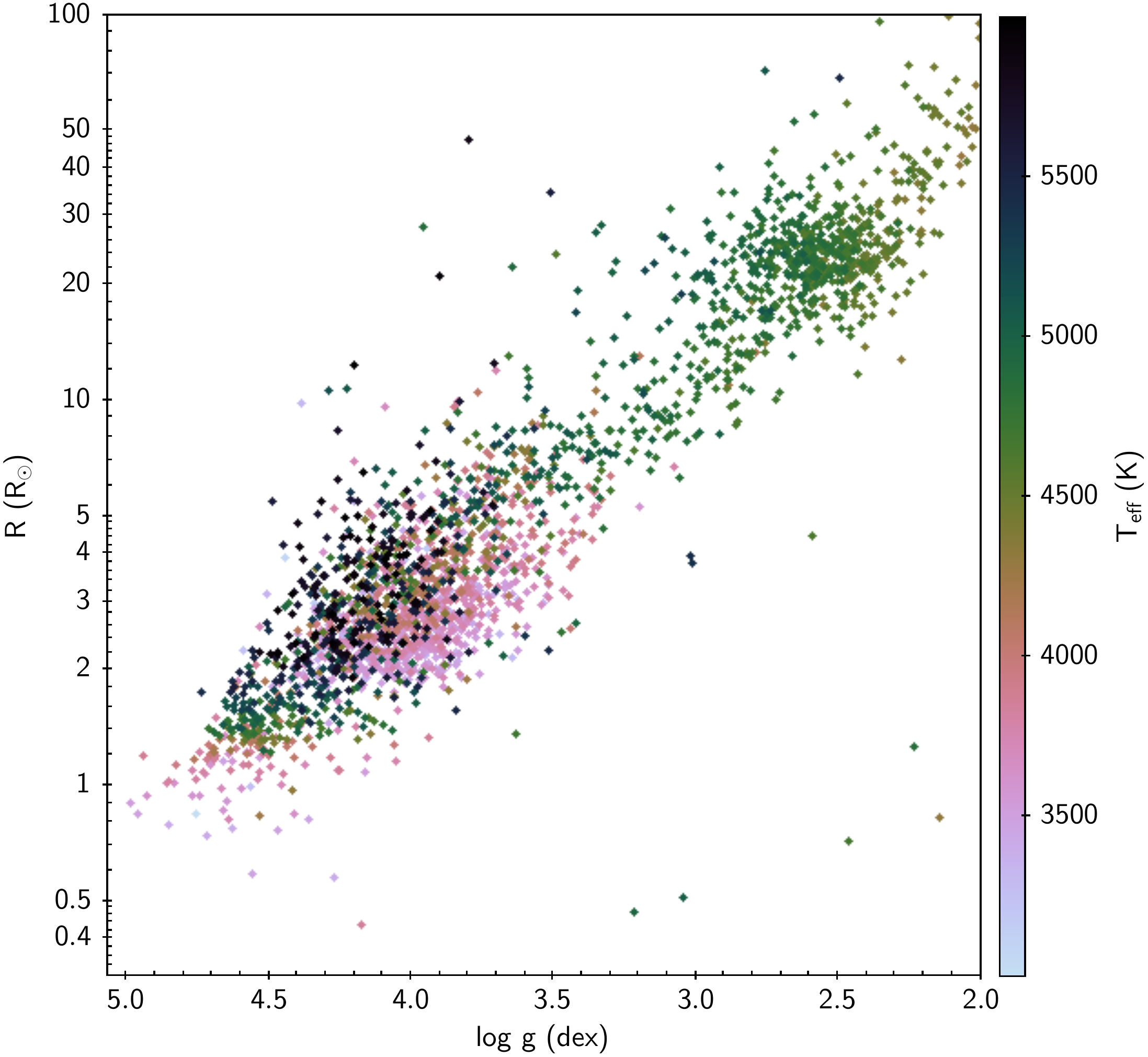}
\plotone{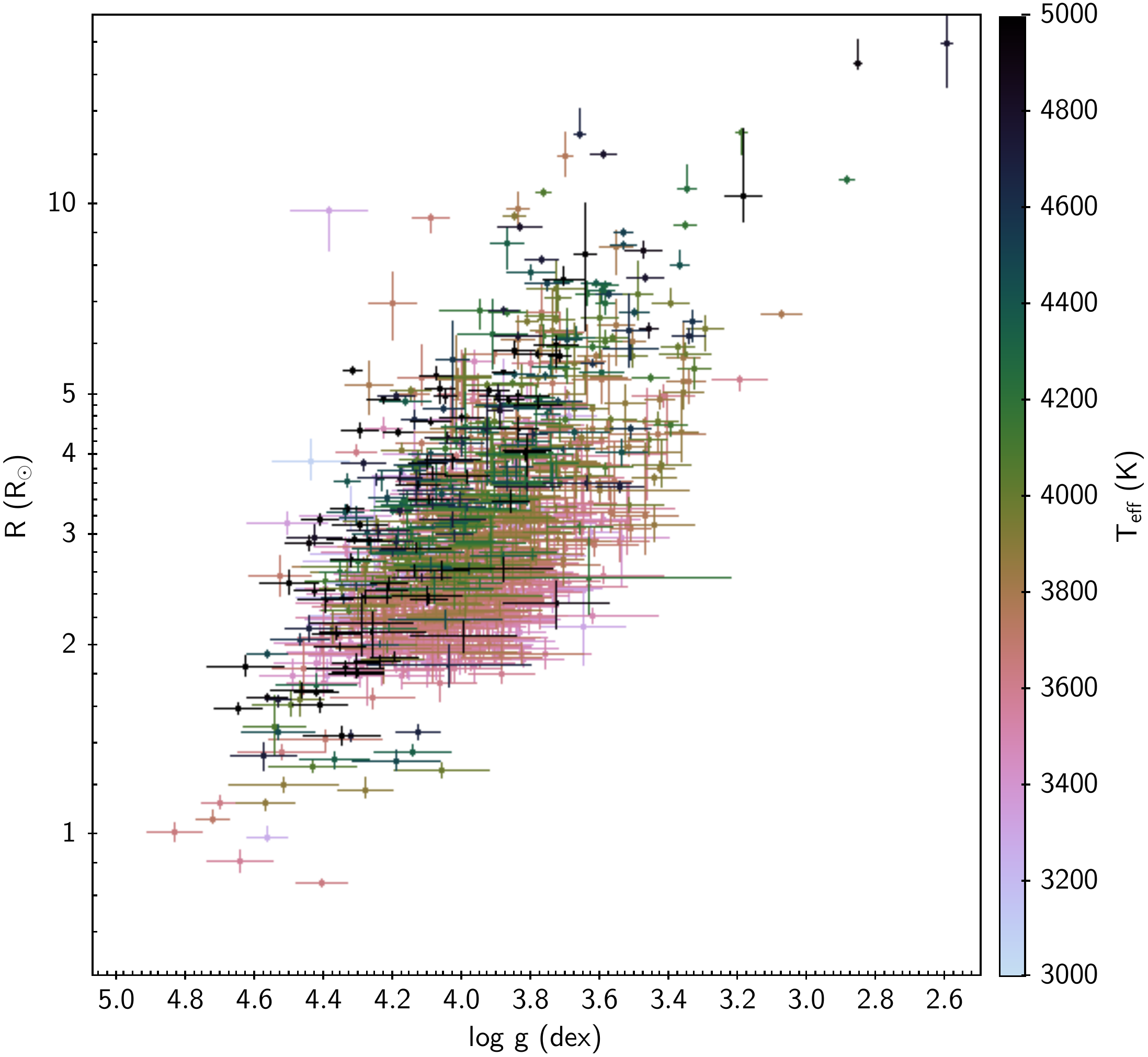}
\plotone{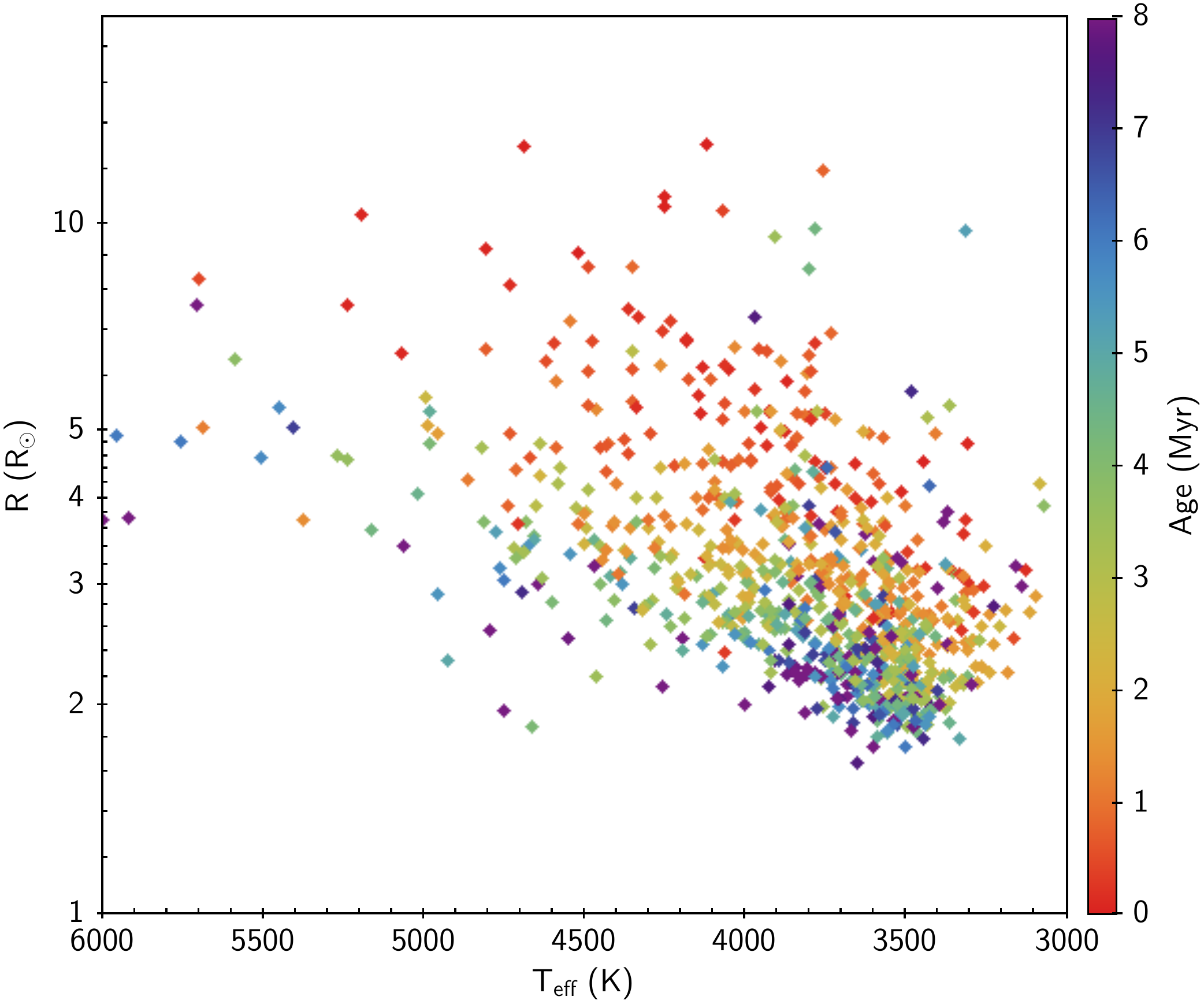}
\caption{\added{Comparison of spectroscopically derived \logg\ and stellar radii measured from the SED fitting from \citet{kounkel2018a}, color coded by \teff. Top panel shows all sources towards Orion, middle panel shows only the YSOs. Bottom panel shows the relation between the stellar radii and \teff, color coded by the photometrically derived ages of YSOs.}
\label{fig:radius}}
\end{figure}

\added{One way to independently validate our spectroscopically determined \logg\ is with the stellar radii that we have previously measured empirically via the bolometric flux from the spectral energy distribution (SED) and parallax in \citet{kounkel2018a} using methods described by \citet{stassun2017}. Since $g$ is proportional to $M/R^2$, we expect that the spectroscopic \logg\ will correlate strongly with the SED-based radius, with some secondary dependence on the stellar mass. And because the young, low-mass stars in our sample are expected to still be contracting along roughly vertical Hayashi tracks in the HR diagram, we can take \teff\ as a proxy for the stellar mass. As expected, Figure \ref{fig:radius} shows that our spectroscopic \logg\ are well correlated with the SED-based radii overall, especially when controlled for \teff. Indeed, stars of a given \teff\ (i.e., mass) show a very strong correlation between \logg\ and radius, and cooler (lower mass) stars show smaller radii for a given \logg.}

\subsection{Metallicity}

\begin{figure}
\epsscale{1.1}
\plotone{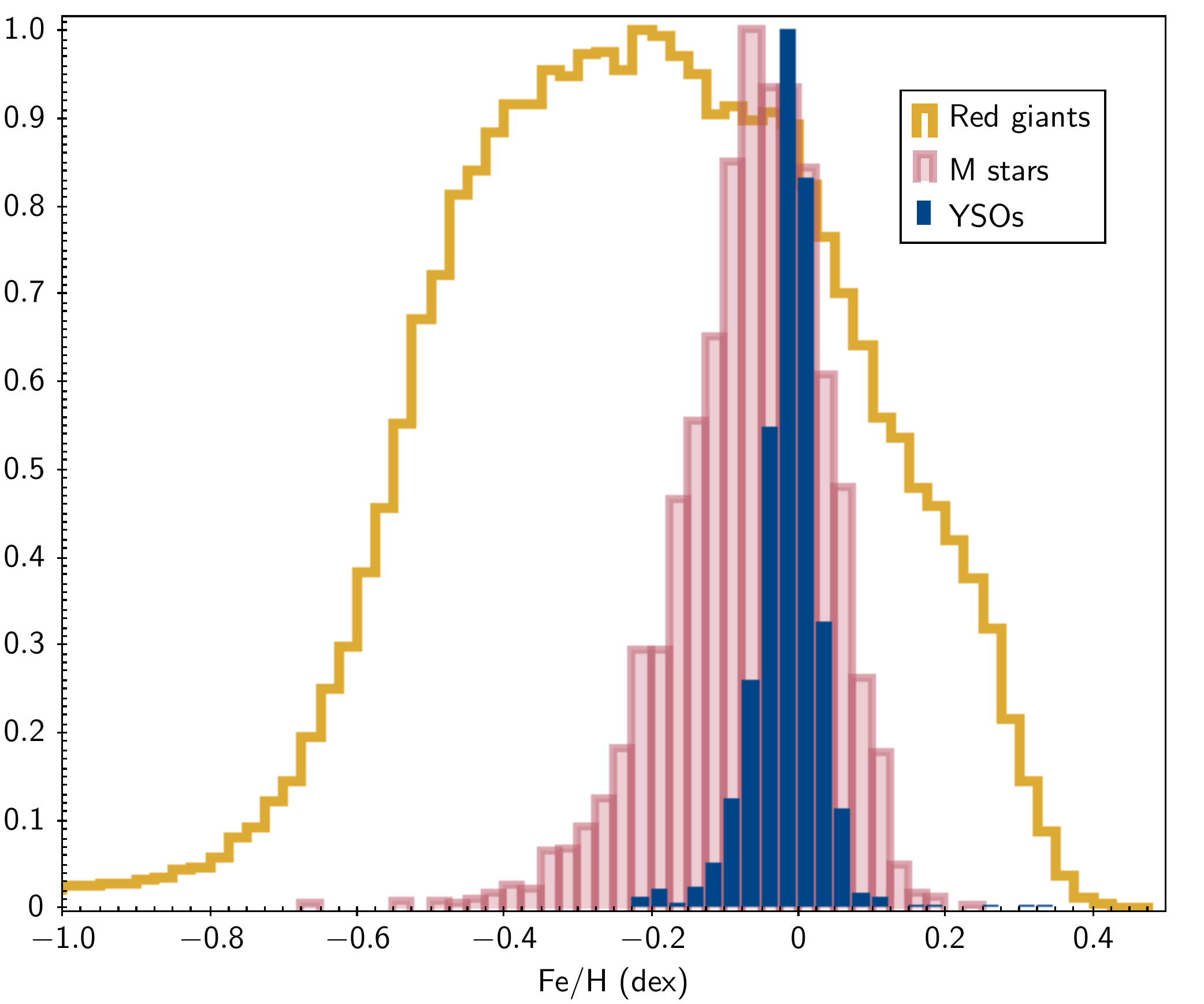}
\plotone{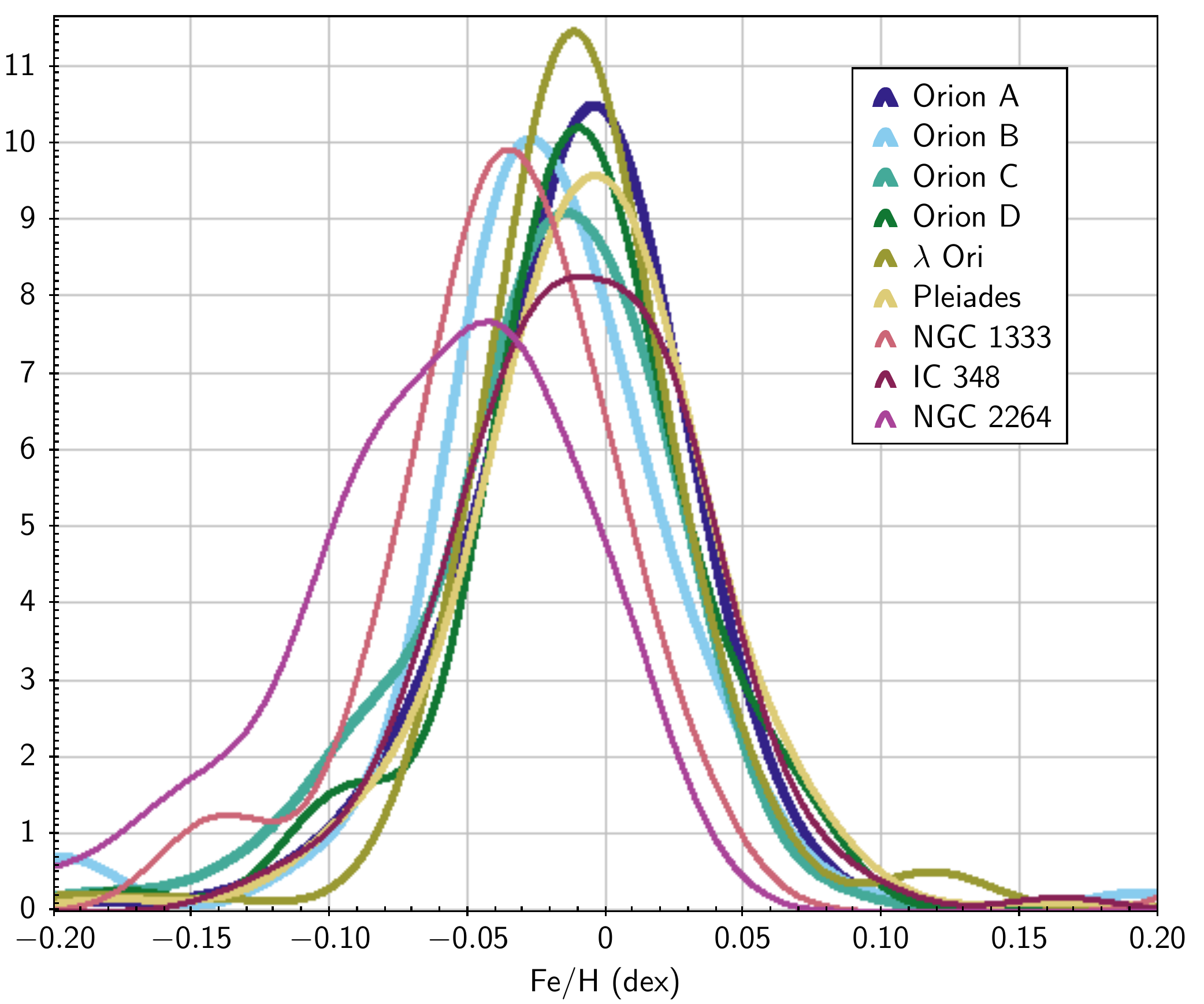}
\plotone{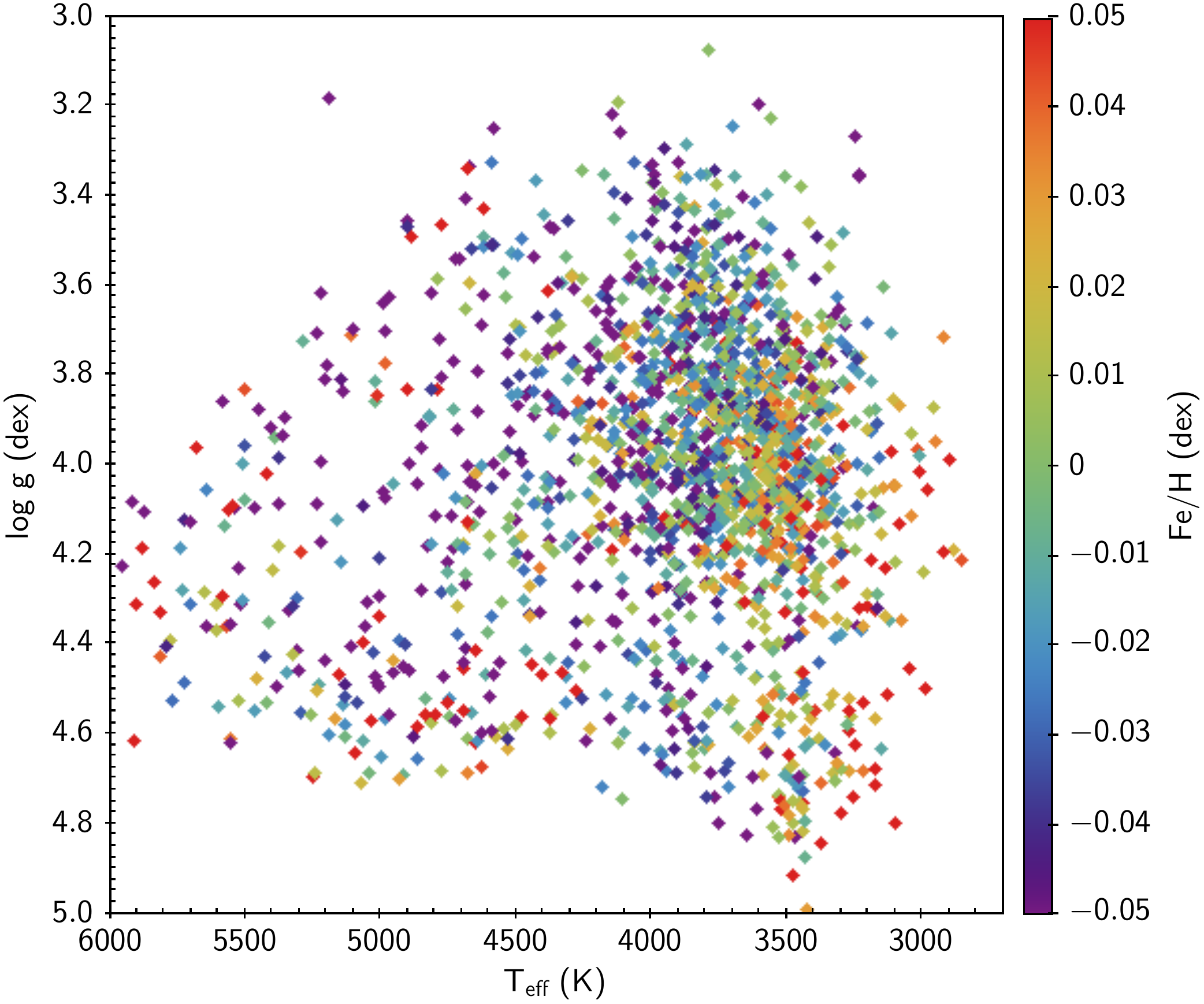}
\caption{\added{Top: distribution of Fe/H in the different samples, with \teff$<$5000 K. Middle: distribution of measured Fe/H for each young population. Bottom: Fe/H of YSOs as function of \teff\ and \logg.}
\label{fig:feh}}
\end{figure}

\added{Stars that form from the same parental cloud form with the same chemical composition; therefore they should have comparable Fe/H. Indeed, GALAH survey did not reveal any chemical inhomogeneity in any of the elements inside the Orion Complex (Kos et al. in prep), to within 0.03 dex. Furthermore, the young populations in the solar neighborhood all have near solar metallicity \citep{Dorazi2009,Dorazi2011}.

Abundances derived by the APOGEE Net do show similar agreement -- the dispersion in Fe/H abundances is within 0.04--0.05 dex in the overall sample of young stars, as well as inside any given population (Figure \ref{fig:feh}). There are slight systematic inhomogeneities, however: hotter stars, as well as stars that have lower \logg\ tend to be marginally more metal poor, whereas cooler dwarfs tend to be marginally more metal rich. The effect is larger than the typical estimated errors, but it is not greater than than a few 0.01 dex.

These systematics slightly affect the average Fe/H between different populations. While most of them have average Fe/H$\sim$0 dex, two of the youngest populations (NGC 1333 and Orion B), appear to be somewhat offset, with average Fe/H$\sim$-0.02 dex. Furthermore, the furthest cluster, NGC 2264, appears to be somewhat metal poor at Fe/H$\sim$-0.06 dex - notably in part because low mass stars cooler than $\sim$3900 K were not targeted to be included in the sample.}

\section{Discussion \label{sec:conclusions}}

Machine learning is an effective approach for classifying stellar spectra, and for the first time we have applied it to spectra of young stars to derive meaningful \teff\ and \logg\ that are interpolatable over the isochrones to determine ages. The performance of these parameters can rival the usage of photometric color-magnitude diagrams, offering two major advantages over them -- spectroscopic parameters are unaffected by extinction, or by the binary sequence. Using both of these approaches in conjunction with one another could allow for a more detailed analysis of the star forming history inside young populations.

The main limiting factor for machine learning is the existence of reliable input labels over which it would be possible to generalize a particular parameter space. However, it is not necessary to derive those labels from scratch a-priori every single time. Instead, through incremental building on previous efforts, it is possible to improve on even coarse estimates. As the parameter space gets further explored, it may be possible to add other properties to the analysis (e.g., other \replaced{metallicity}{abundance} labels that may be more meaningful in analyzing the chemical content of star forming regions, such as $\alpha$/H), and extend the grid to other type of stars (e.g., with \teff$>$8000 K), to produce a fully unified spectral model. Furthermore, if different large spectroscopic surveys with different instruments (having different resolution or different wavelength coverage) have observations of a few sources in common, it would be possible to cross-reference them relative to one another, allowing one to extract stellar parameters in a way that would have fewer systematic offsets.

\acknowledgements M.K. and K.C. acknowledge support provided by the NSF through grant AST-1449476, and from the Research Corporation via a Time Domain Astrophysics Scialog award (\#24217).

Funding for the Sloan Digital Sky Survey IV has been provided by the Alfred P. Sloan Foundation, the U.S. Department of Energy Office of Science, and the Participating Institutions. SDSS acknowledges support and resources from the Center for High-Performance Computing at the University of Utah. The SDSS web site is www.sdss.org. SDSS is managed by the Astrophysical Research Consortium for the Participating Institutions of the SDSS Collaboration including the Brazilian Participation Group, the Carnegie Institution for Science, Carnegie Mellon University, the Chilean Participation Group, the French Participation Group, Harvard-Smithsonian Center for Astrophysics, Instituto de Astrofísica de Canarias, The Johns Hopkins University, Kavli Institute for the Physics and Mathematics of the Universe (IPMU) / University of Tokyo, the Korean Participation Group, Lawrence Berkeley National Laboratory, Leibniz Institut für Astrophysik Potsdam (AIP), Max-Planck-Institut für Astronomie (MPIA Heidelberg), Max-Planck-Institut für Astrophysik (MPA Garching), Max-Planck-Institut für Extraterrestrische Physik (MPE), National Astronomical Observatories of China, New Mexico State University, New York University, University of Notre Dame, Observatório Nacional / MCTI, The Ohio State University, Pennsylvania State University, Shanghai Astronomical Observatory, United Kingdom Participation Group, Universidad Nacional Autónoma de México, University of Arizona, University of Colorado Boulder, University of Oxford, University of Portsmouth, University of Utah, University of Virginia, University of Washington, University of Wisconsin, Vanderbilt University, and Yale University.

This work has made use of data from the European Space Agency (ESA) mission Gaia (https://www.cosmos.esa.int/gaia), processed by the Gaia Data Processing and Analysis Consortium (DPAC, https://www.cosmos.esa.int/web/gaia/dpac/consortium). Funding for the DPAC has been provided by national institutions, in particular the institutions participating in the Gaia Multilateral Agreement.

The authors thank the Nvidia Corporation for their donation of GPUs used in this work.

\appendix

\section{Deep Neural Network Background \label{sec:NeuralNetwork}}
Deep neural networks (DNNs) are parameterized functions that map from an input domain to an output domain, and the parameters of which are learned through a training process. While a DNN can be used for both regression and classification tasks, this work is principally concerned with its utility for performing regressions, as stellar parameters are continuous variables.

We are utilizing a particular architecture of DNN known as a convolutional neural network (CNN). The CNN is particularly well-suited for extracting local patterns in spatial data. It has been widely used to achieve state-of-the-art results in various image recognition tasks \citep{krizhevsky2012imagenet,he2016deep,szegedy2015going,simonyan2014deep}. A CNN typically consists of a sequence of convolutional and pooling layers, followed by a set of fully connected layers. A convolutional layer transforms a subsection of the input, a pooling layer downsamples the input, and a fully connected layer is  a classic DNN or multilayer perceptron. Broadly speaking, the convolutional layers can be interpreted as a feature extractors, and the fully connected layers as the predictors. However, this is just an approximation, as the model prediction error is passed back across all its parameters through backpropagation.

\subsection{Training via backpropagation}
To train a DNN model through backpropagation, it is necessary to use an optimizer and evaluation criteria. In our case we will be using classic stochastic gradient descent for our optimizer. For our evaluation criterion we will be using mean squared error loss (MSE loss). 
    
\begin{equation}
MSE  = \frac{1}{N} \sum_{i=1}^N\sum_{j=1}^A{(Y_{ij}-\hat{Y}_{ij})^2}
\end{equation}
in which N is the number of training data points, A is the number of stellar parameters that are considered, Y is a target label, and $\hat{Y}$ is the corresponding model prediction. 
This loss is then propagated back through the model to produce gradients with respect to each model weight, which are used to update the weights to slightly reduce loss; after enough steps in the direction of the negative gradient, a local minimum is reached.

While the $MSE$ loss sums over all training data points, it is inefficient, and our case infeasible, to load all of the training data into memory at once. Instead, data is usually divided into minibatches, and at each iteration the $MSE$ loss is approximated by summing only over the datapoints in the minibatch.  Using gradient descent with these approximate losses is known as stochastic minibatch gradient descent.  A full pass through the training set is called an epoch. Typically, the training set is shuffled and after each epoch and re-partitioned into minibatches. In addition to the memory consideration, smaller minibatch sizes and shuffling the training set also have the added benefit of helping to improve model generalization; the randomness in the gradient helps to avoid settling into shallow local minima. However, the smaller the minibatch size is, the more often the parameter weights need to be updated, increasing computational expense.

\subsection{Hyperparameters}

Machine learning models typically have design parameters that cannot be learned by the network and must be chosen outside of the training process. These ``hyperparameters'' are typically adjusted by evaluating the model's performance against the held out development data.  The number of datapoints in a minibatch is one such hyperparameter.

Another hyperparameter is the learning rate (also known as step size), the coefficient scaling the gradient used in gradient descent. If the learning rate is too large, the model may fail to converge to a good local minimum. If it is too small, training can be prohibitively slow.

We also treat the dropout rate as a hyperparameter.  Dropout is a technique that helps with model generalization by using a mask to zero out a random subset of hidden units during each gradient calculation. This forces the model to distribute the learned representation of the input across all of its hidden units, yielding a more robust internal representation. A dropout rate of 0.1 implies that on any given training pass, 10\% of the hidden units on the fully connected layer are masked out: they cannot contribute to the prediction, and gradients will not propagate through them.


\section{APOGEE Net Model Code\label{sec:Code}}
\begin{lstlisting}
class APOGEE_Net(nn.Module):

    def __init__(self, num_layers, num_targets, drop_p=0.0):
        super(Net, self).__init__()
        # 3 input channels, 6 output channels,  convolution
        # kernel
        self.conv1 = nn.Conv1d(num_layers, 8, 3, padding=1)
        self.conv2 = nn.Conv1d(8, 8, 3, padding=1)
        self.conv3 = nn.Conv1d(8, 16, 3, padding=1)
        self.conv4 = nn.Conv1d(16, 16, 3, padding=1)
        self.conv5 = nn.Conv1d(16, 16, 3, padding=1)
        self.conv6 = nn.Conv1d(16, 16, 3, padding=1)
        self.conv7 = nn.Conv1d(16, 32, 3, padding=1)
        self.conv8 = nn.Conv1d(32, 32, 3, padding=1)
        self.conv9 = nn.Conv1d(32, 32, 3, padding=1)
        self.conv10 = nn.Conv1d(32, 32, 3, padding=1)
        self.conv11 = nn.Conv1d(32, 64, 3, padding=1)
        self.conv12 = nn.Conv1d(64, 64, 3, padding=1)

        # an affine operation: y = Wx + b
        self.fc1 = nn.Linear(64*133*1, 512)
        self.fc1_dropout = nn.Dropout(p=drop_p)
        self.fc2 = nn.Linear(512, 512)
        self.fc3 = nn.Linear(512, num_targets)

    def forward(self, x):
        # Max pooling over a (2) window

        x = F.max_pool1d(F.relu(self.conv2(F.relu(self.conv1(x)))), 2)
        x = F.max_pool1d(F.relu(self.conv4(F.relu(self.conv3(x)))), 2)
        x = F.max_pool1d(F.relu(self.conv6(F.relu(self.conv5(x)))), 2)
        x = F.max_pool1d(F.relu(self.conv8(F.relu(self.conv7(x)))), 2)
        x = F.max_pool1d(F.relu(self.conv10(F.relu(self.conv9(x)))), 2)
        x = F.max_pool1d(F.relu(self.conv12(F.relu(self.conv11(x)))), 2)
        x = x.view(-1, self.num_flat_features(x))
        x = F.relu(self.fc1_dropout(self.fc1(x)))
        x = F.relu(self.fc1_dropout(self.fc2(x)))
        x = self.fc3(x)
        return x

    def num_flat_features(self, x):
        size = x.size()[1:]  # all dimensions except the batch dimension
        num_features = 1
        for s in size:
            num_features *= s
        return num_features

\end{lstlisting}




\end{document}